\documentstyle[prd,aps,epsf]{revtex}

\begin{document}

\title{Zoom \& Whirl: Eccentric equatorial orbits around spinning 
black holes and their evolution under gravitational radiation reaction.}

\author{Kostas Glampedakis}
\address{ Department of Physics and Astronomy, Cardiff University
 P.O. Box 913, Cardiff, CF24 3YB, UK}

\author{Daniel Kennefick}

\address{SOCAS, 50 Park Place, Cardiff University, Cardiff, CF1 3AT, UK \\
and}

\address{Theoretical Astrophysics, California Institute of Technology, 
Pasadena, Ca 91125, USA\\
{\rm e-mail: glampedakis@astro.cf.ac.uk ; danielk@tapir.caltech.edu}}

\date{\today}

\maketitle

\begin{abstract}

We study eccentric equatorial orbits of a test-body around a Kerr black
hole under the influence of gravitational radiation reaction. 
We have adopted a well established two-step approach: assuming that
the particle is moving along a geodesic (justifiable as long as the orbital 
evolution is adiabatic) we calculate numerically the fluxes of energy and 
angular momentum radiated to infinity and to the black hole horizon, via the 
Teukolsky-Sasaki-Nakamura formalism. We can then infer the rate of change of 
orbital energy and angular momentum and thus the evolution of the orbit.      
The orbits are fully described by a semi-latus rectum $p$ 
and an eccentricity $e$. We find that 
while, during the inspiral, $e$ decreases until shortly before the orbit
reaches the separatrix 
of stable bound orbits (which is defined by $ p_{s}(e)$), in many 
astrophysically relevant cases
the eccentricity will still be significant in the last stages of
the inspiral. In addition, when a critical 
value $p_{crit}(e)$ is reached, the eccentricity begins to increase as
a result of continued radiation induced inspiral. The 
two values $p_{s}$,$~p_{crit}$ (for given $e$) move closer to each other,
in coordinate terms, as 
the black hole spin is increased, as they do also for fixed 
spin and increasing eccentricity. Of particular interest are moderate and high 
eccentricity orbits around rapidly spinning black holes, 
with $ p(e) \approx p_{s}(e) $. We call these
``zoom-whirl'' orbits, because of their characteristic behaviour involving 
several revolutions around the central body near periastron. 
Gravitational waveforms produced by such 
orbits are calculated and shown to have a very particular signature. 
Such signals 
may well prove of considerable astrophysical importance for the future LISA 
detector.   
 
\end{abstract}


\section{\textbf{Introduction}}

\subsection{\textbf{Background}}

Binary star systems, 
consisting of compact objects such as black holes and neutron stars, 
are relatively strong sources of 
gravitational radiation and are expected to be prime sources 
for the terrestrial network of kilometer-sized interferometric gravitational
wave detectors, which will soon be fully operational,
or for space-based detectors such as the proposed LISA mission \cite{lisa}. 
In order to detect gravitational radiation and subsequently study the 
physics of these sources it is absolutely necessary to have a prior 
theoretical knowledge of their dynamics. This is especially true 
because of the method of matched filtering (see \cite{mfilter} for a recent
review) that is likely to be employed in order to identify
true gravitational wave signals ``buried'' inside the detector's noisy output. 
The success of this method depends on the use of an accurate template of the 
incoming waveform.      

This paper will focus on the case of extreme mass ratio systems,
modelling a massive central object which is a spinning (Kerr) black hole 
while the orbiting body is ``light'' and compact enough to be considered 
as a test-particle moving in the gravitational field of its companion. 
There are two important reasons for studying such a model.

Firstly, due to the extreme mass ratio, the motion of the small mass can be
accurately approximated by a geodesic trajectory (which is well known
\cite{chandrabook}) and the system's gravitational radiation 
is well described by first-order black hole perturbation theory
techniques. The celebrated Teukolsky formalism \cite{teuk1} has proven 
particularly successful for this task. One thus has the opportunity to make 
a detailed study of a fully relativistic celestial system. 
For this reason, black hole perturbative 
studies can  be used as a test for numerical relativity
simulations of two-body systems (and vice versa!) \cite{numrel}. 

Secondly, in recent years there has been an accumulation of evidence of the 
existence of supermassive black holes (of mass range $ 10^6-10^9 M_{\odot}$) in 
galactic nuclei (including our own Milky Way) \cite{mbhs}. 
It is expected that scattered
stellar-mass $\sim 1-10 M_{\odot}$ compact objects from the surrounding stellar 
population will be captured by the central black hole as a result of   
two-body encounters and interactions with the inhomogeneities of the 
background gravitational
potential. The same scenario can of course work equally well for normal stars; 
however they will soon be tidally  disrupted as they approach the black hole 
\cite{rees}, \cite{sigur2} \cite{sigur}. 

Once in a bound orbit, the compact object will slowly inspiral towards the 
central black hole due to the emission of gravitational radiation. As the 
frequency of the emitted waves scales as $1/M$  (where $M$ denotes the central 
black hole's mass), they will potentially lie in the low-frequency band 
($10^{-5} - 10^{-1} Hz$) where LISA will have its peak sensitivity. 
The (still uncertain) estimate for the number of such events is around 
$1/\mbox{year}$, or better, out to a distance of $1$Gpc and they should be 
detectable by LISA, with typical signal to noise ratios of $10-100$, 
\cite{rees}, assuming the use of some optimal filtering technique, such as 
matched filtering \cite{mfilter}. 

A huge payoff from direct observations of 
such events is to be expected, \textit{provided} we have an accurate a priori
description of the emitted waveform. In principle, for instance, the black 
hole parameters (masses,spins) can be measured to a high accuracy. Similarly, 
information on the mass-function of compact stellar populations in galactic 
nuclei could be provided. Because the total luminosity of the source depends 
only on its mass it may be possible to work out the distance to the 
source, which would be very useful to cosmologists. Moreover, one might be able
to identify
the massive object as a Kerr black hole, as opposed to some other, 
more speculative object (for example a boson star \cite{boson}). 
This was demonstrated by Ryan \cite{ryan1} who showed in detail how the 
massive body's multipole moments are encrypted in the waveform emitted by an 
orbiting particle. 
In the near term, precise numerical results in the low-mass-ratio limit
will be useful for testing the accuracy of post-Newtonian (PN) derived 
templates aimed at ground based detectors like 
LIGO \cite{chapter}, \cite{pade}.

LISA will monitor the last year of inspiral of a compact body into a massive black
hole by tracking the phase of the emitted waveform. It has been suggested that for
astrophysically likely scenarios, drag forces operating on the orbiting body due
to gas accreting onto the black hole, will operate on a timescale much longer than
the radiation reaction timescale of the particle \cite{narayan}, \cite{narayan1}.
Based on requirements that the initial highly eccentric orbit in which the 
particle finds itself as the result of some scattering event should have a 
small enough periastron so that the radiation reaction timescale is shorter 
than the timescale
for a second scattering event at apastron, we expect that
the initial periastron  should be rather close, so that $r_{p} < 20M $ 
while the apastron will extend to a distance 
$10^4 -10^6 M$ \cite{sigur}, \cite{phinney},\cite{freitag}. 
Newtonian order estimates suggest that although radiation reaction will 
considerably reduce this enormous 
initial eccentricity during the course of the inspiral, the
eccentricity will remain finite and non-negligible when the particle enters the 
strong-field region of interest to this paper (see section VB below). Exactly
how much eccentricity remains will depend critically on the initial periastron
distance and is largely insensitive to the initial apastron distance (and thus
to the initial eccentricity).
   
We can thus argue that for a sufficiently bound orbit the system of the 
massive black hole and the orbiting compact object will evolve under its 
own spacetime dynamics. This tends to justify our 
``black hole plus particle'' model.
However, even in this simplified picture there are problems. 
The particle, in general, will move along a non-equatorial eccentric orbit 
(as the galactic central stellar population is almost spherically symmetric,
capture orbits of arbitrary inclination are to be expected). The Teukolsky
formalism cannot, at present, deal with such orbits, for reasons discussed
in \cite{scott_circ}, in particular the problem of determining the rate of change
of the ``Carter constant'' of the motion due to the emission of gravitational
waves (much effort, towards this goal, is being focused on building a framework
for calculating the gravitational self-force acting on the orbiting particle
\cite{force}). For this reason we restrict our attention to equatorial
orbits around the central body. In such a case the rate of change of the
orbital parameters can be deduced by reading the gravitational wave fluxes
for the energy and angular momentum at infinity and the black hole horizon.

There is, however, a factor that cannot be accounted for by the previous
flux-balance argument which lies at the heart of our approach. 
As has been observed recently \cite{pfenning}, \cite{lior} the gravitational 
self-force contains a conservative piece which is not associated with any 
radiation emission. Although the effect of this conservative force is 
negligible (scaling as $\sim \mu^2$) over short timescales (say, one orbital 
period), it is conceivable that the same will not be true for the accumulated 
effect after $10^4-10^5$ orbits \cite{lior} (this is, roughly, the number of 
orbits that LISA will record).

Another issue that has risen recently concerns the possible difficulty in 
defining the notion of adiabaticity and averaged flux for generic, i.e. 
eccentric and non-equatorial, orbits in Kerr spacetime. This is related to the 
belief that generic orbits have no well-defined orbital periods as they show an 
apparently non-periodic behaviour. For that reason, it has been suggested 
\cite{scott_proc} that an ``ergodicity'' criterion would 
be more appropriate. However, recent work suggests \cite{wolfram} that it is 
possible, after all, to rigorously define  (by means of Hamilton-Jacobi theory) 
a triplet of fundamental frequencies for generic Kerr orbits. Consequently, one 
may still be able to define adiabaticity for these orbits too.

Serious complications can also arise at the level of free motion, where 
radiation reaction is neglected. In general, the small body will have its own 
intrinsic spin. In such a case, due to the coupling of the particle's spin with 
the background gravitational field, the motion is no longer geodesic. Although 
the (specific) spin magnitude is small, i.e. $S \sim {\cal O}(\mu) $,  
spin-induced effects could become important over timescales much longer than, 
say, one orbital period. A particularly dramatic possibility is that when the 
test-particle is allowed to have spin, ``chaotic'' features may appear in the 
orbital motion \cite{chaos}. Presently it is unclear whether chaotic behaviour 
will be important for extreme mass ratio systems likely to be observed by LISA. 

When radiation reaction is ``switched-on'' in the spinning particle case, 
one finds, not surprisingly, that the 
radiative fluxes at infinity and the horizon are inadequate for determining the
evolution of the orbit. This is, in part, due to the fact that there is no 
known analog of the Carter constant (so there is one less constant of motion 
available), and also due to the existence of additional spin-degrees of freedom. 
A Newtonian order, weak-field estimation for the radiative change of the spin 
has been worked out by Apostolatos {\em et.al.} \cite{apostolatos}. Some 
speculations of what could happen to circular orbits under strong field 
conditions can be found in  \cite{tanaka},\cite{kgspin}. For generic orbits, 
most likely only a self-force calculation will be able to describe the full 
orbital evolution. 

As we are still far away from dealing with all of these challenges we make two 
major simplifications for this paper, that the orbiting particle has no spin and 
that
it always remains in an equatorial orbit.
Although Ryan has shown \cite{ryan3}, for orbits in the weak field region,
that non-equatorial orbits are forced by radiation reaction towards
becoming retrograde equatorial orbits, the effect is small. The effect remains
small even in the strong field region, as was recently shown by Hughes 
\cite{scott_circ}.
Precisely equatorial orbits, pro- or retrograde, will remain 
equatorial under radiation reaction. Therefore it is reasonable to expect that
detectors such as LISA will actually observe signals from particles in near
equatorial orbits.

Previous studies have shown that
slightly eccentric orbits of particles around Schwarzschild \cite{apostolatos2} 
and Kerr black holes \cite{dk2} 
and arbitrarily eccentric orbits around Schwarzschild black holes \cite{cutler} 
decrease in their eccentricity  until shortly before the innermost stable 
circular orbit (ISCO) when a point is reached after which the eccentricity
begins to increase.
The present work comes as an additional piece to this series of papers. 
Specifically, we 
consider equatorial eccentric orbits of particles around a Kerr black hole and 
study 
their evolution under gravitational radiation reaction. This class of orbits is
not exactly what we would expect in reality, but it is an important step towards
a more realistic view of gravitational waves from this type of low-frequency
source, because it includes two very important features which we know will be 
present in all or most sources, black hole spin and orbital eccentricity.
Our study is, in this respect, a useful companion piece to Hughes' 
discussion of non-equatorial (but circular) 
orbits \cite{scott_circ}. Eccentric equatorial orbits were first investigated by 
Shibata \cite{shibata_ecc} who calculated fluxes and waveforms, without, 
however, discussing the 
impact of radiation reaction on the orbital motion. Our approach is similar to
previous papers investigating eccentric orbits around non-spinning black holes 
\cite{cutler}, and nearly circular orbits around spinning black holes
\cite{dk2} and the results are qualitatively similar to those of both
papers. In addition, we compute gravitational waveforms produced by moderate/high 
eccentricity, strong-field orbits (not discussed in Shibata's study 
\cite{shibata_ecc}) 
which we call ``zoom-whirl''  orbits. We find that these waveforms are
a very characteristic, though complex, signal that might be important from an 
observational point of view for the planned  LISA space antenna. 

In this paper we focus on the final part of the inspiral, when the particle is
at small radii, relatively close to the last stable bound orbit. In
consequence we deal with orbits with moderate eccentricities, between 0.1 and
0.7. In a future paper \cite{paperII} we intend to study the full
inspiral, thus expanding our scope to cover orbits with large radii and
larger eccentricities, on the order of 1. 
In that paper we plan to present wavetrains and spectra associated with a
long stretch of the inspiral, covering many orbital periods,
along the lines of \cite{scott_insp}.

Our results in this paper can be summarised as follows. Moderate eccentricities
will be a feature of the signals from many inspiralling compact binaries right
up to the final plunge.  Immediately before plunge there
will be an eccentricity increasing phase in all cases, particularly noticeable
for retrograde orbits.  The total amount of eccentricity gained in this phase
will generally be small, on the order of 10\% or less for low-eccentricity ($e<0.1$)
prograde orbits, but perhaps as much as 50\% for low-eccentricity retrograde orbits.
Orbits with moderate eccentricities will gain much less in eccentricity.
Where $e>0.3$ and the orbit is prograde, zoom-whirl 
features will be prominent in the waveform in the very last stages of the 
inspiral. Where these orbits are observed from a position away from the
polar axis of the source there will relatively strong high-frequency component
to the signal due to beaming of higher multipoles in the radiation in the
direction of the orbiting particle's motion. One expects that
these signals will present particular problems for signal analysis, a
situation which may be ameliorated when a positive detection of the source
has been made during the earlier part of the inspiral when the waveform,
though highly eccentric, will be less complex.


\subsection{\textbf{Organisation of the paper}}

The remainder of this paper is organised as follows. In Part II we discuss the 
geodesic motion of eccentric equatorial orbits (Sections IIA \& IIB), paying 
particular attention to the so-called ``zoom-whirl'' orbits (Section IIC). 
In those sections we define useful orbital parameters such as the semi-latus 
rectum $p$ and the eccentricity $e$. Some analytic approximations on the orbital 
periods and number of revolutions for a particle in an orbit  close to becoming 
dynamically unstable, are presented in Section IID. Sections IIIA \& IIIB of 
Part III  contain a review of the 
Teukolsky-Sasaki-Nakamura formalism for the calculation of gravitational 
waveforms and fluxes.
In Section IIIC, we give a preliminary discussion on the orbital evolution under 
radiation reaction (definition of adiabaticity, general formulae for the rate of 
change of orbital parameters). Part IV is entirely devoted to analytic results.
Section IVA discusses the weak-field limit for the orbital parameter's rates of 
change.In Section IVB we derive an approximate formula relating the energy flux 
to the angular momentum flux, emitted by orbits close to becoming unstable. 
We subsequently use this formula to find strong-field approximate expressions 
for the rate of change of $p$ and $e$. In Section IVC we study the particularly 
interesting family of (equatorial) horizon-skimming orbits that can exist 
around a rapidly rotating black hole. The main (numerical) results of this paper 
are contained in Part V. In Section VA, we sketch the methods used in our 
numerical code and, moreover, give estimates for the various introduced errors. 
In Section VB we give results on the averaged rate of change of the parameters 
$p,e$ (which determine the evolution of any given orbit). This allows us to draw 
conclusions for the ``global''  behaviour of bound equatorial orbits under the 
influence of radiation reaction. Section VC contains calculations of 
waveforms generated from some zoom-whirl orbits. Part V ends with a presentation 
of the spectral content of the radiation emitted at infinity and at the black 
hole horizon (Section VD). Our conclusions are summarised in Part VI, where we 
also discuss prospects for future work. Tables with samples of our numerical 
data can be found at the end of the paper's main body. Three Appendices are 
devoted to some technical details. Throughout this paper we have adopted 
geometrised units ($c=G=1$).      
 
\section{\textbf{Geodesic motion}}

\subsection{\textbf{Equations of motion}}

\begin{figure}[tbh]
\centerline{\epsfysize=5cm \epsfbox{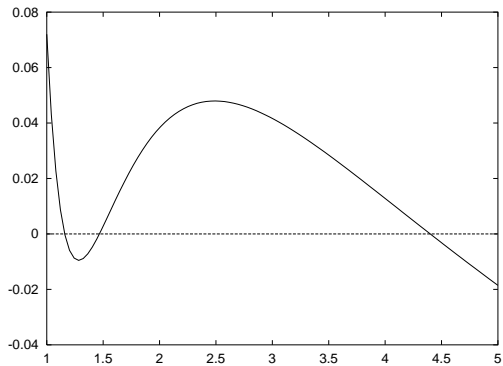}}
\vspace{0.3cm}
\caption{The radial potential $V_r$ (in units of $M^{-4}$) as a function
of $r$ (in units of $M$) for $p=2.2M,~ e=0.5$. The black hole spin is $a=0.99M$.  
Motion is permissible at the regimes where $V_r \geq 0$. It is easy to 
distinguish the apastron at $r_a = 4.4M$ and the periastron at $r_p= 1.47M$. 
The event horizon is at $r_{+}= 1.141M$.}
\label{Vr}
\end{figure}

We start by considering a test body moving in a Kerr gravitational field. 
For the moment, we neglect any radiation reaction effects and focus on 
purely geodesic motion. Working in the usual Boyer-Lindquist coordinate frame, 
the equations of motion, specialised for an equatorial orbit, are given by 
\cite{bardeen}, 
\begin{eqnarray}
r^2 \frac{dr}{d\tau}&=& \pm (V_r)^{1/2} \;, 
\label{eom1}
\\
\nonumber \\
r^2 \frac{d\phi}{d\tau}&=& V_{\rm \phi} \equiv -(aE -L) + \frac{aT}{\Delta} \;,
\label{eom2}
\\
\nonumber \\
r^2 \frac{dt}{d\tau}&=& V_{\rm t} \equiv -a(aE -L) + \frac{(r^2 +a^2)T}{\Delta} \;,
\label{eom3}
\\
\nonumber \\
\theta(\tau)&=& \pi/2 \;,
\end{eqnarray}
where 
\begin{math}
T=E(r^2+a^2) -La , \ \ 
V_r= T^2 -\Delta(r^2 + (L -aE)^2 ),  \ \
\Delta= r^2 -2Mr + a^2 .
\end{math}
The two constants of motion $E$, $L$ denote the orbit's specific energy and 
$z$-component of angular momentum (for notational simplicity we drop the subscript
$z$ for the angular momentum). We have prograde (retrograde) orbits 
according to whether $L > 0 ~(< 0)~$ (note that at certain points, where there is 
no danger of
confusion, we shall label retrograde orbits by a negative value for the spin 
parameter $a$). Moreover, since we shall be discussing 
bound orbits, $ 0<E<1 $. A general bound equatorial orbit can be equivalently 
described \cite{chandrabook} either by the constants $E$ and $L$ or by a 
semi-latus rectum $p$ and an eccentricity $e$ 
(with $ 0 \leq e < 1$). The restriction on the values of $p$ is discussed 
below. We define these parameters in terms of the two turning points of the 
orbit ($r_p$ is the periastron and $r_{a}$ the apastron, see Fig.~\ref{Vr} for
a typical illustration),
\begin{equation}
r_p = \frac{p}{1+e}, \ \
r_a = \frac{p}{1-e} \;.
\label{tpoints}
\end{equation}
A turning point $r_{\rm o}$ by definition satisfies $V_r(r_{\rm o})=0$, 
or explicitly, 
\begin{equation}
(E^2 -1)r^3 + 2Mr^2 -( x^2 + a^2 + 2aEx)r + 2Mx^2 =0 \;,
\label{turneq}
\end{equation}
where we have further defined $ x= L -aE$.
Writing this polynomial in the form $ (E^2-1)(r-r_p )(r-r_{a})(r-r_3)$ 
we can immediately write an expression for the energy,
\begin{equation}
E= \left[ 1 -\left (\frac{M}{p}\right ) (1-e^2) \left \{  
1 -\frac{x^2}{p^2}(1-e^2) \right  \} \right ]^{1/2} \;.
\label{energy}
\end{equation}
Similarly, the third root $r_3$ of (\ref{turneq}) is found to be
\begin{equation}
r_3= \frac{2M(1-e^2)x^2}{p^2 (1-E^2) } \;.
\label{r3}
\end{equation}
It then follows that,
\begin{equation}
x^2= \frac{ -N(p,e) \mp \Delta^{1/2}_x(p,e) }{2F(p,e)} \;.
\label{x2}
\end{equation}
The explicit forms of the functions $N,F$ and $\Delta_x $ 
are given in Appendix A.
In this expression, the upper (lower) sign corresponds to a prograde
(retrograde) orbit. The same convention will be followed throughout
the paper.

The radial coordinate can be parameterised as
\begin{equation}
r(\chi)= \frac{p}{1+e\cos\chi} \;,
\label{rfunc}
\end{equation}
where $\chi$ is a monotonically varying parameter, running from $\chi=0$ 
(at $r=r_{p}$) to $\chi=\pi$ (at $r=r_{a}$) and finally up to
$\chi= 2\pi$ (back to $ r= r_{p}$ ). The radial motion can be separated 
into two distinct branches, namely, the motion from $r_{p}$ to 
$r_{a}$, and the ``inverse'' motion from $r_{a}$ back to 
$r_{p}$ again.
Integration of (\ref{eom3}) gives
\begin{equation}
t(r)= \left \{\begin{array}{ll}
              \hat{t}(r) & \mbox{first branch} \\
              T_{r} - \hat{t}(r)  & \mbox{second branch}
             \end{array} 
     \right.
\label{teq1} 
\end{equation}
where
\begin{equation}
\hat{t}(r)= \int_{r_1}^{r} \frac{1}{r^2} \left (\frac{dr}{d\tau} \right)^{-1}
\left [  ax + \frac{r^2 +a^2}{\Delta}(Er^2 -ax) \right ]dr 
\label{teq2}
\end{equation}
We have also denoted as $T_r$ the period of the radial motion. 
For the $\phi$-motion we similarly write,
\begin{equation}
\phi(r)= \left \{ \begin{array}{ll}
               \hat{\phi}(r) & \mbox{first branch} \\
               \Delta \phi - \hat{\phi}(r)  & \mbox{second branch} \;,
              \end{array}  
      \right.
\label{pheq1} 
\end{equation}
where
\begin{equation}
\phi(r)= \int_{r_1}^{r} \frac{1}{r^2} \left (\frac{dr}{d\tau} \right )^{-1}
\left [ x + \frac{a}{\Delta}(Er^2 -ax) \right ]dr \;.
\label{phieq1}
\end{equation}
and $\Delta\phi$ is the change of $\phi$ during an interval $T_r$.
Both integrands in (\ref{teq1}), (\ref{pheq1}) are (unphysically) divergent 
at the turning points, an undesirable feature in a numerical 
calculation. This difficulty can be avoided by choosing $\chi$ as the 
integration parameter. Using
\begin{equation}
\frac{dr}{d\tau}=  \frac{e\sin\chi}{p}[ x^2 + a^2 +2xaE
 -\frac{2Mx^2}{p}(3 +e\cos\chi)]^{1/2}  \;,
\end{equation}
we get 
\begin{equation}
\phi(\chi)= \int_{0}^{\chi} d\chi^{\prime} \frac{ \tilde{V}_{\rm \phi}
(\chi^{\prime},p,e)}{J(\chi^{\prime},p,e) \tilde{V}_{\rm r}^{1/2}
(\chi^{\prime},p,e)}  \;,
\label{phichi}
\end{equation}
\begin{equation}
t(\chi)=  \int_{0}^{\chi} d\chi^{\prime} \frac{ \tilde{V}_{\rm t}
(\chi^{\prime},p,e)}{J(\chi^{\prime},p,e) \tilde{V}_{\rm r}^{1/2}
(\chi^{\prime},p,e)}  \;,
\label{tchi}
\end{equation}
where 
\begin{eqnarray}
\tilde{V}_{\rm r}(\chi, p, e) &=& x^2 + a^2 + 2axE -\frac{2Mx^2}{p}
(3 + e\cos\chi) \;,
\label{tilVr}
\\
\tilde{V}_{\rm \phi}(\chi,p,e) &=& x + aE -\frac{2Mx}{p}(1+e\cos\chi) \;,
\\
\tilde{V}_{\rm t}(\chi,p,e) &=& a^2 E  -\frac{2aMx}{p}(1+e\cos\chi) +
\frac{Ep^2}{(1+e\cos\chi)^2} \;,
\\  
J(\chi,p,e) &=& 1 -\frac{2M}{p}(1+e\cos\chi) +\frac{a^2}{p^2}(1+e\cos\chi)^2 \;.
\end{eqnarray}
The integrand quantities in (\ref{phichi}), (\ref{tchi}) are well behaved
and, moreover, these equations are valid for both branches of the radial 
motion. The radial period is simply given by $ T_r= t(2\pi)= 2t(\pi) $, and 
similarly, $\Delta\phi \equiv \phi(2\pi)=2\phi(\pi) $.

A general bound equatorial orbit is the combination of two separable motions:
the radial motion which is, strictly speaking, periodic (in the sense
that the radial coordinate returns to its original value after a certain
time interval $T_r$ has elapsed) and the azimuthal motion which is not purely 
periodic (in the sense that the $\phi$-coordinate monotonically increases but,
nevertheless, the orbit returns to the same configuration after $\phi$ has increased
by some value $ \Delta\phi$). The former motion is known in classical mechanics 
\cite{goldstein} as ``libration'' while the latter motion is called 
``rotation''. For such a combination of motions, it is generally known that 
there is a fundamental period (the period of libration) which fully describes
the motion (see Appendix B for further details). 
We shall, therefore, call $T_r$ the {\it orbital period}. The fact that the 
orbit is periodic in a strict sense will enable us to rigorously define 
adiabaticity  when radiative effects are to be included.

In line with the foregoing discussion, we define the 
{\it orbital frequency} to be $\Omega_r = 2\pi/T_r$. We can similarly 
refer to the frequency of the $\phi$-motion as 
$\Omega_{\phi} = \Delta\phi/T_r $. The gravitational waves emitted 
by our systems will have frequencies which depend on these orbital frequencies. 
Below we shall see how they form a spectrum of discrete frequencies 
parameterised by the following wave numbers: $\ell$, which identifies the 
multipole of the emitted waves ($\ell=2$ for quadrupole, for instance), 
$m$ which runs from $-\ell$ to $+\ell$ and $k$ which counts the harmonics 
created by the linear composition of the two orbital frequencies. 
The frequency of the waves emitted by a given harmonic $k$ of a given 
multipolar contribution $m$ is

\begin{equation}
\omega=k \Omega_r + \frac{m \Delta\phi}{2\pi} \Omega_r .
\end{equation}

In the calculational scheme to be outlined below we will evaluate the fluxes of
energy and angular momentum which are carried by waves of a given frequency 
(that is, a given multipole and harmonic of the frequency spectrum) and sum the
fluxes for all frequencies of the discreet spectrum to get the total radiated
fluxes of these quantities.


\subsection{Separatrix curve}

\begin{figure}[tbh]
\centerline{\epsfysize=5cm \epsfbox{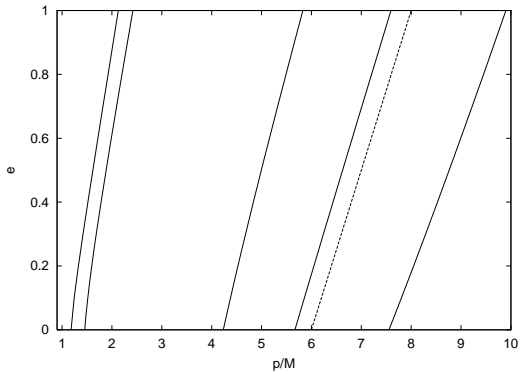}}
\vspace{0.3cm}
\caption{Separatrices on the ($p,e$) plane for a variety of black hole
spins. From left to right: $ a/M=  0.999, 0.99, $ 
$ 0.5, 0.1, 0 (\mbox{dashed}) $, $-0.5$. 
As $a \to M $ the prograde separatrix goes to the limiting value 
$ p_{s} \to M(1+e) $.}
\label{sepx}
\end{figure}

In general eqn. (\ref{turneq}) has three distinct real roots. The
case with $r_p=r_3$ corresponds to a marginally stable orbit: once at the 
periastron, the particle  will enter into a circular orbit of radius  
$r_{\rm isbo}=r_p=r_3$ (ISBO stands for Innermost Stable Bound Orbit). 
At this stage the orbit has become unstable, 
so that a slight inwards ``push'' will drive the particle to 
catastrophically plunge into the black hole. Therefore, stable bound orbits 
should satisfy $r_3 < r_p$.  This translates to the inequality,  
\begin{equation}
x^2(1+e)(3-e) < p^2 \;.
\label{sepax}
\end{equation}
We can imagine a division of the $(p,e)$  plane into regions of stable
and unstable orbits. The boundary curve $p_{s}(e)$ satisfying the equality
in (\ref{sepax}), defines the separatrix of bound orbits.  In Fig.~\ref{sepx}
we illustrate separatrices for a variety of black hole spins. A sample of
numerical data used to generate this figure can be found in Table~\ref{tab_sepax}. 
As one might have anticipated, spinning up the black hole will cause the 
separatrix curve  for prograde (retrograde) orbits to move to the left (right) 
with respect to the Schwarzschild curve $p_{s}(a=0)= (6+2e)M$ \cite{cutler}.  
This behaviour can be seen most easily by a slow rotation approximation to 
(\ref{sepax}). At leading order we find,
\begin{equation}
p_{s}=(6+2e)M \mp 8a \left [ \frac{1+e}{6+2e} \right ]^{1/2} + {\cal O}(a^2) \;.
\label{slowsepax}
\end{equation}
On the other hand, as can be verified by direct substitution in 
(\ref{sepax}), for extreme rotation ($a=M$) the prograde separatrix 
becomes $ p_{s}(e)= M(1+e)$, i.e. for all eccentricities, the periastron
``descends'' into the black hole ``throat'' at $r=M$, but is still
separated by a finite proper distance from the horizon itself \cite{bardeen}.

\begin{figure}[tbh]
\centerline{\epsfysize=7cm \epsfbox{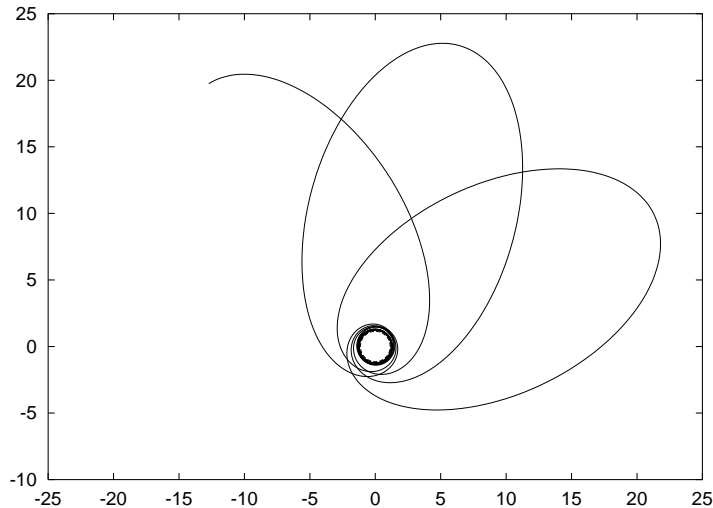}}
\vspace{0.3cm}
\caption{A zoom-whirl orbit with $p=2.35M, e=0.9$ around an $a=0.99M$ Kerr
black hole. In this figure, the particle has performed more than twenty 
revolutions in less than three orbital periods. The periastron is at 
$r_p= 1.237M $, located close to the hole's event horizon at 
$r_{+}= 1.141M $ (denoted by the dashed line). The ISBO radius 
is $r_{\rm isbo}= 1.216M$. }
\label{zoom}
\end{figure}


\subsection{\textbf{Zoom-Whirl orbits}}

From the short discussion in the previous Section one can imagine that 
as the orbit gradually approaches the separatrix, the particle will spend
a considerable amount of its orbital ``life''  close to the periastron 
(see Fig.~\ref{zoom}). An approximation for $T_{\rm r}$ as $p \to p_{s}$,
derived in the following Section, gives
\begin{equation}
T_{\rm r} \sim -\ln(p-p_{s}) \;,
\label{nearTr}
\end{equation}
which shows that the period will grow (and eventually diverge) as the 
separatrix is approached. In that region, the particle will trace a 
quasi-circular path before being reflected back to the apastron. 
Such behaviour will be particularly prominent for high eccentricity
orbits: the particle will ``zoom in'' from its apastron position, and perform
a certain number of quasi-circular revolutions (``whirls'') reaching the 
periastron (which should have a value close to 
$r_{\rm isbo}(e)= p_{s}(e)/(1+e) $). Finally, the particle will be reflected 
and ``zoom out'' towards the apastron again. We shall heuristically 
(but quite descriptively) name these orbits  ``Zoom-Whirl'' orbits.
They resemble a set of orbits known in the literature as homoclinic orbits 
\cite{rachel}. 
Zoom-whirl orbits can exist in both Kerr and Schwarzschild geometries, and
their potential significance for the detection of gravitational waves by space-based
instruments was first pointed out some years ago by Curt Cutler 
and Eric Poisson\footnote{The name ``Zoom-Whirl'' originated with the work of these
two at Caltech. It may have been suggested by Kip Thorne.}, 
who concluded that the small number of whirls in the Schwarzschild case made
the phenomenon less interesting for spinless central bodies.
But as we shall shortly see, they are more pronounced in 
the case of near-extreme Kerr black holes, for prograde orbits. 
A typical example of such an orbit is illustrated 
in Fig.~\ref{zoom}, for the case of a rapidly spinning ($a=0.99M$) black hole.     

\begin{figure}[tbh]
\centerline{\epsfysize=5cm \epsfbox{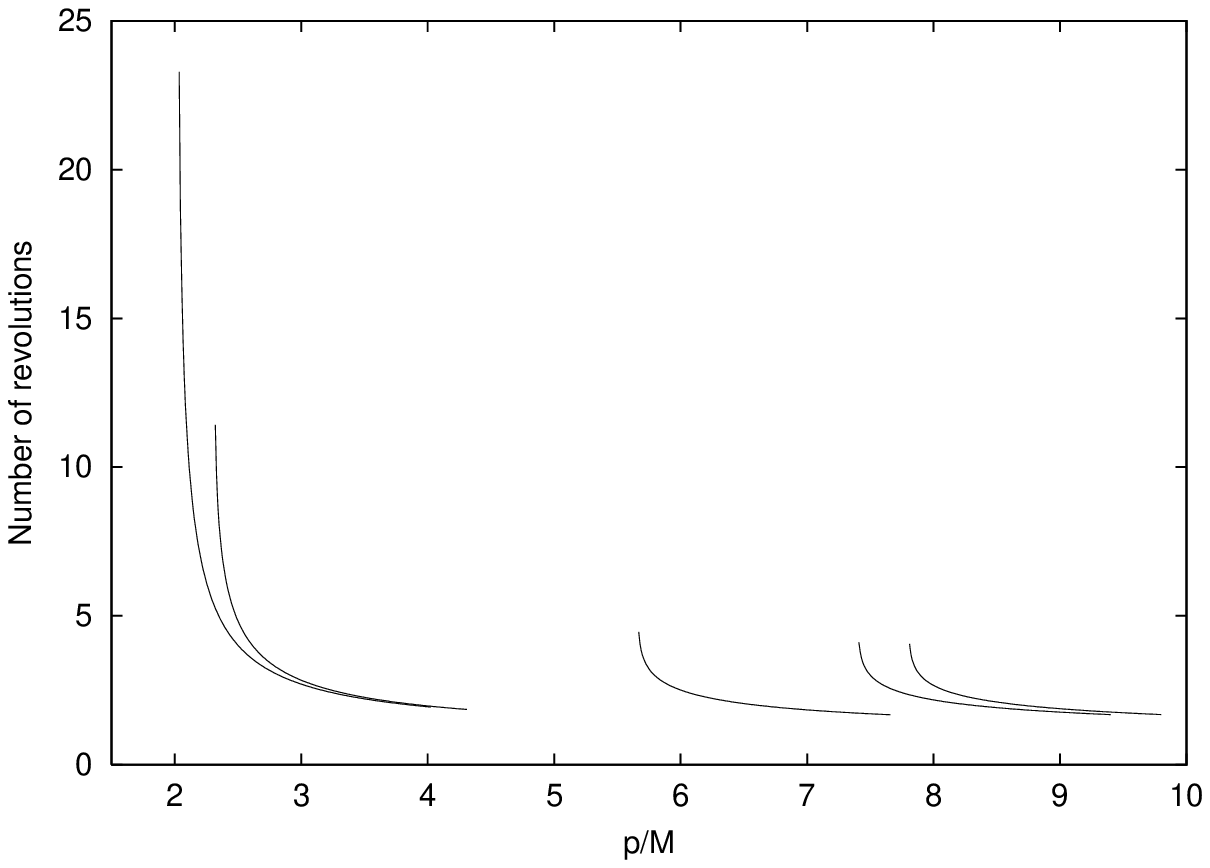}  }
\centerline{\epsfysize=5cm \epsfbox{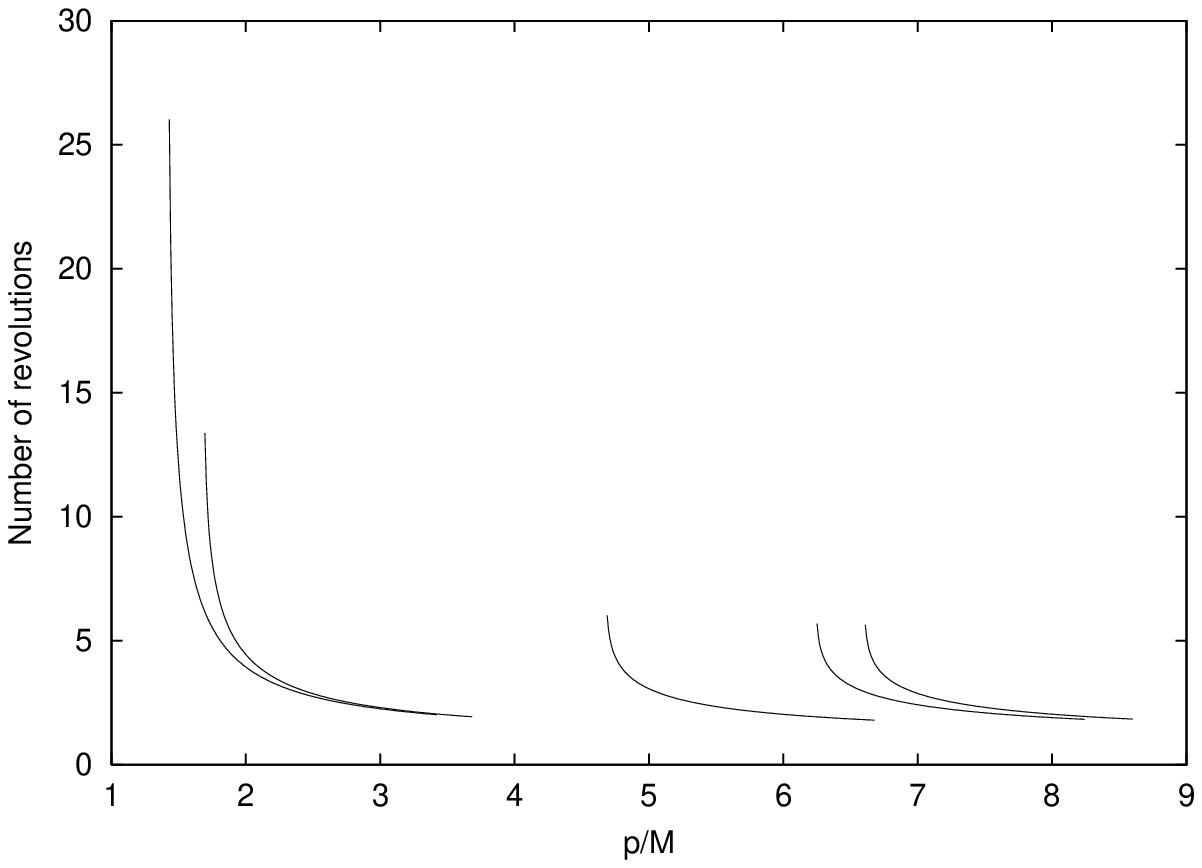}}
\vspace{0.3cm}
\caption{Number of revolutions as a function of the semi-latus
rectum $p$ for fixed eccentricity $e=0.9$ (top frame) and $e=0.3$ (bottom frame). 
The black hole spin is, from right to left, $a/M= 0, 0.1, 0.5, 0.99, 0.999$. 
Each curve terminates at a point located $\delta p=0.01M$  away from the 
respective separatrix value. Evidently, zoom-whirl orbits are expected to be more 
pronounced for rapidly rotating black holes.}
\label{revol}
\end{figure}

It is straightforward to calculate the total number of azimuthal revolutions 
$N_{\rm r}= \Delta\phi/2\pi$ during one orbital period, by numerically 
integrating (\ref{phichi}). Results obtained by such a calculation are presented 
in Fig.~\ref{revol}. In this figure we have considered orbits of a given 
eccentricity ($e=0.9$ and $e=0.3$) and for a variety of black hole spins. For all 
depicted cases, the smallest value of $p$ resides at the same distance $\delta p$ from
the corresponding separatrix value $p_s(e)$. As can be seen, the number
of revolutions increases as the separatrix is approached, in agreement with
our intuitive expectations. In fact, an approximate formula (valid for $p \to 
p_{s}$) derived in Section IID shows that, 
\begin{equation}
N_{\rm r} \sim -\ln(p-p_{s}) \;.
\label{Nrev1}
\end{equation}
We can furthermore deduce that the ``whirling'' of the particle near the
separatrix becomes more pronounced as the black hole spin increases.
Although for small and moderate spins $N_r$ stays close to the corresponding
Schwarzschild  value, it grows rapidly as $a \to M$, basically due to the intense
``frame-dragging'' induced by the black hole's rotation in the very strong 
field region close to the horizon which can be reached by particles in prograde 
orbits. The overall behaviour can 
be understood as an extreme example of perihelion advance (as in the celebrated
case of the planet Mercury).

In principle, as (\ref{Nrev1}) suggests, the number of revolutions can be
made arbitrarily large irrespective of the black hole spin, provided the
particle approaches the separatrix sufficiently closely. However, as we 
discuss in Section IIIC  the adiabatic assumption upon which our formalism
relies breaks down in this regime.
Sufficiently close to the separatrix, radiation reaction makes a 
significant correction to the
particle's motion in each orbital period. Before long this causes the particle 
to cross the separatrix and plunge into the black hole. 
These transition/plunging regimes have 
been studied recently by Ori and Thorne \cite{ori} for the case of 
circular equatorial orbits 
in the Kerr geometry. More relevant to the present discussion is the work of 
O'Shaughnessy 
and Thorne \cite{o's} which concerns the transition regime of zoom-whirl orbits. 
They show that for the case of an extreme Kerr black hole and 
eccentricity close to unity, the particle may experience more than 20 whirls per 
orbit before plunging. These have to be added to the number of whirls performed 
during the adiabatic phase of the orbit. 

In a realistic scenario, we should not expect 
to find (apart from chance cases where the particle enters a 
near-separatrix orbit as a result of its initial scattering)
very high eccentricity zoom-whirl orbits, as it is well known that
the orbit has a general tendency to circularise \cite{peters}. However, despite
the decrease in eccentricity over the greater part of the inspiral, a substantial 
amount of eccentricity will survive, in many cases, up to the point where the 
orbit is about to plunge. These orbits will probably become zoom-whirl orbits, 
especially when a rapidly spinning black hole is involved and especially for
prograde orbits. Keep in mind that
many scattered particles will be in highly non-equatorial orbits. Zoom-whirl
behaviour should also be seen in these cases as there is still a separatrix 
present, close to which the particle can spend a considerable amount of time.  

A compact body in a zoom-whirl orbit will spend a considerable fraction of the 
orbital period in strong field regions (it can even travel close to the event 
horizon if the central black hole is spinning rapidly enough) and hence will 
radiate strongly. 
Our numerical results together with analytic approximations, reveal 
that a good fraction of the averaged flux is radiated during the motion near 
the periastron. As the orbit approaches the separatrix it tends to radiate as 
if it was a circular orbit of angular frequency $\Omega_{\phi}$ 
(see also \cite{cutler} for a similar statement in the Schwarzschild case). 
This is clear evidence that most of the radiation is coming from the whirl part 
of the orbit, during which the radius hardly changes and there is a single 
dominant frequency characterised 
by the azimuthal ($\phi$-dependent) orbital period. However, the most important 
feature of a zoom-whirl orbit is the characteristic form of the gravitational
wave it emits, which is a series of rapid ``quasi-circular'' oscillations separated 
by relatively ``quiet'' intervals. In Section VC  below we calculate some 
waveforms of this type. 

\subsection{\textbf{Approximations near the separatrix (I)}}

Orbits that reside near the separatrix of the $(p,e)$ plane are amenable to
analytic approximation, basically due to the fact that the turning point 
$r_{p}$ is close to a local minimum of the radial potential $V_r$.
In this Section we derive approximate expressions for $T_{\rm r}$ and $\Delta\phi$. 
We already know that (see eqn. (\ref{tchi})), 
\begin{eqnarray}
T_{\rm r} &=& 2 \int_{0}^{\pi} d\chi \frac{ \tilde{V}_{\rm t}
(\chi,p,e)}{J(\chi,p,e) \tilde{V}_{\rm r}^{1/2}(\chi,p,e)} \;,
\label{Tr}
\\
\nonumber \\
\nonumber \\
\Delta\phi &=& 2\int_{0}^{\pi} d\chi \frac{ \tilde{V}_{\rm \phi}
(\chi,p,e)}{J(\chi,p,e) \tilde{V}_{\rm r}^{1/2}(\chi,p,e)} \;.
\label{Dphi}
\end{eqnarray}
We take the ``distance'' $\epsilon= p - p_{s}$ from the separatrix  to
be small, i.e. $\epsilon/M \ll 1$. Also, we shall  exclude small eccentricity 
orbits (more precisely, orbits with $e \lesssim \epsilon/M$), or marginally bound 
orbits ($e \to 1$). The former case of nearly circular orbits has already been 
discussed in \cite{dk2}. In what follows, quantities with an ``s'' subscript 
are to be evaluated exactly at the separatrix.
From (\ref{tilVr}) we get,
\begin{equation}
\tilde{V}_{\rm r}(\chi,p,e) = \frac{M}{p_{s}} \left \{ 1 + 
{\cal O}(\epsilon) \right \} \left [ \epsilon S + 2e x_{\rm s}^2(1 -\cos\chi) 
+ {\cal O}( \epsilon^2, \epsilon(1-\cos\chi) \right ]  \;,  
\end{equation}
where  
\begin{equation}
S= 2p_{s} - (1+e)(3-e) \left ( \frac{\partial x^2}{\partial p} 
\right)_{p=p_{s}}  \;.
\end{equation}
We see that $ \tilde{V}_{\rm r} \to 0$ as $\epsilon \to 0 $ and $\chi=0$
(i.e. the periastron ``touches'' the separatrix). At the same limit, 
$\tilde{V}_{\rm t}$ and $J$ remain nonzero. We can write then, at leading 
order in $\epsilon$,
\begin{eqnarray}
T_{\rm r} &\approx& \left ( \frac{p_{s}}{M} \right )^{1/2}
\int_{0}^{\pi} d\chi \frac{ A_{\rm t}(1-\cos\chi) }{ \left [ \epsilon S
+ 2ex^{2}_{s} (1-\cos\chi) \right ]^{1/2} }  \;,
\label{Tr2}
\\
\nonumber \\
\nonumber \\
\Delta\phi &\approx&  \left ( \frac{p_{s}}{M} \right )^{1/2}
\int_{0}^{\pi} d\chi \frac{ A_{\rm \phi}(1-\cos\chi)}{ \left [ \epsilon S
+ 2ex^{2}_{s} (1-\cos\chi) \right ]^{1/2} } \;.
\label{Dphi2}
\end{eqnarray}
We have defined the functions 
\begin{eqnarray}
A_{\rm t}(y) &=& \frac{ \left [ a^2 E_{\rm s} (1+e -ey)^2 - 2aMx_{\rm s}
(1+e -ey)^{3}/p_{s} + E_{\rm s} p_{s}^{2} \right ] }{(1+e-ey)^{2} 
\left [ 1 -2M(1+e -ey)/p_{s} + a^{2}(1+e -ey)^{2}/p_{s}^2 \right ]} \;,
\label{fnc1}
\\
\nonumber \\
\nonumber \\
A_{\rm \phi}(y) &=& \frac{ \left [ x_{\rm s} + aE_{\rm s} -2Mx_{\rm s}
(1+e-ey)/p_{s} \right ] }
{\left [ 1 -2M(1+e -ey)/p_{s} + a^{2}(1+e -ey)^{2}/p_{s}^2 \right ]} \;,
\label{fnc2} 
\end{eqnarray}
with argument $y= 1-\cos\chi$.
In order to isolate the divergent pieces in the integrals (\ref{Tr2}), 
(\ref{Dphi2}) we split the functions (\ref{fnc1}),(\ref{fnc2}) 
\begin{equation}
A_{\rm t, \phi}(y) = A_{\rm t, \phi}(0) + B_{\rm t, \phi}(y) \;.
\label{splitz}
\end{equation}
These expressions are just Taylor expansions around the regular point $y=0$ 
(with $B_{t,\phi}$ containing the first and all higher derivatives of $A_{t,\phi}$). 
Not surprisingly, both functions $ B_{\rm t, \phi}(y) $ take the form
\begin{equation}
B_{\rm t, \phi}(y)= ey \tilde{B}_{\rm t, \phi}(y) \;.
\end{equation}
Although we don't write the functions $\tilde{B}_{\rm t, \phi}(y)$ explicitly
here (as they don't take a simple form and they are not needed in what follows)
we have verified that  $\tilde{B}_{\rm t, \phi}(0) \neq 0 $.
It follows that the contribution to the integrals from 
$B_{\rm t, \phi}(1-\cos\chi)$ is finite when $ \epsilon, \chi \to 0 $. 
On the other hand, the contribution from $ A_{\rm t, \phi}(0) $ is found to be 
divergent at the same limit,
\begin{equation}
\int_{0}^{\pi} \frac{d\chi}{\left [ \epsilon S + 2ex_{\rm s}^{2}
(1 -\cos\chi) \right ]^{1/2} } = \frac{1}{2} (e x^{2}_{\rm s} )^{-1/2}
\ln \left [  \frac{64 e x^{2}_{\rm s}}{\epsilon S} \right ]
+ {\cal O} \left ( \frac{\epsilon}{e} \ln \left [ \frac{e}{\epsilon} \right ]  \;. 
\right )
\end{equation}
Hence at leading order in $\epsilon $ (therefore close to the
separatrix),
\begin{eqnarray}
T_{\rm r} &\approx & A_{\rm t}(0) \left [ \frac{(1+e)(3-e)}{eMp_{s}}
\right ]^{1/2} \ln \left [ \frac{ 64ep_{s}^{2} }{\epsilon S (1+e)(3-e)}
\right ] \;,
\label{Tr3}
\\
\nonumber \\
\nonumber \\
\Delta\phi &\approx& A_{\rm \phi}(0) \left [ \frac{(1+e)(3-e)}{eMp_{s}}
\right ]^{1/2} \ln \left [ \frac{ 64ep_{s}^{2} }{\epsilon S (1+e)(3-e)}
\right ] \;.
\label{Dphi3} 
\end{eqnarray}

The divergence of $T_{\rm r}$ and $\Delta\phi$ at the separatrix is the 
result of the particle being trapped in an unstable circular orbit at 
the location of the minimum of the radial potential $V_{\rm r}$.

    
\section{\textbf{RADIATION REACTION: formulation of the problem}}

\subsection{\textbf The Teukolsky formalism}

In this paper, we shall employ Teukolsky's formalism \cite{teuk1} for the
calculation of gravitational fluxes and waveforms. His eponymous equation 
describes the evolution of linearised radiative 
perturbative fields in a Kerr geometry 
background. In particular, instead of dealing directly with metric 
perturbations, the Teukolsky formalism considers perturbations on the Weyl 
curvature scalar $\psi_4$. This quantity is a result of the projection of the 
Weyl tensor on the null vectors $n^{\alpha}$, $\bar{m}^{\beta}$ which are 
members of the Newman-Penrose null tetrad \cite{newman}, that is 
$\psi_4= -C_{\alpha\beta\gamma\delta} n^{\alpha} \bar{m}^{\beta} n^{\gamma} 
\bar{m}^{\delta} $.  The feature that makes this formalism 
attractive to our problem is that the radiative fluxes (at infinity and at the 
horizon) as well as the two wave polarisations $h_{\rm +}$,$~h_{\rm x}$ can all
be extracted from $\psi_4$. The ``master'' perturbation equation is separable 
in the Fourier domain by means of a decomposition
\begin{equation}
\psi_4(t,r,\theta,\phi)= \rho^{-4}\sum_{\ell m}\int d\omega 
e^{-i\omega t+i m \varphi} \ _{-2}S_{\ell m}^{a\omega}(\theta)
R_{\ell m\omega}(r) \;,
\label{ps4}
\end{equation}
where $ \rho= r -ia\cos\theta$. The radial function $R_{\ell m\omega}(r)$ 
satisfies the Teukolsky equation
\begin{equation}
\Delta^2{d\over dr}\left({1\over \Delta}{dR_{\ell m\omega}\over dr}
\right)
-V(r) R_{\ell m\omega}=T_{\ell m\omega} \;.
\label{radTeuk}
\end{equation}
The potential $V(r)$ is given by 
\begin{equation}
V(r)=-{K^2+4i(r-M)K \over \Delta}+8i\omega r+\lambda,
\label{potential}
\end{equation}
where $K=(r^2+a^2)\omega-ma$ and $\lambda= E_{lm} + a^2\omega^2 -2a m\omega$. 
The angular functions $ _{-2}S_{lm}^{a\omega}(\theta)$ are $s=-2$ spin-weighted spheroidal 
harmonics \cite{goldberg} which satisfy the following eigenvalue equation,
\begin{eqnarray}
&& \left [ {1 \over \sin\theta}{d \over d\theta}
 \left \{ \sin\theta {d \over d\theta} \right \}
+ a^2\omega^2 \cos^2\theta -\frac{m^2}{\sin^2\theta} + 4a\omega\cos\theta \right.
\nonumber \\
 && \left. + \frac{4m\cos\theta}{\sin^2\theta} -4\cot^2\theta -2 + E_{\ell m} \right ] 
{}_{-2}S_{\ell m}^{a\omega}=0.
\label{slm}
\end{eqnarray}
We have adopted the following normalisation for the spheroidal harmonics 
(hereafter we drop the subscript $-2$ for notational simplicity),  
\begin{equation}
\int_0^{\pi} |S_{\ell m}^{a\omega}|^2 \sin\theta d\theta=1.
\label{snorm}
\end{equation}    
The source term $T_{\ell m\omega}$ present in (\ref{radTeuk}) is constructed
directly from the particle's energy-momentum tensor and this is the point where
the particle's motion enters explicitly in the perturbation equation.
Its explicit form is given below. Let us now return to the radial 
equation (\ref{radTeuk}). A particular solution of this equation can be found 
in terms of two independent solutions $R^{\rm in}_{\ell m\omega}$, 
$R^{\rm up}_{\ell m\omega}$ of the homogeneous equation, 
\begin{equation}
R_{\ell m \omega}(r)= \frac{R^{\rm up}_{\ell m\omega}(r)}{W}
\int^{r}_{r_{+}} dr' \frac{T_{\ell m\omega}(r')R^{\rm in}_{\ell m\omega}(r')}
{\Delta^{2}(r')} +  \frac{R^{\rm in}_{\ell m\omega}(r)}{W}
\int^{+\infty}_{r} dr' \frac{T_{\ell m\omega}(r')R^{\rm up}_{\ell m\omega}(r')}
{\Delta^{2}(r')} \;,
\label{Rfield}
\end{equation}
where $W$ the (constant) Wronskian 
$ W[ \Delta^{-1/2} R^{\rm in}_{\ell m\omega}, \Delta^{-1/2} 
R^{\rm up}_{\ell m\omega} ] $. 
The solutions $R^{\rm in}_{\ell m\omega}$, $R^{\rm up}_{\ell m\omega}$ are
chosen such as to have, respectively,  purely ingoing behaviour at the horizon, 
and purely outgoing behaviour at infinity. Explicitly, 
\begin{equation}
R^{\rm in}_{\ell m\omega} \to\cases{
\Delta^2  e^{-ikr^*}\,&
for $r\to r_+$ \cr
r^3 B^{\rm  out}e^{i\omega r^*}+
r^{-1}B^{\rm in}e^{-i\omega r^*}\,&
for $r\to +\infty,$ \cr}
\label{Rin} 
\end{equation}
\begin{equation} 
R^{\rm up}_{\ell m\omega} \to\cases{ 
C^{\rm  out} e^{ik r^*}+ 
\Delta^2 C^{\rm in} e^{-ik r^*}\,& 
for $r\to r_+,$ \cr r^3  e^{i\omega r^*}\,& 
for $r\to +\infty$\cr } \;,
\label{Rup} 
\end{equation} 
where $k=\omega-ma/2Mr_+$, $~r_{+}= M + (M^2-a^2)^{1/2}$ is the 
outer event horizon, and $r_{\ast}$ is the usual tortoise coordinate defined by 
$dr_{\ast}/dr= (r^2+a^2)/\Delta $. From these expressions we have that
$ W= 2i\omega B^{\rm in}$. 
The solution (\ref{Rfield}) describes ingoing waves at the horizon and 
outgoing waves at infinity as it should be required on physical grounds.
That is,
\begin{eqnarray}
R_{\ell m\omega}(r\to r_+) & \to & 
{ \Delta^2 e^{-i k r^*} \over 
2i\omega B^{\rm in}}
\int^{\infty}_{r_+}dr' {T_{\ell m\omega}(r') R^{\rm up}_{\ell m\omega}(r') 
\over\Delta^{2}(r')} 
\equiv Z^{\rm \infty}_{\ell m\omega} \Delta^2(r) e^{-i k r^*} 
\\
\nonumber \\
\nonumber \\
R_{\ell m\omega}(r\to\infty) & \to &
{r^3e^{i\omega r^*} \over 2i\omega B^{\rm in}}
\int^{\infty}_{r_+}dr'{T_{\ell m\omega}(r') \;, 
R^{\rm in}_{\ell m\omega}(r') 
\over\Delta^{2}(r')} \equiv  Z_{\ell m\omega}^{\rm H}r^3e^{i\omega r^*} \;.
\label{asymptotics}
\end{eqnarray}
The source term $T_{\ell m\omega}$ is given by \cite{chapter} 
\begin{equation} 
T_{\ell m\omega}
 =4\int d\Omega dt\rho^{-5}{\bar \rho}^{-1}(B_2'+B_2'^*)
e^{-im\varphi+i\omega t}{_{-2}S^{a\omega}_{\ell m} \over
\sqrt{2\pi}},
\label{source}
\end{equation}
where
\begin{eqnarray}
B_2'&=&-{1 \over 2}\rho^8{\bar \rho}L_{-1}[\rho^{-4}L_0
(\rho^{-2}{\bar \rho}^{-1}T_{nn})]  \nonumber\\
&-&{1 \over 2\sqrt{2}}\rho^8{\bar \rho}\Delta^2 L_{-1}[\rho^{-4}
{\bar \rho}^2 J_+(\rho^{-2}{\bar \rho}^{-2}\Delta^{-1}
T_{{\bar m}n})], 
\\
\nonumber \\
\nonumber \\
B_2'^*&=&-{1 \over 4}\rho^8{\bar \rho}\Delta^2 J_+[\rho^{-4}J_+
(\rho^{-2}{\bar \rho}T_{{\bar m}{\bar m}})] \nonumber\\
&-&{1 \over 2\sqrt{2}}\rho^8{\bar \rho}\Delta^2 J_+[\rho^{-4}
{\bar \rho}^2 \Delta^{-1} L_{-1}(\rho^{-2}{\bar \rho}^{-2}
T_{{\bar m}n})] \;. 
\label{eq:B2}\\
\nonumber
\end{eqnarray}
We have defined the operators
\begin{eqnarray}
L_s&=&\partial_{\theta}+{m \over \sin\theta} -a\omega\sin\theta +s\cot\theta, 
\\
\nonumber \\
\nonumber \\
J_+&=&\partial_r+{iK/\Delta}  \;.
\\
\end{eqnarray}

The quantities $T_{nn}$, $T_{\bar{m}n},T_{{\bar m}{\bar m}}$  are the result
of the projection of the particle's energy-momentum tensor $T^{\mu \nu}$ 
on the tetrad vectors, i.e. $T_{nn}=T_{\mu\nu}n^\mu n^\nu$ etc.
The energy-momentum tensor for a particle in an arbitrary orbit 
$(t,r(t),\theta(t),\phi(t))$ is given by
\begin{equation}
T^{\mu\nu}= \mu \frac{u^{\mu}u^{\nu}}{\Sigma \sin\theta u^{t} } \delta(r -r(t))
\delta(\theta -\theta(t)) \delta(\phi -\phi(t)) \;,
\label{emtensor}
\end{equation}
where $ u^{\mu}= dx^{\mu}/d\tau $ and $ \Sigma= r^2 + a^2\cos\theta$.
We obtain for the individual projections \cite{chapter}, 
\begin{eqnarray}
T_{nn}&=&\mu{C_{nn} \over \sin\theta}
\delta(r-r(t)) \delta(\theta-\theta(t)) \delta(\phi-\phi(t)),
\\
\nonumber \\
\nonumber \\
T_{{\bar m}n}&=&\mu{C_{{\bar m} n} \over \sin\theta}
\delta(r-r(t)) \delta(\theta-\theta(t)) \delta(\phi-\phi(t)),
\\
\nonumber \\
\nonumber \\
T_{{\bar m}{\bar m}}&=&\mu{C_{{\bar m}{\bar m}} \over \sin\theta}
\delta(r-r(t)) \delta(\theta-\theta(t)) \delta(\phi-\phi(t)) \;,
\end{eqnarray}
with
\begin{eqnarray}
C_{nn}&=&{1\over 4\Sigma^3 }(u^{\rm t})^{-1}\left[E(r^2+a^2)-a L
+\Sigma u^{\rm r} \right]^2,
\nonumber\\
C_{{\bar m}n}&=& -{\rho \over 2\sqrt{2}\Sigma^2 }(u^{\rm t})^{-1}
\left[E(r^2+a^2)-aL +\Sigma u^{\rm r} \right]
\left[i\sin\theta\Bigl(aE-{L \over \sin^2\theta}\Bigr) +
\Theta(\theta) \right], 
\nonumber \\
C_{{\bar m}{\bar m}}&=& {\rho^2 \over 2\Sigma  }(u^{\rm t})^{-1}
\left[i\sin\theta \Bigl(aE-{L_z \over \sin^2\theta}\Bigr) + 
\Theta(\theta) \right]^2 \;.
\label{Cq}
\end{eqnarray}
The quantity $ \Theta(\theta) $ represents the effective latitudinal potential,
i.e, $ (\Sigma u^{\theta})^2= \Theta(\theta) $. 
 Substitution in (\ref{source}) yields
\begin{eqnarray}
T_{\ell m\omega}&=&{4\mu\over\sqrt{2\pi}}\int^{\infty}_{-\infty}
dt\int_{0}^{\pi} d\theta e^{i\omega t-im\varphi(t)} \nonumber\\
&\times&
\Bigl[-{1 \over 2}L_1^{\dag} \bigl\{ \rho^{-4}L_2^{\dag}(\rho^3 S_{\ell 
m}^{a\omega})
\bigr\}
C_{nn}\rho^{-2}{\bar \rho}^{-1}\delta(r-r(t))
\delta(\theta-\theta(t)) \nonumber\\
&+&{\Delta^2 {\bar \rho}^2 \over \sqrt{2} \rho}
\bigl(L_2^{\dag} S_{\ell m}^{a\omega} + ia({\bar \rho}-\rho)\sin\theta 
S_{\ell m}^{a\omega}\bigr) J_+ \bigl\{ C_{{\bar m}n}\rho^{-2}{\bar \rho}^{-
2}\Delta^{-1} \delta(r-r(t))\delta(\theta-\theta(t)) \bigr\} \nonumber\\
&+&{1 \over 2\sqrt{2} }
L_2^{\dag}\bigl\{ \rho^3 S_{\ell m}^{a\omega} ({\bar \rho}^2 \rho^{-
4})_{,r} 
\bigr\} C_{{\bar m}n}\Delta \rho^{-2}{\bar \rho}^{-2}
\delta(r-r(t))\delta(\theta-\theta(t)) \nonumber\\
&-&{1 \over 4}\rho^3 \Delta^2 S_{\ell m}^{a\omega} J_+\bigl\{\rho^{-4}
J_+\bigl({\bar \rho} \rho^{-2}C_{{\bar m}{\bar m}}
\delta(r-r(t))\delta(\theta-\theta(t))\bigr) \bigr\}
\Bigr],
\label{source2}\\
\nonumber
\end{eqnarray}
where
\begin{equation}
L_s^{\dag}=\partial_{\theta}-{m \over \sin\theta}
+a\omega\sin\theta+s\cot\theta \;. 
\end{equation}
The $\theta$-integration can be performed directly to give 
\begin{eqnarray}
T_{\ell m\omega}&=&\mu\int^{\infty}_{-\infty}dt 
e^{i\omega t-i m \varphi(t)}
\Delta^2\Bigl[(A_{nn0}+A_{{\bar m}n0}+
A_{{\bar m}{\bar m}0})\delta(r-r(t)) \nonumber\\
&+&\left\{(A_{{\bar m}n1}+A_{{\bar m}{\bar m}1})
\delta(r-r(t))\right\}_{,r}
+\left\{A_{{\bar m}{\bar m}2}
\delta(r-r(t))\right\}_{,rr}\Bigr]_{\theta= \theta(t)} \;,
\label{source3} 
\end{eqnarray}
where 
\begin{eqnarray}
A_{nn0}&=&{-2 \over \sqrt{2\pi}\Delta^2}
C_{nn}\rho^{-2}{\bar \rho}^{-1}
L_1^+\{\rho^{-4}L_2^+(\rho^3 S_{\ell m}^{a\omega})\},
\\
\nonumber \\
\nonumber \\
A_{{\bar m}n0}&=&{2 \over \sqrt{\pi}\Delta} 
C_{{\bar m}n}\rho^{-3}
\Bigl[\left(L_2^+S_{\ell m}^{a\omega}\right)
\Bigl({iK \over \Delta}+\rho+{\bar \rho}\Bigr) 
\nonumber \\
& &\qquad-a\sin\theta(t) S_{\ell m}^{a\omega} {K \over \Delta}
({\bar \rho}-\rho)\Bigr], 
\\
A_{{\bar m}{\bar m}0}
&=&-{1 \over \sqrt{2\pi}}\rho^{-3}{\bar \rho}
C_{{\bar m}{\bar m}}S\Bigl[
-i\Bigl({K \over \Delta}\Bigr)_{,r}-{K^2 \over \Delta^2}+
2i\rho {K \over \Delta}\Bigr],
\\
A_{{\bar m}\,n\,1}&=&{2\over \sqrt{\pi}\Delta }
\rho^{-3}C_{{\bar m}n}
[L_2^{+}S_{\ell m}^{e\omega} +ia\sin\theta(t)({\bar \rho}-\rho)
S_{\ell m}^{a\omega}], 
\\
A_{{\bar m}{\bar m}1} &=&-{2 \over \sqrt{2\pi}} \rho^{-3}{\bar \rho}
C_{{\bar m}{\bar m}}S_{\ell m}^{a\omega}
\Bigl(i{K \over \Delta}+\rho\Bigr), 
\\
A_{{\bar m}{\bar m}2}
&=&-{1\over \sqrt{2\pi}}\rho^{-3}{\bar \rho}
C_{{\bar m}{\bar m}}S_{\ell m}^{a\omega}. 
\label{Aq} 
\end{eqnarray}
Note that all functions of $\theta$ are evaluated at $\theta= \theta(t)$. 
The amplitudes $Z_{\ell m\omega}^{\infty,H}$ defined in (\ref{asymptotics})
can be written as
\begin{eqnarray}
Z^{H}_{\ell m\omega} &=&
{\mu\over2i\omega B^{\rm in}}
\int^{\infty}_{-\infty}dt e^{i\omega t-i m \varphi(t)}
{\cal I}^{H}_{\ell m \omega}(r(t),\theta(t)) \;,
\label{Zh} 
\\ 
\nonumber \\
\nonumber \\
 Z^{\infty}_{\ell m\omega} &=&
{\mu\over2i\omega B^{\rm in}}
\int^{\infty}_{-\infty}dt e^{i\omega t-i m \varphi(t)}
{\cal I}^{\infty}_{\ell m \omega}(r(t),\theta(t))  \;,
\label{Zinf}
\end{eqnarray}
where 
\begin{eqnarray}
{\cal I}^{\rm H}_{\ell m\omega}&=&
\Bigl[R^{\rm in}_{\ell m\omega}\{A_{nn0}+A_{{\bar m}n0}
+A_{{\bar m}\,{\bar m}\,0}\} 
\nonumber\\
& &-{dR^{\rm in}_{\ell m\omega} \over dr}\{ A_{{\bar m}n1}
+A_{{\bar m}{\bar m}1}\}
 +{d^2 R^{\rm in}_{\ell m\omega} \over dr^2}
 A_{{\bar m}{\bar m}2}\Bigr]_{r=r(t),\theta=\theta(t)} \;,
\\
\nonumber \\
\nonumber \\
{\cal I}^{\rm \infty}_{\ell m\omega}&=&
\Bigl[R^{\rm up}_{\ell m\omega}\{A_{nn0}+A_{{\bar m}n0}
+A_{{\bar m}{\bar m}0}\}
\nonumber\\
& &-{dR^{\rm up}_{\ell m\omega} \over dr}\{ A_{{\bar m}n1}
+A_{{\bar m}{\bar m}1}\}
 +{d^2 R^{\rm up}_{\ell m\omega} \over dr^2}
 A_{{\bar m}{\bar m}2}\Bigr]_{r=r(t),\theta=\theta(t)} \;.
\label{Iotas}
\end{eqnarray}

Up to this point, all expressions listed in this Section are valid for an 
arbitrary orbit. We now specialise our discussion to equatorial orbits by 
setting $\theta(t)= \pi/2$.  In this case, $ {\cal I}^{\infty,H}_{\ell m\omega}$ 
are functions of $r(t)$ only. As discussed in detail in Appendix B, the 
quantities
\begin{equation}
\alpha^{\infty,H}(t)= {\cal I}^{\infty, H}(r(t))~e^{-im[\phi(t) -
\Omega_{\phi}t]} \;,
\end{equation}
are periodic functions of time (with period equal to $T_r$). Consequently, 
they can be expanded in a Fourier series   
\begin{equation}
\alpha^{\infty,H}(t)= \sum_{k=-\infty}^{+\infty}  
\alpha^{\infty,H}_{k}e^{-ik\Omega_r t} \;,
\label{fourier}
\end{equation}
with $\Omega_{\rm r}= 2\pi/T_{\rm r}$. Inverting, we obtain the Fourier 
coefficients
\begin{equation}
\alpha^{\infty,H}_{k}= \frac{1}{T_r} \int_{0}^{T_r}dt ~
\alpha^{\infty,H}(t) e^{ik\Omega_r t} \;.
\end{equation}
Using the Fourier series (\ref{fourier}) in (\ref{Zh}),(\ref{Zinf})  
we arrive at
\begin{equation}
Z^{\infty, H}_{\ell m \omega} = \sum_{k=-\infty}^{+\infty}
Z^{\infty, H}_{\ell m k} \delta(\omega -\omega_{mk}) \;,
\label{Zlmk1}
\end{equation}
where $\omega_{mk}= m\Omega_{\phi} + k\Omega_r $ and
\begin{equation}
Z^{\infty,H}_{\ell m k}= \frac{\mu\Omega_r}{2i\omega_{mk} B^{in}}
\int_{0}^{T_r} dt ~ {\cal I}^{\infty, H}(r(t))~ e^{i\omega_{mk}t -im\phi(t)} \;.
\label{Zlmk2}
\end{equation}
Due to symmetries of the Teukolsky equation (\ref{radTeuk}) we have the
following conjugation relation,
\begin{eqnarray}
Z^{\infty, H}_{\ell, -m, -\omega}&=& (-1)^{\ell} 
\bar{Z}^{\infty,H}_{\ell, m, \omega} \;,
\\
Z^{\infty, H}_{\ell, -m, -k}&=& (-1)^{\ell} \bar{Z}^{\infty,H}_{\ell, m, k} \;,
\label{conj}
\end{eqnarray}
where an overbar denotes complex conjugate. 

We proceed by writing (\ref{Zlmk2}) as an integral over $\chi$,
\begin{equation}
Z^{\infty,H}_{\ell m k}= \frac{\mu\Omega_r}{2i\omega_{mk}B^{in}}
\int_{0}^{2\pi} d\chi \frac{\tilde{V}_{\rm t}(\chi)}
{J(\chi)~ \tilde{V}_{\rm r}^{1/2}(\chi) } 
{\cal I}^{\infty, H}_{\ell m\omega}(r(\chi))~ e^{i\omega_{mk}t(\chi) 
-im\phi(\chi)} \;.
\label{Zlmk3}
\end{equation}
As in the case of eqns. (\ref{phichi}), (\ref{tchi}), this expression is
well-behaved at the orbital turning points. Moreover, noting that the 
$\chi$-dependence of the integrand in (\ref{Zlmk3}) appears in terms of 
the form 
$\cos\chi$ (in terms with $r(\chi)$)) and $\sin\chi$ 
(in terms with $u^{\rm r}$) we can write,
\begin{eqnarray}
Z^{\infty,H}_{\ell m k} &=& \frac{\mu\Omega_r}{2i\omega_{mk}B^{in}}
\int_{0}^{\pi} d\chi \frac{\tilde{V}_{\rm t}(\chi)}
{J(\chi)~ \tilde{V}_{\rm r}^{1/2}(\chi) } \left [ 
{\cal I}^{\infty, H}_{\ell m\omega~(+)}(r(\chi))~ e^{i\omega_{mk}t(\chi) 
-im\phi(\chi)}  \right.
\nonumber \\
&+& \left. {\cal I}^{\infty, H}_{\ell m\omega~(-)}(r(\chi))~ e^{-
i\omega_{mk}t(\chi) +im\phi(\chi)} \right ] \;.
\label{Zlmk4}
\end{eqnarray}
The subscripts $(\pm)$ mean ``$\sin\chi \to \pm \sin\chi $'' in the
functions ${\cal I}_{\ell m\omega}^{\infty, H}$. The numerical calculation
of the amplitudes $Z^{\infty,H}_{\ell m k}$ is the ``backbone'' of
our radiation reaction code (see Section V).   
Finally, we express the $A$ and $C$ amplitudes (\ref{Cq}),(\ref{Aq}), 
in terms of $\chi, p,e$,
\begin{eqnarray}
C_{nn}(\chi,p,e)&=& \frac{J(\chi,p,e)}{4p^4 \tilde{V}_{\rm t}(\chi,p,e)}
\left [ p^2 E -ax(1+ e\cos\chi)^2 + ep\sin\chi 
\tilde{V}^{1/2}_{\rm r}(\chi, p,e) \right ]^2  \;,   
\\
\nonumber \\  
C_{\bar{m}n}(\chi,p,e) &=& \frac{ix J(\chi,p,e)}{2\sqrt{2} p^3 
\tilde{V}_{\rm t}(\chi,p,e)}(1+ e\cos\chi) \left [ p^2 E -ax(1+e\cos\chi)^2
+ ep\sin\chi \tilde{V}_{\rm r}^{1/2}(\chi,p,e) \right ] \;, 
\\
\nonumber \\ 
C_{\bar{m}\bar{m}}(\chi,p,e) &=&  -\frac{x^2 J(\chi,p,e)}{2p^2 
\tilde{V}_{\rm t}(\chi,p,e)} (1+e\cos\chi)^2 \;, 
\label{Cq2} 
\end{eqnarray}
and
\begin{eqnarray}
A_{\bar{m}n0}(u)&=& \frac{2}{\sqrt{\pi}}
\frac{C_{\bar{m}n}}{u(1 -2Mu +a^2 u^2)^2}
\left [ 2a^2 u^3 + \{ia(a\omega -m) -4M\}u^2 + 2u + i\omega \right ]
\nonumber \\
&& \times \left [ \frac{\partial S_{\ell m}^{a\omega}}{\partial\theta}(\pi/2) + 
(a\omega -m)S_{\ell m}^{a\omega}(\pi/2) \right ]  \;,
\\
\nonumber \\
\nonumber \\ 
A_{\bar{m}\bar{m}0}(u)&=& \frac{1}{\sqrt{2\pi}}
\frac{C_{\bar{m}\bar{m}}S_{\ell m}^{a\omega}(\pi/2)}{u^2(1 -2Mu +a^2u^2)^2} 
\left [ -2ia^3(a\omega -m) u^5 + a(a\omega -m)\{ 6iM + a(a\omega -m)\} u^4 
\right.
\nonumber \\
&& \left. -4ia(a\omega -m) u^3 + 2\omega\{iM + a(a\omega -m)\} u^2 
-2i\omega u + \omega^2   \right ] \;, 
\\ 
\nonumber \\
\nonumber \\
A_{\bar{m}n1}(u)&=& \frac{2}{\sqrt{\pi}}\frac{C_{\bar{m}n}}{ u(1-2Mu 
+a^2u^2)} \left [ \frac{\partial S_{\ell m}^{a\omega}}{\partial\theta}(\pi/2) 
 + (a\omega -m)S_{\ell m}^{a\omega}(\pi/2) \right] \;,
\\ 
\nonumber \\
\nonumber \\
A_{\bar{m}\bar{m}1}(u) &=& -\sqrt{\frac{2}{\pi}}
\frac{C_{\bar{m}\bar{m}}S_{\ell m}^{a\omega}(\pi/2)}
{u^2 (1-2Mu + a^2u^2)} \left [ a^2u^3 + \{ia(a\omega -m) -2M\} u^2 + u 
+ i\omega \right ]  \;,
\\ 
\nonumber \\
\nonumber \\ 
A_{\bar{m}\bar{m}2}(u) &=& -\frac{1}{\sqrt{2\pi}}
\frac{C_{\bar{m}\bar{m}} S_{\ell m}^{a\omega}(\pi/2)}{ u^2} \;,
\\
\nonumber \\
\nonumber \\ 
A_{nn0}(u) &=& -\sqrt{\frac{2}{\pi}}\frac{C_{nn}}{(1 -2Mu +a^2u^2)^2}
\left [ -2ia \left ( \frac{\partial S_{\ell m}^{a\omega}}{\partial \theta}
(\pi/2) + (a\omega -m)S_{\ell m}^{a\omega}(\pi/2) \right )u 
\right.
\nonumber \\
\nonumber \\
&&\left.  + \frac{\partial^2 S_{\ell m}^{a\omega}}{\partial \theta^2}(\pi/2)
+ 2(a\omega -m) \frac{\partial S_{\ell m}^{a\omega}}{\partial \theta}(\pi/2)
+ \{(a\omega -m)^2 -2\}S_{\ell m}^{a\omega}(\pi/2) \right ] \;,
\end{eqnarray}
where $ u(\chi, p,e)= (1+e\cos\chi)/p$. 
By means of (\ref{Zlmk1}) one can obtain the following expressions for 
$\psi_{4}$ at infinity and on the horizon, 
\begin{equation}
\psi_{4}(t,r,\theta, \phi) \to 
\cases{ \rho(r_{+})^{-4} \sum_{\ell m k} \psi_{\ell m k}^{H} 
\,& for $r\to r_+$ 
\cr
\cr
 r^{-1} \sum_{\ell m k} \psi_{\ell m k}^{\infty}  \,&
for $r\to +\infty,$ \cr}
\label{ps4_2} 
\end{equation}
where
\begin{equation}
\psi_{\ell m k}^{H, \infty}= \frac{1}{\sqrt{2\pi}} Z^{\infty, H}_{\ell mk} 
S_{\ell m}^{a\omega_{mk}}(\theta) e^{-i\omega_{mk}(t -r_{\ast}) +im\phi} \;.
\label{ps4_3}
\end{equation}

Once the Weyl scalar $\psi_4 $ is known, 
we can immediately relate it to the two
polarisation components $ h_{\rm +}, h_{\rm x} $ of the transverse-traceless 
metric perturbation as $r \to \infty $ \cite{teuk1},
\begin{equation}
\psi_4 \approx \frac{1}{2} \left ( \frac{\partial^2 h_{\rm +}}{\partial t^2} 
- i \frac{\partial^2 h_{\rm x}}{\partial t^2} \right ) \;.
\label{wave1}
\end{equation}
It follows from (\ref{wave1}), (\ref{ps4_3}) that 
$ h_{\rm +,x}(t,r,\theta,\phi) $ are given by
(here the coordinates $t,r,\theta, \phi$ are referred to the observation
point), 
\begin{equation}
h_{\rm +} -i h_{\rm x} = \frac{2}{r} \sum_{\ell mk} \frac{Z^{\rm H}_{\ell mk}}
{\omega^{2}_{mk}} \frac{S_{\ell m}^{a\omega_{mk}}(\theta)}{\sqrt{2\pi}} 
e^{-i\omega_{mk}(t -r_{\ast}) + im\phi} \;.
\label{wave2}
\end{equation} 
Note that the gravitational waveform is exclusively radiated at harmonics of the
two orbital frequencies $\Omega_r$, $\Omega_{\phi}$.  
The gravitational wave energy and angular momentum flux at infinity can
be found in terms of the Landau-Lifschitz pseudotensor \cite{landau},
\begin{eqnarray}
\left ( \frac{dE}{dt} \right )_{GW}^{\infty} &=& \frac{1}{16\pi} \int 
\left \{ \left ( \frac{\partial h_{\rm +}}{\partial t} \right )^2 +
\left ( \frac{\partial h_{\rm x}}{\partial t} \right )^2 \right \} r^2 d\Omega \;,
\label{LL1}
\\
\nonumber \\
\left ( \frac{dL}{dt} \right )_{GW}^{\infty} &=& -\frac{1}{16\pi}
\int \left \{ \frac{\partial h_{\rm +}}{\partial t} \frac{\partial 
h_{\rm +}}{\partial \phi}
+ \frac{\partial h_{\rm x}}{\partial t} \frac{\partial h_{\rm x}}
{\partial \phi} \right \}
r^2 d\Omega \;.
\label{LL2}
\end{eqnarray}
We define the averaged (over one orbital period) fluxes to be $(C= E,L)$, 
\begin{equation}
\dot{C}_{GW} \equiv \frac{1}{T_r} \int_{0}^{T_r} dt 
\left ( \frac{dC}{dt}  \right )_{GW}^{\infty} \;.
\label{average}
\end{equation}
With the help of (\ref{LL1}), (\ref{LL2}), (\ref{wave2}) we arrive at 
\cite{Press}
\begin{eqnarray}
\dot{E}_{GW}^{\infty} &=& \sum_{\ell,m,k} \frac{|Z^{\rm H}_{\ell mk}|^2}
{4\pi \omega^2_{mk}} \;,
\label{GWflux1}
\\
\dot{L}_{GW}^{\infty} &=& \sum_{\ell,m,k} \frac{m|Z^{\rm H}_{\ell mk}|^2}
{4\pi \omega^3_{mk}} \;. 
\label{GWflux2}
\end{eqnarray}
The calculation of the respective fluxes at the black hole horizon is a more
complicated issue as it is not possible to use expressions such as (\ref{LL1}),
(\ref{LL2}). Despite this difficulty, Teukolsky and Press \cite{Press} were
able to derive formulae for the horizon fluxes using the approach of
Hawking and Hartle \cite{hawking} who studied the deformation of the hole's 
event horizon under the influence of infalling radiation. These formulae 
are,
\begin{eqnarray}
\dot{E}_{GW}^{H} &=& \sum_{\ell,m,k} \alpha_{\ell mk} 
\frac{|Z^{\infty}_{\ell mk}|^2}{4\pi \omega^2_{mk}} \;,
\label{GWflux3}
\\
\dot{L}_{GW}^{H} &=& \sum_{\ell,m,k} \alpha_{\ell mk} 
\frac{m|Z^{\infty}_{\ell mk}|^2}{4\pi \omega^3_{mk}} \;, 
\label{GWflux4}
\end{eqnarray}
where
\begin{equation}
\alpha_{\ell mk}=  \frac{ 256(2Mr_{+})^5 p_{mk} (p_{mk}^2 + 4\epsilon^2 )
(p_{mk}^2 + 16\epsilon^{2} )\omega_{mk}^{3}}{C_{\ell mk}} \;,
\end{equation}
with $ \epsilon= \sqrt{M^2 -a^2}/4Mr_{+} ~$ and
\begin{eqnarray}
C_{\ell mk} &=& [ (\lambda +2)^2 + 4am\omega_{mk} -4a^2 \omega_{mk}^2 ] 
( \lambda^2 + 36am\omega_{mk} -36a^2 \omega^{2}_{mk} ) 
\nonumber \\
&& + (2\lambda +3)(96 a^2 \omega^2_{mk} -48am \omega_{mk} ) + 144 \omega_{mk}^{2}
(M^2 -a^2) \;,
\end{eqnarray}
is the so-called Starobinsky constant.
Note that eqns. (\ref{GWflux1})-(\ref{GWflux4}) have to be divided by 
$\mu$ in order to convert them to fluxes of specific energy and angular momentum.
Moreover, we can exploit the conjugation relations (\ref{conj}) in the numerical 
calculation of the amplitudes $ Z_{\ell mk} $ and reduce by one half the required
computational time.


\subsection{\textbf{The Sasaki-Nakamura equation}}

From eqn. (\ref{Zlmk4}) it is obvious that in order to calculate the amplitudes 
$Z^{\infty,H}_{\ell mk}$, which will give us 
the gravitational waveform and fluxes (\ref{wave2}), (\ref{GWflux1})-
(\ref{GWflux4}), we need to evaluate the quantity $ B^{\rm in} $. 
In principle, one could numerically integrate the Teukolsky equation 
(\ref{radTeuk}) from the horizon out to ``infinity'' and extract the amplitudes 
$B^{\rm in,out}$. But this is a poor strategy, since the effective potential 
$V(r)$ is long-ranged and the $B^{\rm in}$ term drops off towards infinity 
much faster than the $B^{\rm out}$ term and can only be extracted with very 
low accuracy \cite{detweiler}. A way to circumvent this difficulty is 
to integrate, instead, the Sasaki-Nakamura equation \cite{chapter}, \cite{SN} 
\begin{equation}
\frac{d^2 X}{dr_{\ast}^2} -F(r) \frac{dX}{dr_{\ast}} - U(r) X= 0 \;.
\label{sneq}     
\end{equation}
The ``potentials'' $ F(r), U(r)$ are given in Appendix C. 
The solutions of this equation are related to the solutions of the Teukolsky
equation via the transformation,
\begin{equation}
R_{\ell m \omega}(r)= \frac{1}{\eta} \left [ \left ( \alpha +
\frac{\beta_{,r}}{\Delta} \right ) 
\frac{\Delta X_{\ell m \omega}}{(r^2 + a^2)^{1/2} } - \frac{\beta}{\Delta} 
\frac{d}{dr} \left ( \frac{\Delta X_{\ell m \omega}}{(r^2 + a^2)^{1/2} }
\right ) \right ]  \;.
\label{transf}
\end{equation}
The functions $\alpha(r), \beta(r) $ are also given in Appendix C.
The key property of (\ref{sneq}) is that it encompasses a short-range potential.
This can be demonstrated more easily if we shift to the function,
\begin{equation}
Y(r)= \eta^{1/2}(r) X(r) \;.
\end{equation}
Then, eqn. (\ref{sneq}) transforms into the Schr\"{o}dinger-type equation,
\begin{equation}
\frac{d^2 Y}{dr^2_{\ast}} + Q Y= 0 \;,
\label{sneq2}
\end{equation}
with the effective potential,
\begin{equation}
Q= -U -\frac{1}{4} F^2 + \frac{\Delta}{2\eta (r^2 + a^2)^2} \left \{
\Delta \eta_{,rr} - \frac{\Delta}{\eta} (\eta_{,r})^2 + 2M\eta_{,r} 
\left ( \frac{r^2 -a^2}{r^2 + a^2} \right ) \right \} \;.
\label{effpot}
\end{equation}
The functions $F(r)$, $U(r)$ have the following behaviour at infinity and 
at the horizon,
\begin{equation}
F(r) \to \cases{ 0 + {\cal O}(r-r_{+}) \,& for $r\to r_+$ 
\cr
 -r^{-2} c_{1}/c_{0} + {\cal O}(r^{-3}) \,& for $r\to +\infty,$ \cr}
\label{F} 
\end{equation}
\begin{equation}
U(r) \to 
\cases{ -k^2 + {\cal O}(r-r_{+})   \,& for $r\to r_+$ 
\cr
 \ -\omega^2 + r^{-2}\left [ \lambda + 2(1+ am\omega -a^2\omega^2)
 -i\omega c_{1}/c_{0} \right ] + {\cal O}(r^{-3})
,& for $r\to +\infty \;.$ \cr}
\label{U} 
\end{equation}
It follows that,
\begin{equation}
Q(r_{\ast}) \to  
\cases{ \omega^2  - r_{\ast}^{-2} \left [ \lambda + 
2(1+ am\omega -a^2\omega^2 -i\omega c_1 /c_{0} \right ] + 
{\cal O}( r_{\ast}^{-3} \ln r_{\ast} ) \,& 
for $ r_{\ast} \to +\infty$ 
\cr
 k^2 + {\cal O}( e^{c r_{\ast}})  \,& for $r_{\ast} \to -\infty,$ \cr}
\label{Qasymp} 
\end{equation}
where $ c= (r_{+} -r_{-})/2M $ is a positive constant. From this expression is 
obvious that $Q$ is short-ranged. Consequently, equation (\ref{sneq}) admits a 
solution (``in'' mode) which is purely ingoing at the horizon and a mixture of 
ingoing/outgoing waves at infinity:  
\begin{equation}
X^{\rm in} \to  
\cases{ A^{\rm down} e^{-ikr_{\ast}} \,& for $r\to r_+$ 
\cr
A^{\rm in} e^{-i\omega r_{\ast}} + A^{\rm out}e^{i\omega r_{\ast}} 
\, & for $r\to +\infty,$ \cr}
\label{Xin} 
\end{equation}
Another useful independent solution to (\ref{sneq}) is the ``up'' mode,
 \begin{equation}
X^{\rm up} \to  
\cases{ D^{\rm in} e^{-ikr_{\ast}} + D^{\rm out} e^{ikr_{\ast}}  \,& for $r\to r_+$ 
\cr
D^{\rm up} e^{i\omega r_{\ast}}  \, & for $r\to +\infty.$ \cr}
\label{Xup} 
\end{equation}
The relation between the asymptotic amplitudes appearing in  (\ref{Rin}) and
(\ref{Xin}) can be deduced from (\ref{transf}),
\begin{eqnarray}
B^{\rm in} &=& -\frac{1}{4\omega^2} A^{\rm in} \;,
\\
B^{\rm out} &=& -\frac{4\omega^2}{c_{0}} A^{\rm out} \;, 
\label{amplitrans} 
\end{eqnarray}
where the constant $c_0$ is given in Appendix C. 
Hence, we can simply integrate equation (\ref{sneq}) instead of (\ref{radTeuk})
and easily identify the ingoing and outgoing waves and evaluate their 
respective amplitudes. We can then simply find the desired amplitudes 
$B^{\rm in/out}$ from (\ref{amplitrans}). Similarly, knowledge of the 
wavefunction $X(r)$ and its derivative at a given point immediately leads to 
the Teukolsky radial function $R(r)$ and its derivative via the rule 
(\ref{transf}). In conclusion, all the quantities (apart from the 
spheroidal harmonics) required for the calculation of the gravitational flux 
and waveform, can be obtained by numerical integration of the Sasaki-Nakamura 
equation (\ref{sneq}).


\subsection{\textbf{Orbital evolution: adiabaticity and flux balance}}

Due to the emission of gravitational radiation the orbit of a particle
around a black hole will slowly evolve in time and the orbital constants 
$E,L$ (or equivalently $p,e$) will no longer be conserved. 
Radiation reaction effects become noticeable on a timescale that scales as 
$ \sim M^{2}/\mu $, i.e. they are always tiny in a timescale $\sim {\cal O}(M)$, 
provided the system's mass ratio is sufficiently small. 
We can define as the radiation reaction timescale, 
\begin{equation}
T_{\rm RR}= \mbox{min} [ T_{p}, T_{e} ] \;,
\label{RRscale}
\end{equation}
where $ T_{e}= e/|\dot{e}| $ and $ T_{p}= p/|\dot{p}| $ are the radiative 
timescales for $p$ and $e$ respectively (approximate expressions for these 
timescales are given in the following Section).
We will then say that an orbit evolves adiabatically if
\begin{equation}
T_{\rm RR} \gg T_{\rm r} \;.
\label{adiab}
\end{equation}
In other words, it is a good approximation to assume the motion
of the particle to be geodesic, as long as we are interested in 
timescales much shorter than $T_{\rm RR}$. On the other hand,
by making this simplification we ``freeze'' the evolution of the orbit, 
as if there was no radiation reaction.
That is, within the adiabatic approximation, we cannot 
know the exact evolution of the functions $E(t), L(t)$ (or of 
$p(t), e(t)$). It is still possible, however, to calculate an
averaged rate of change of such quantities. This can be done by 
assuming the following  ``flux-balance'' relation
\begin{equation}
\dot{C}= -\dot{C}_{\rm GW}= -( \dot{C}_{\rm GW}^{\infty} +
\dot{C}_{\rm GW}^{H} ) \;,
\end{equation}
where $C= E, L$. We have separately denoted the gravitational wave
fluxes at infinity and down to the horizon by 
$ \dot{C}_{\rm GW}^{\infty}, \dot{C}_{\rm GW}^{H} $ respectively. The overdot
symbol stands for the averaged (over one orbital period) rate, 
see Section IIIB. We can equally well describe an orbit by means of 
the parameters $(p,e)$ and calculate the relevant averaged rates of change 
of those quantities. Since $E=E(p,e)$  and $L= L(p,e)$ we have that 
(commas denote partial derivatives),
\begin{eqnarray}
\dot{E}&=& E_{,p} \dot{p} + E_{,e}\dot{e} \;,
\\
\dot{L}&=& L_{,p} \dot{p} + L_{,e}\dot{e} \;.
\end{eqnarray}
These relations can be inverted to obtain,
\begin{eqnarray}
\dot{p} &=& H^{-1} ( -E_{,e} \dot{L} + L_{,e} \dot{E} ) \;,
\label{pdot1}  
\\ 
\dot{e} &=& H^{-1} ( E_{,p} \dot{L} - L_{,p} \dot{E} ) \;, 
\label{edot1} 
\end{eqnarray}
with  $H= E_{,p} L_{,e} - E_{,e} L_{,p} $.
 Eventually, all partial derivatives of $E$ and $L$ can be 
found in terms of the corresponding partial derivatives of 
the functions $F, N$ and  $ \Delta_{\rm x}$ which are given explicitly 
in Appendix A. However, the resulting formulae are quite messy so we do 
not present them here.

Note that although the formalism adopted in our analysis offers only a 
``local'' information on the radiative orbital evolution, it can be 
further manipulated in order to obtain additional information. 
A recipe for ``evolving'' orbits under radiation reaction, using the
known averaged rates of change of the relevant orbital constants, was
given recently by Hughes \cite{scott_insp} in the context of circular
non-equatorial orbits. In effect, one is able to construct a series of
``snapshots'' of the radiation-induced inspiral, and make predictions
of the evolution of the emitted waveform close to the point where
the orbit becomes unstable (that is until the adiabatic condition
no longer holds).

As we have already mentioned, adiabaticity will eventually break down
near the separatrix, no matter how small the mass ratio is. This can be 
immediately seen from (\ref{adiab}) and recalling that $T_r \to \infty$
at the separatrix, as predicted by (\ref{Tr3}). For an order-of-magnitude 
estimation, we can use the quadrupole approximation for the fluxes 
(see next Section) and translate (\ref{adiab}) into a constraint on the 
mass ratio,
\begin{equation}
\frac{\mu}{M} \ll \frac{5}{128\pi} \left ( \frac{p}{M} \right )^{5/2}
f_{3}^{-1}(e)  \left [ 1 \pm \frac{a}{M} \left ( \frac{M}{p} \right )^{3/2}
f_{3}^{-1}(e) \left ( \frac{169}{12} + \frac{185}{12} e^2 + 
\frac{223}{96} e^4 \right ) \right ]  \;.
\label{massrat1} 
\end{equation}
The function $f_{3}(e)$ is defined in the following Section.
Equation (\ref{massrat1}) is accurate to leading order 
in $M/p$ and $a/M$, and to derive it we have used the 
corresponding order expression for the orbital period 
\begin{equation}
T_{r}= 2\pi M (1-e^2)^{-3/2}  \left ( \frac{p}{M} \right )^{3/2}
\left [ 1 \mp \frac{3a}{M} \left (\frac{M}{p} \right )^{3/2}
(1-e^2) \right ] \;.  
\label{weakTr} 
\end{equation}
The mass-ratio constraint (\ref{massrat1}) is automatically 
satisfied as the black hole perturbation scheme we employ 
requires $ \mu/M \ll 1$. 

On the other hand, in the strong-field regime near the separatrix we find
(using results derived in Section IVB),
\begin{equation}
\frac{\mu}{M} \ll  \delta H
\left ( \ln \left [ \frac{64ep_{s}^2} {\epsilon S(1+e)(3-e)} 
\right ] \right )^{-1} \;,
\label{massrat2} 
\end{equation}
where $\delta$ is a combination of $A_t(0), A_{\phi}(0), E_{, p/e}, L_{, p/e}$ 
and is of order unity. 
As we  discuss in Section IVB, the quantity $H$ also becomes zero when
$\epsilon \to 0 $ (unless $a=M$, in which case it remains finite).
This is clearly the most severe restriction for the mass ratio.
Fortunately, the real astrophysical systems we are trying to model are 
typically characterised by a mass ratio $\mu/M \sim 10^{-6}$. Therefore we 
can approach the separatrix closely, probably to the point where the physical 
body would begin its plunge into the black hole, in the cases which 
interest us.


\section{\textbf{Analytical results}}

\subsection{\textbf{Weak-field approximations}}

Orbits with $p \gg M$ are well described by weak-field approximate
results. In particular, the energy and angular momentum fluxes should be
given with sufficient accuracy by the quadrupole-order formulae 
as given in \cite{ryan3},\cite{shibata_ecc}. However, 
these authors make use of a different set of orbital parameters. 
For example, Ryan's semi-major axis $\bar{a}$ and
eccentricity $\bar{e}$  \cite{ryan3} are related to our parameters via
the transformation,
\begin{eqnarray}
1-e^2 &=& (1-\bar{e}^2) \left [ 1- \frac{4a}{M} 
\left ( \frac{M}{p} \right)^{3/2} e^2 \cos\iota  \right ] \;, 
\\
p &=& \bar{a} (1-e^2) \left [ 1- \frac{2a}{M} 
\left ( \frac{M}{p} \right )^{3/2} e^2 \cos\iota  \right ]  \;.
\end{eqnarray}
Note that at a Newtonian level the two sets are consistent to each other.
Rewriting Ryan's fluxes in terms of our parameters we obtain :
\begin{eqnarray}
\dot{E}_{quad} &=& -\frac{32}{5}\frac{\mu^2}{M^2} 
\left ( \frac{M}{p} \right )^{5} (1-e^2)^{3/2} 
\left [ f_{1}(e)  - \frac{a}{M} \left ( \frac{M}{p} \right )^{3/2} 
 f_{2}(e) \right ] \;,
\label{PN1}         
\\
\dot{L}_{quad}  &=& -\frac{32}{5}\frac{\mu^2}{M} 
\left (\frac{M}{p} \right )^{7/2}
(1-e^2)^{3/2} \left [  f_{3}(e)  
 + \frac{a}{M} 
\left ( \frac{M}{p} \right )^{3/2}
(f_{4}(e) - f_{5}(e)) \right ] \;,
\label{PN2} 
\end{eqnarray}
where
\begin{eqnarray}
f_{1}(e) &=& 1+ \frac{73}{24}e^2 + \frac{37}{96} e^4 \;,
\\
f_{2}(e) &=&  \frac{73}{12} + \frac{823}{24}e^2 + 
\frac{949}{32} e^4 + \frac{491}{192} e^6 \;,
\\
f_{3}(e) &=&  1+ \frac{7}{8}e^2
\\
f_{4}(e) &=& \frac{61}{24} + \frac{63}{8} e^2 + \frac{95}{64} e^4 \;,
\\
f_{5}(e) &=& \frac{61}{8} + \frac{91}{4}e^2 + \frac{461}{64} e^4 \;.
\end{eqnarray}

The formulae (\ref{PN1}),(\ref{PN2}) can be utilised
for order of magnitude estimations even in the strong field regime though
becoming increasingly inaccurate with decreasing $p/M$ (this has been verified
by comparing them to the fully numerical results ).
We can now estimate the timescales of radiative evolution for $p$,$~e$.
For $ p \gg M $ the energy and angular momentum, at leading order
in $M/p$ and $a/M$, are given by,
\begin{eqnarray}
E &\approx& 1 -\frac{M}{2p} (1-e^2) \mp \frac{a}{M} (1-e^2)^2 
\left ( \frac{M}{p} \right )^{5/2} \;,
\\
L &\approx & \pm \sqrt{Mp} -\frac{aM}{p} (3+e) \;.
\end{eqnarray}
Accordingly, eqns. (\ref{pdot1}), (\ref{edot1}) become,
\begin{eqnarray}
\dot{p} &=& -\frac{64}{5} \frac{\mu}{M} (1-e^2)^{3/2}
\left ( \frac{M}{p} \right )^{3} \left [ f_{3}(e)  
\mp \frac{a}{4M} \left ( \frac{M}{p} \right)^{3/2} 
f_{6}(e)  \right ] \;,
\label{pdot5}
\\
\nonumber \\
\dot{e} &=& -\frac{304}{15} \frac{\mu}{M^2} e(1-e^2)^{3/2}
\left ( \frac{M}{p} \right )^4 \left [  f_{7}(e)
\mp \frac{a}{M} \left (\frac{M}{p} \right )^{3/2} f_{8}(e) \right ] \;,
\label{edot5}
\end{eqnarray}
where
\begin{eqnarray}
f_{6}(e) &=& \frac{133}{12} + \frac{379}{24}e^2 + \frac{475}{96} e^4 \;,
\\
f_{7}(e) &=& 1+ \frac{121}{304}e^2 \;,
\\
f_{8}(e) &=& \frac{879}{76} + \frac{699}{76} e^2 + \frac{1313}{608} e^4 \;.
\end{eqnarray}  
The equations above demonstrate the well known fact \cite{peters} 
that, in the weak-field regime, eccentric orbits tend to circularise under 
radiation reaction (while they slowly shrink towards the central body).
For the associated timescales one finds,
\begin{equation}
\frac{T_p}{T_e} = \frac{19}{12} \left ( 1 + \frac{7}{8} e^2 \right )^{-1}
\left ( 1 + \frac{121}{304} e^2 \right ) 
\left [ 1 \mp \frac{a}{M} \left ( \frac{M}{p} \right )^{3/2}
f_{3}^{-1}(e) \left ( \frac{55}{114} + \frac{6431}{1824} e^2 +
\frac{9593}{1824} e^4 + \frac{9191}{4864} e^6 \right )  
\right ] \;.
\label{wtimescales}
\end{equation}
According to this equation, in the weak-field regime the 
eccentricity decays faster than the size of the orbit \cite{cutler}.
The leading order spin term furthermore implies that this behaviour is
more pronounced for retrograde orbits.


\subsection{\textbf{Approximations near the separatrix (II)}}

The previous Section discussed results which are already
familiar from the existing literature \cite{cutler}. 
We now present new results regarding
strong-field orbits which reside near the separatrix. 

The analysis of Section IIIA has shown that the gravitational wave spectrum 
will essentially contain harmonics of $ \Omega_{\rm r},~\Omega_{\rm \phi}$.
We can use the approximate expressions (\ref{Tr3}),(\ref{Dphi3}) to deduce 
that for orbits near the separatrix, i.e. $ p - p_{s}(e) =\epsilon \ll M$ 
(and as long as $e$ is not close to zero and is not unity),
\begin{eqnarray}
\Omega_{\rm r} &\approx& \frac{2\pi}{A_{\rm t}(0)} \left [\frac{eM p_{s}}
{(1+e)(3-e)} \right ]^{1/2}  \left ( \ln \left [ 
\frac{64ep_{s}^2}{\epsilon S(1+e)(3-e)} \right ] \right )^{-1} \;,
\label{Omegr}
\\
\nonumber \\
\nonumber \\
\Omega_{\rm \phi} &\approx& \frac{A_{\rm \phi}(0)}{A_{\rm t}(0)} \;.
\label{Omegphi}
\end{eqnarray} 
Hence, for $\epsilon \to 0 $ we have $\Omega_r \to 0 $ and in effect, the spectrum 
becomes almost ``circular'':
\begin{equation}
\omega_{\rm mk} \approx m \Omega_{\rm \phi} \;.
\end{equation}
Furthermore, by substitution in (\ref{GWflux1}), (\ref{GWflux2}) we get
\begin{equation}
\dot{E}_{GW}^{\infty, H} \approx  \Omega_{\rm \phi} \dot{L}_{GW}^{\infty, H} \;.
\label{circflux}
\end{equation}
We conclude that orbits near the separatrix radiate energy and
angular momentum at rates so that the ratio $\dot{E}_{GW}/\dot{L}_{\rm GW}$ 
is almost equal to the respective ratio of a circular orbit with the same 
$\Omega_{\rm \phi}$. The effective radius of this fiducial orbit is
\begin{equation}
r_{\rm eff}= M^{1/3} \left ( \frac{A_{t}(0)}{A_{\phi}(0)} \mp a \right )^{2/3} \;.
\end{equation}
For example, for the prograde orbit $p=2.11M, e=0.7$ we find 
$r_{\rm eff}=  1.88M > r_{p}= 1.24M $ while for the orbit 
$ p=2.35M, e=0.9$ we find $r_{\rm eff}= 3.90M > r_{p}= 1.24M $ 
(for both cases we have taken $a=0.99M$ ).  

Equation (\ref{circflux}) suggests that particles in zoom-whirl orbits lose 
most of their energy and angular momentum while they revolve near the periastron, 
which is what we would intuitively expect.

We next discuss approximations for $\dot{p}$ and $\dot{e}$ near the separatrix. 
Unfortunately, the lack of a simple analytic expression for $p_{s}(e)$ makes
such a task difficult, and the resulting formulae are quite cumbersome with
little analytic transparency. Nevertheless, we can follow a much simpler path 
and still gain some significant insight. For $p \approx p_{s}$ and using 
eq. (\ref{circflux}), eqns. (\ref{pdot1}), (\ref{edot1}) become,
\begin{eqnarray}
\dot{p} & \approx & \left [ H^{-1} ( L_{,e} \Omega_{\phi} -E_{,e} ) \dot{L} 
\right ]_{p \approx p_{s}} \;,
\label{pdot4}
\\
\nonumber \\
\dot{e} & \approx & \left [ H^{-1} ( -L_{,p} \Omega_{\phi} + E_{,p} ) \dot{L} 
\right ]_{p \approx p_{s}} \;.
\label{edot4}
\end{eqnarray}
By direct substitution of (\ref{sepax}), it can be shown that the function 
$H(p,e)$ becomes exactly zero at the separatrix. On the other hand, and as 
long as $a \neq M$, one can verify numerically that the numerators in 
(\ref{pdot4}), (\ref{edot4}) remain finite near and at the separatrix. 
It follows that for non-extreme Kerr holes both $\dot{p}$ and $\dot{e}$ diverge
at the location of the separatrix. This pathological behaviour signals the 
breakdown of the adiabaticity assumption upon which our method stands. 
A proper discussion of this transition regime should take into account the 
rapid radiative evolution of the orbit (which now varies in a timescale 
comparable to the orbital period).

Moving on, we divide (\ref{edot4}) into (\ref{pdot4}) to get,
\begin{equation}
\frac{\dot{e}}{\dot{p}} \approx \left [  \frac{( -L_{,p} \Omega_{\phi} + E_{,p} )}
{( L_{,e} \Omega_{\phi} -E_{,e} )} \right ]_{p \approx p_{s}} \;.
\label{epratio}
\end{equation}
Exploring the numerical value of this quantity for numerous very near-separatrix 
orbits and black hole spins $a < M $, we have found it to be always negative 
and {\em finite}. This means that $\dot{e}$ and $\dot{p}$ have opposite signs 
near the separatrix. Since the latter is always negative 
(the orbit always shrinks) we conclude that very close to the separatrix 
$ \dot{e} >0 $, i.e. the orbit {\it gains} eccentricity,
(as previously found, in less general cases, in \cite{apostolatos2}, 
\cite{tanaka2}, \cite{dk2}, \cite{cutler}).
Since weak field orbits always lose eccentricity, there must be a critical 
curve 
$p_{crit}(e) $ on the $(p,e)$ plane at which $\dot{e}=0$.  
As eqn. (\ref{epratio}) is formally accurate (within the constraints imposed
by the adiabaticity condition) not only at the separatrix but also in its 
vicinity,
we can actually study the behaviour of $\dot{e}$ in a thin zone near the separatrix.
For a given small or moderate black hole spin, we find that the ratio 
(\ref{epratio}) is again negative. However, for high eccentricities 
$e \approx 1$ we initially get a positive
value which gradually passes from zero and becomes negative as $ p \to p_{\rm e} $.
With increasing $a/M$ we observe the same behaviour at even lower eccentricities,
provided we are considering prograde orbits. The opposite behaviour is observed
for retrograde orbits. For $ a \approx M $, (\ref{epratio}) becomes 
negative only very close to the separatrix for all eccentricities. These results 
suggest that, at least for $ e \approx 1$, the critical  curve $p_{crit}(e)$  is 
located close to $ p_{s}(e)$ (this has been shown to be true for $a=0$ 
\cite{cutler}), 
and that (for prograde orbits) $ p_{crit}(e) -p_{s}(e) \to 0 $ as $ a \to M $ 
(which resembles the situation for nearly circular equatorial Kerr orbits 
\cite{dk2}). 
All of our (semi)analytic predictions are fully supported by the numerical results 
that are presented in Section VB.   

Prograde orbits near the separatrix of an extreme Kerr hole are discussed
separately, and fully analytically, in the following Section. Here we should
emphasise once again that all the approximations presented in this Section are
valid provided $ e \gg \epsilon/M $. This excludes nearly circular orbits, which
have been explored in detail in \cite{dk2}.

We can now write approximate expressions for the timescales $T_e, T_p$ 
for an orbit close to the separatrix.
\begin{equation}
\frac{T_e}{T_p} \approx  \left [\frac{e}{p} \frac{|  
\Omega_{\phi} L_{,e} - E_{,e} |}{| E_{,p} -\Omega_{\phi} L_{,p} |} 
\right ]_{p \approx p_{s}} \;.
\label{stimescales} 
\end{equation}
For example, for a $a=0.99M$ Kerr hole, we have
$ T_e/T_p = 0.81 $ for $p=1.7M, e=0.3$ while for
$p=2.11M, e=0.7 $ we get $ T_e/T_p= 4.5 $ (for both orbits, eqn. (\ref{stimescales})
is a good approximation). This situation is typical for non-extreme holes. 
As we move along the separatrix keeping a fixed distance from it, the ratio 
$T_e/T_p$ tends to increase (and become larger than unity) with increasing 
eccentricity. In comparison, the corresponding weak-field  timescales ratio, 
eqn. (\ref{wtimescales}), is relatively insensitive to variations of 
eccentricity.


\subsection{\textbf{Horizon-skimming orbits}}
A particularly interesting class of prograde strong-field orbits are those that 
potentially ``graze'' the black hole horizon. These orbits can only exist provided 
the black hole is near extremally rotating, $a \approx M $ 
(this can be deduced from Fig.~\ref{sepx}). Circular, non-equatorial 
horizon-skimming orbits were first studied by Wilkins \cite{wilkins} and more 
recently by Hughes \cite{scott_hs}. Here, on the other hand, we discuss 
equatorial horizon-skimming orbits of arbitrary (but not equal to unity or 
close to zero) eccentricity around an extreme Kerr black hole. 

As the separatrix for these orbits takes the very simple form 
$ p_{s}(e)= M(1+e) $ we can duplicate the analysis of the
previous section following a purely analytical path.
Expanding (\ref{x2}) around $p= p_{s}= M(1+e) $ we find that  
\begin{equation}
x^2= M^2 \left ( \frac{1+e}{3-e} \right ) + {\cal O}(\epsilon) \;.
\end{equation}
We then get for the energy and angular momentum,
\begin{eqnarray}
E &=& \sqrt{ \frac{1+e}{3-e} } + g(e) \frac{\epsilon}{M} + {\cal O}(\epsilon^{2}) \;,
\\
\\
L &=& 2M \sqrt{ \frac{1+e}{3-e} } + f(e) \frac{\epsilon}{M} + {\cal O}(\epsilon^{2}) \;.
\end{eqnarray}
The explicit form of the functions $ g(e), f(e) $ is not required for
the following analysis.
We use these equations (together with (\ref{circflux}), noting that 
$\Omega_{\phi}= 1/2M $ on the separatrix for the orbits under discussion)  
to obtain,
\begin{eqnarray}
H & \approx & \frac{2 [ 2M g(e) -f(e) ]}{(1+e)^{1/2} (3-e)^{3/2} } \;,
\\
\\ \nonumber
 -E_{,e} \dot{L} + L_{,e} \dot{E}  &\approx& M [ 2M g(e) -f(e) ] \dot{E} \;, 
\\
\\ \nonumber
 E_{,p} \dot{L} - L_{,p} \dot{E}  & \approx& [ 2M g(e) -f(e) ] \dot{E}  \;.
\end{eqnarray}
Here, unlike the non-extreme case, the function $H$ remains
finite as $\epsilon \to 0$.
The above formulae, as well as the following ones, have a fractional error 
$ {\cal O}( \epsilon/eM )$. 
Hence for horizon skimming orbits,
\begin{eqnarray}
\dot{p} &\approx& \frac{1}{2}M (1+e)^{1/2} (3-e)^{3/2} \dot{E} \;,
\label{hspdot}
\\
\dot{e} & \approx & \frac{1}{2} (1+e)^{1/2} (3-e)^{3/2} \dot{E} \;.
\label{hsedot}
\end{eqnarray}
We see that both rates are finite all the way down to the separatrix, 
unlike the $ a< M$ case. However, the adiabaticity condition (\ref{adiab}) 
is still invalidated at $ p= p_{s} $.
 
More interesting is the behaviour of the ratio of the rates (\ref{hspdot}),
(\ref{hsedot}),
\begin{equation}
\frac{\dot{e}}{\dot{p}}= \frac{1}{M} + {\cal O}(\epsilon/eM) \;.
\end{equation}
This is always positive, which means that the $ \dot{e} > 0$ region that exists
for $ a < M $ shrinks to zero for extreme Kerr black holes. In other words, the
critical curve $ p_{crit}(e) $ has the same value of the Boyer-Lindquist
coordinate as the separatrix itself. This
conclusion completes the discussion of the previous subsection.
For the ratio of the respective timescales we get,
\begin{equation}
\frac{T_e}{T_p}= \frac{e}{1+e} + {\cal O}(\epsilon/eM) \;.
\label{hscales}
\end{equation}
As (\ref{hscales}) predicts, the eccentricity always evolves 
more rapidly than the semi-latus rectum, which again is contrary 
to the situation with weak-field orbits.


\section{\textbf{Numerical results}} 

\subsection{\textbf{Method and error estimates}}

Numerical solution of the Teukolsky equation or related equations has been a
minor industry for nearly thirty years, since the pioneering work of Press and 
Teukolsky \cite{Press}, Detweiler \cite{detweiler} and Sasaki and Nakamura
\cite{SN}. Our method is based on a numerical algorithm outlined in \cite{sussman} 
and employs subroutines found in \cite{Recipes}. It involves the 
use of  Bulirsch-Stoer integration to solve the Sasaki-Nakamura 
equation. Our code is a direct descendent of the codes used in \cite{cutler} 
and \cite{dk2}, since we deal with both arbitrarily eccentric orbits and black 
holes of non-zero spin. Romberg integration is used both to integrate Eqns. 
(\ref{phichi}) and (\ref{tchi}) and to integrate Eqn. (\ref{Zlmk4}). 
To calculate the spheroidal harmonic functions $S_{\ell m}^{a\omega}$ we use the 
``spectral decomposition'' method described in \cite{scott_circ}. The reliability 
of these methods in general is well known.

We were able to check our numerical results for circular equatorial orbits with 
those of \cite{scott_circ}, where our agreement was good to 5 or 6 significant 
digits, and with the codes on which this code was based \cite{cutler}, 
\cite{dk2}, for eccentric orbits in Schwarzschild and for nearly circular orbits 
in Kerr, and again our agreement was good to 4 to 6 significant digits. 
A similarly good agreement was achieved by comparison with the published results 
of Tanaka {\em et.al.} \cite{tanaka2} for equatorial eccentric orbits in 
Schwarzschild and with those of Shibata \cite{shibata_circ} for circular equatorial
orbits in Kerr. We were also able to compare our results with those given by 
Shibata \cite{shibata_ecc} for equatorial eccentric orbits in Kerr. In this 
case, however, we did find some disagreement of about $ \sim 1 \% $. 
The cause of this disagreement is not apparent, while it stays roughly at the
same level for moderate and high eccentricities. The disagreement does not seem 
to be due to the problems of maintaining accuracy with the long runtimes and 
large number of harmonics required for moderate/large 
eccentricities. The error introduced by the truncation of the 
$ \ell, k$ sums in the flux calculation does not seem to be the source of the
disagreement. We cannot say at present which code might be at fault.

Finally, we have also been able to compare our code with some results for circular 
orbits in Kerr from \cite{finn}, and again our agreement is good to several 
significant figures. Similarly, comparison with post Newtonian results for 
eccentric Kerr orbits (as found, for example, in \cite{ryan3} ) reveals 
good agreement in the weak-field regime. In Table~\ref{tab_comp} we compare some 
sample results.

In view of the lack of any check for strong field orbits with high/moderate 
eccentricities and high spins, it is of obvious importance that we present some 
estimate of the likely error in our numerical results. The main sources of 
numerical error in our code are as follows:

1. Inaccuracy in the Bulirsch-Stoer integration routines, from \cite{Recipes}. 
We set the relative accuracy parameter $eps$, which governs the convergence of 
the final result, at  $10^{-6}$. 

2. Inaccuracies in the Romberg integrator, also from \cite{Recipes}. We set 
$eps=10^{-6}$ for the routines which integrated Eqs. (\ref{tchi}) and  
(\ref{phichi}). This parameter governs the level of convergence which the 
routine demands in the final result, before it stops iterating. However, in 
the case of the Romberg routine which governed the main program loop, i.e. 
the integration of Eq. (\ref{Zlmk4}), we typically set $eps=10^{-5}$ in many 
cases in order to achieve large savings in computing times. In a few runs 
designed to produce data for illustration of waveforms only 
(not numerical data on flux quantities), we used  $eps=10^{-4}$.

3. Our method requires that we calculate the quantity $B^{\rm in}$ 
(see Eqn. (\ref{Zlmk4}) above). To do this we integrated the Sasaki-Nakamura 
equation (\ref{sneq})  out to $r=100/\omega_{mk}$ and then successively doubled 
the limit of integration, until our Richardson extrapolator told us we had 
achieved convergence to a relative accuracy of $10^{-5}$.

4. Our method for calculating spheroidal harmonic functions $S_{\ell m}^{a\omega}$ 
involved writing them as an expansion of an infinite set of spherical harmonic 
functions. Fortunately this expansion can be truncated at 30 terms and remain 
very precise in most cases, but for high black hole spins, $a$ and high angular
frequencies, $\omega_{mk}$ we were obliged to use 40 terms to avoid truncation 
errors causing small high frequency ripples in the wave forms. However, in our 
numerical results of flux rate and orbital evolution this source of error 
appears to be considerably less than $10^{-6}$.

5. In principle our calculation of fluxes must be summed over an infinite number 
of harmonics in each of the integers $\ell$,$m$ and $k$. In practice truncating 
these sums for the $\ell$ and $m$ harmonics was not difficult. Fluxes for a 
sequence of these harmonics usually monotonically decrease after a peak at some 
value of $\ell$ and $m$ and so we demanded that the loop through these variables
halt once fluxes went below a factor of $10^{-5} $ times the peak contribution.
However the spectrum of fluxes in the $k$ harmonic was much more complicated, 
typically involving several peaks (see figures in Section VD) before 
finally monotonically decreasing after a number of peaks which increased for 
increasingly eccentric orbits. We examined 
spectra in $k$ by hand to confirm the machine's results and experimented widely 
to convince ourselves that we had caught all significant contributions to the
total flux from different frequencies.

Clearly there are several significant independent sources of error, so that we 
can only offer our best judgement of the total relative error in our code in 
those cases where we have no independent check on our results. 
As we have addressed every systematic source of error that we encountered, and 
as we are confident that the code is running correctly in all of the cases dealt 
with in this paper, we estimate the relative error for numerical results quoted 
in this paper as no greater than $ 10^{-3}- 10^{-4}$, 
in the case of fluxes, $\dot{E}$ and $\dot{L}$, 
and no greater than $10^{-2} - 10^{-3}$ for quantities such as $\dot{e}$ 
and $\dot{p}$  because cancellations between terms when converting flux numbers 
into orbital evolution quantities tends to increase the size of the relative 
errors. 
This is especially true near the critical point where the rate of 
change of the eccentricity becomes zero, due to the complete cancellation
of these terms. 
As an illustration of this, we will note in passing that the mysterious "bump"
seen in Fig. 2 of Ref. \cite{cutler} turns out to be due to a rare case where flux
errors which appear insignificant themselves are greatly magnified when the
numerical flux data is combined to produce $\dot{e}$ and $\dot{p}$.

It is useful, in this context, to mention that in comparing the results
from our code to the code in \cite{dk2}, the flux quantities for radiation
emitted toward infinity agree to about 1 part in $10^{-6}$, the flux quantities
for radiation towards the horizon agree to about 1 part in $10^{-4}$ and
the position of the critical curve, as calculated by the two codes, can
disagree by about 1\%. 
This suggests that the only way in which our numerical error is large enough to 
make a visible difference in our figures would be as a slight change in the position 
of the critical curve in Fig.~\ref{planes} (the retrograde case). 


\subsection{\textbf{Backreaction on the orbit}}

\begin{figure}[tbh]
\centerline{\epsfysize=8cm \epsfbox{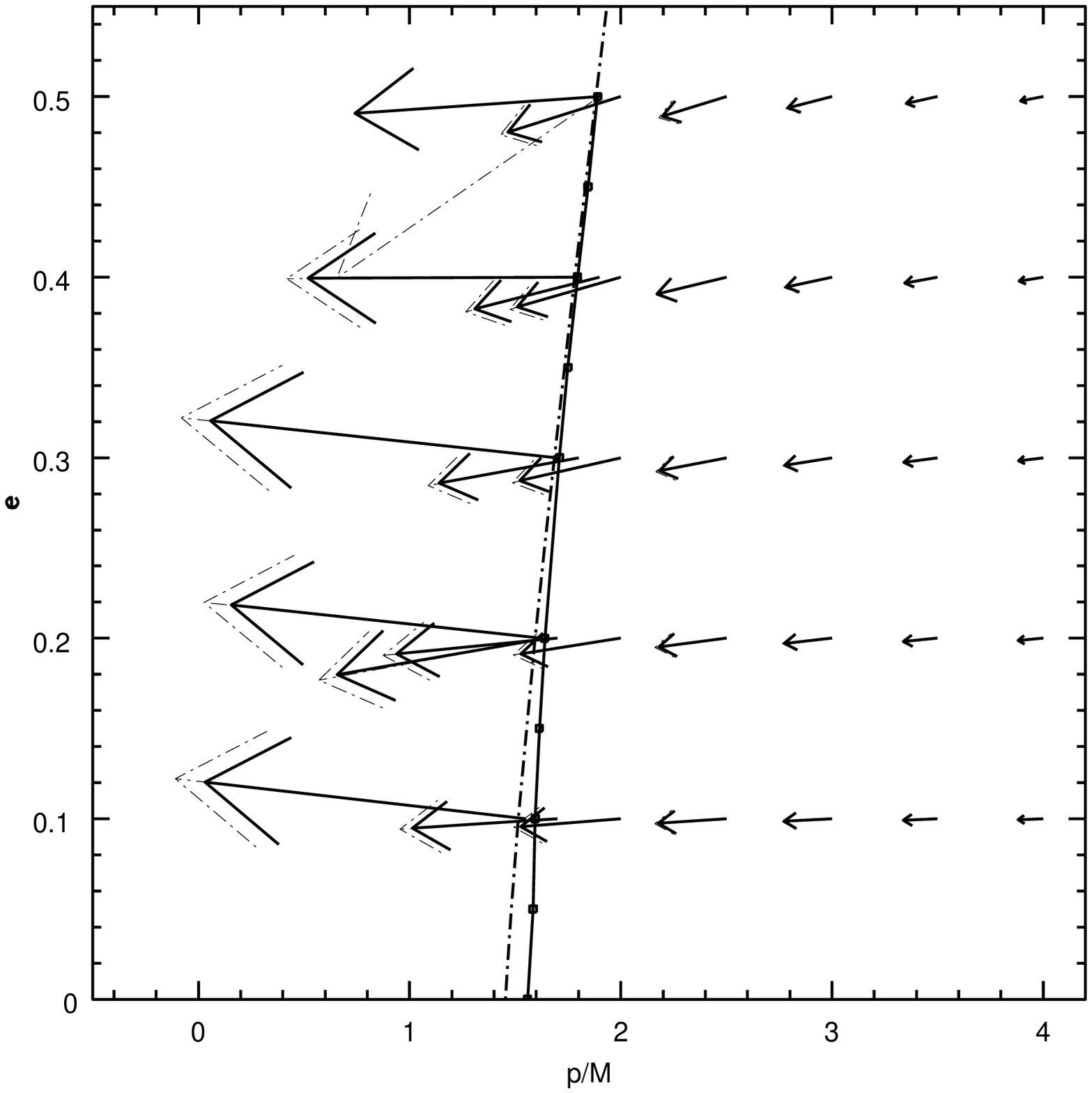}}
\centerline{\epsfysize=8cm \epsfbox{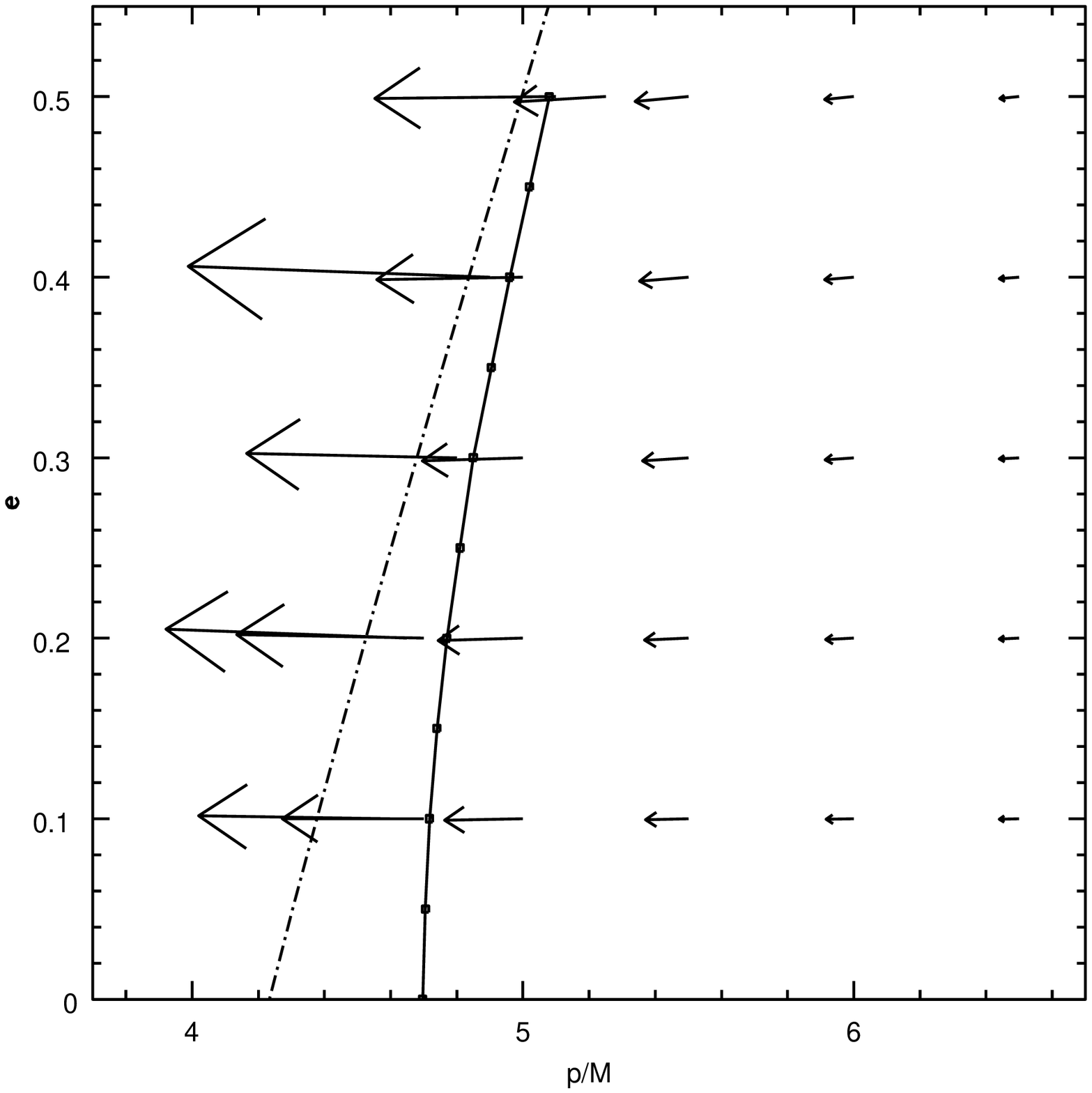}}
\centerline{\epsfysize=8cm \epsfbox{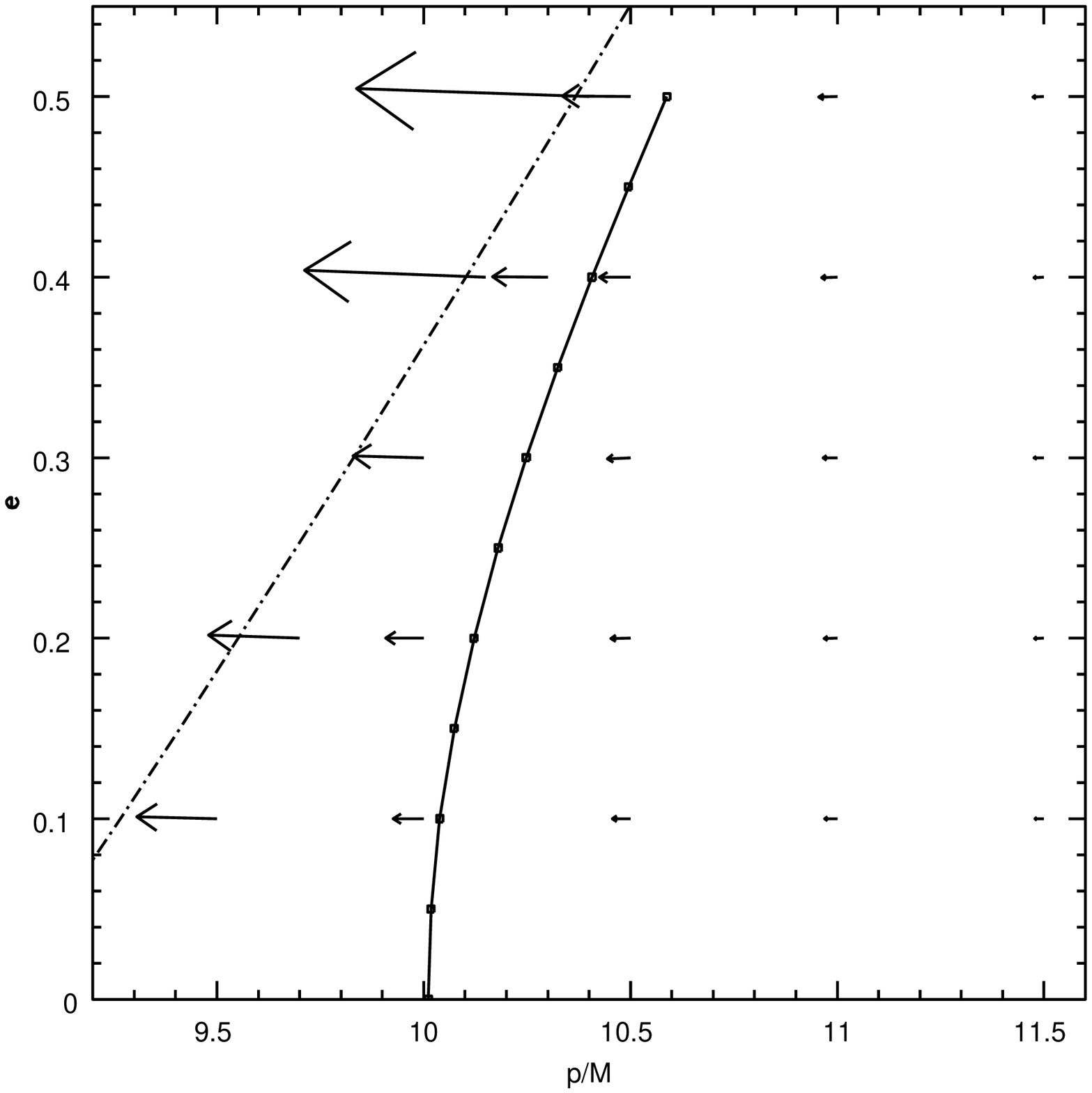}}
\caption{The evolution of a family of eccentric equatorial orbits, illustrated on
the $(p,e)$ plane. The black hole spin is $a=0.99M$ (top graph, prograde orbits;
bottom graph, retrograde orbits) and  $a=0.5M$ (graph in the middle; prograde 
orbits). The dashed curves represents the separatrix of stable orbits, while the 
solod curves represents the critical $\dot{e}= 0 $ curve. Each orbit 
corresponds to a point in the graph, and its (adiabatic) radiative evolution 
is represented by a vector with components 
$ (M/\mu)(\dot{p}, M \dot{e})$. Solid and dashed arrows represent the 
orbital evolution respectively with and without including the fluxes at the 
black hole horizon (the difference between these arrows is visible
only in the $a=0.99M$ case). When the black hole is rapidly spinning the
horizon flux effectively represents gain of energy for the orbiting particle,
an effect attributed to superradiance, see discussion in the main
text for more details. As a result, the inspiral of the body 
is stalled and the critical curve is slightly pushed outwards. This is the
reason for the strong misalignment between the solid and dashed vectors
at the point $p=1.9, e=0.5$ of the $a= 0.99M$ plane.
Note the much more pronounced orbital evolution for 
the prograde $a=0.99M$ case (a consequence of the particle's motion in very 
strong field regions) and the approach (diverge) of the relative positions 
of the separatrix and the critical curve between the 
$a=0.5M$ case and the prograde (retrograde) $a=0.99M$ case.}
\label{planes}
\end{figure}

In this Section we present numerical results on the evolution of bound equatorial
orbits in terms of $p$ and $e$. This pair of parameters is preferable to
the equivalent set $E,L$, because of their clearer geometrical meaning. 
We have calculated the averaged rates $\dot{p}$,$\dot{e}$ for a number of
prograde and retrograde orbits and for two different black hole spins: 
$a=0.5M$, and $a=0.99M$. A representative part of our numerical results is 
presented in Fig.~\ref{planes}. Each individual orbit is represented as 
a point on the $(p,e)$ plane. At each one of those points we have attached a 
vector with components $ (M/\mu)(\dot{p},M\dot{e}) $ 
that indicates the direction at which the orbit adiabatically evolves under 
radiation reaction.
Moreover, all orbits shown are chosen so as to be strongly adiabatic for the 
typical mass ratio $10^{-6}$ (the most severe constraint is $\mu/M \ll 10^{-2} $ ).
These figures (together with the analytic approximations of Sections IVA, IVB and 
IVC) lead to the following conclusions :

\begin{itemize}

\item The semi-latus rectum $p$ always decreases (the orbit is shrinking).
      In fact, $|\dot{p}|$ grows monotonically, and finally diverges, 
      as the separatrix is reached. This divergence is an artificial feature of our
      formalism, associated to the breakdown of the adiabatic approximation.  

\item The eccentricity $e$ shows a more complicated behaviour. For sufficiently
      large $p$, we always find $\dot{e}<0$. However, as the orbit approaches 
      the separatrix, $\dot{e}$ changes sign and becomes positive, i.e near the 
      separatrix, eccentricity increases. As in the case of $\dot{p}$, $\dot{e}$ 
      will also diverge at the separatrix, due to the failure of adiabaticity.   

\item As the black hole spins up, in the case of prograde orbits the critical 
      radius after which  $\dot{e}>0$ moves closer to the separatrix (in coordinate 
      terms) 
      for a given $e$. The same is true for a fixed black hole spin, but for 
      increasing $e$. For retrograde orbits, spinning up the black hole tends
      to move the $\dot{e}=0$ curve away from the separatrix. 

\item In a sense, the increasing eccentricity regime is a precursor of
      orbital instability and plunging. This is hinted by the proximity of the 
      critical curve $p_{crit}$, where $\dot{e}$ flips sign and becomes positive, 
      to the
      separatrix curve $p_{s}(e) $ which is the boundary between stable and
      unstable bound orbits. Qualitively speaking, at this stage of the inspiral, 
      the radial  potential $V_r$ is quite ``flat''  and as a consequence the 
      particle has more room to move radially, even as it continues to ``sink'' 
      towards the bottom of the potential well (the behaviour which is responsible 
      for the characteristic ``circularizing'' tendency).

\end{itemize}

These results agree with, and at the same time generalise previous results
concerning bound orbits (of arbitrary $e$) around Schwarzschild black holes
\cite{cutler} and slightly eccentric equatorial orbits around Kerr black holes 
\cite{dk2}. As we have discussed in Section VC, some of the conclusions above must 
be modified when the black hole is extreme ($a=M$).
 
In Table~\ref{tab_num} we give a sample of our numerical data, for the energy and 
angular momentum fluxes as well as for $\dot{p},~\dot{e}$, for some of the orbits 
presented in Fig.~\ref{planes}. As we have discussed, we believe that these numbers 
have fractional accuracy at least $10^{-3}$.

Another important result concerns the significance of the horizon fluxes  on the  
evolution of orbits with relatively small periastrii. Specifically, we have 
encountered very-strong field orbits for which 
$|\dot{E}_{GW}^{\rm H}| \sim 0.1 |\dot{E}_{GW}^{\infty}| $. 
However, the most intriguing property of the 
horizon fluxes is that they assist, in a sense, the orbiting body. 
This is most easily illustrated by plotting the evolution of the set 
of orbits of the top graph of Fig.~\ref{planes}, without including the horizon flux
(represented by dashed arrows in Fig.~\ref{planes}; the solid arrows represent the
total rates). For very strong field orbits, when the horizon fluxes are taken 
into account, the shrinking of the orbit is noticeably stalled. 
Typically, when $\dot{E}_{GW}^{\rm  H} $ is non-negligible it also happens that it 
represents energy gain instead of energy loss (in other words the fluxes 
$ \dot{E}_{GW}^{\rm H} ,\dot{E}_{GW}^{\infty} $ 
have opposite signs). The orbiting particle is effectively draining energy from 
the black hole itself. This is just a manifestation of the so-called superradiance 
phenomenon, well known in black hole physics \cite{Press},\cite{superr}: 
waves scattered within the black hole's ergoregion and having frequencies 
(as measured at infinity) that lie in the interval  $ 0 < \omega < m \omega_{+} $, 
effectively appear (for a distant observer) as emerging from the horizon, and 
amplified at the expense of the hole's rotational energy. The outgoing reflected 
waves ``push'' the particle outwards, and this interaction
is manifested as a gain of orbital energy and angular momentum.    
Our result can be easily understood if we recall that the ergoregion is growing for 
increasing black hole spin. At the same time, because the boundary of instability
moves in to lower radii with increasing spin, a particle can enter regions with
much stronger fields
and therefore emit a substantial amount of radiation towards the ergoregion 
(see Fig.~\ref{sepx}). Hence
we find a significant negative (superradiant) horizon flux. 
An alternative way of viewing this phenomenon is as
an exchange of energy and angular momentum via tidal coupling analogous to tidal 
friction in the Earth-Moon system (and elsewhere). For an exposition of this 
intuitively instructive viewpoint see \cite{scott_insp}, and references therein.

It was recognised long ago \cite{Teuknature} that if the superradiance effect ever 
became large enough a floating orbit would result when it balanced the energy 
loss due to radiation emitted towards infinity. Our results confirm earlier work 
in collaboration with Scott Hughes, suggesting that even for orbits 
very close to the horizon of very rapidly rotating black holes, the gain in energy 
from superradiance is only 10\%   
of the energy lost by the system as a whole. For further details, 
see \cite{scott_insp}.

\begin{figure}[tbh]
\centerline{\epsfysize=7cm \epsfbox{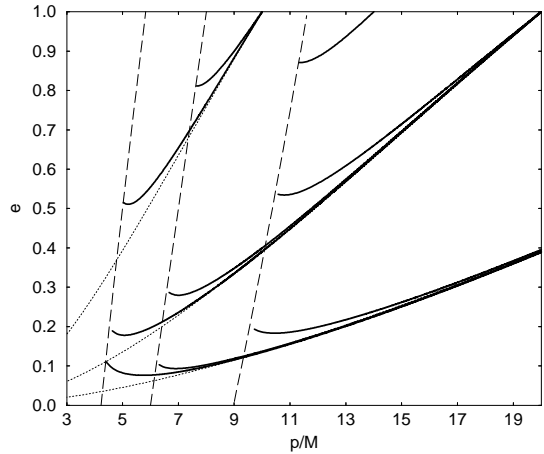}}
\vspace{0.3cm}
\caption{The radiative inspiral of a set of equatorial eccentric orbits with
initial parameters (solid curves from left to right) $r_p= 5, 10, 20M$ and 
$r_a= 10^6 M$, and for black hole spins $a=0, 0.5M$. An additional set of 
retrograde orbits with $r_p= 14, 20, 40M$ and same apastron, and for $a=0.99M$ is also 
shown (solid curves on the right side). The three dashed lines represent the separatrices 
for corresponding spins. Dotted curves are the Newtonian-order predictions, while the 
solid curves are the result of a  more accurate calculation discussed in the main text. 
Note the significant qualitive difference between the two calculations at the vicinity of 
each separatrix.}
\label{inspfig}
\end{figure}

The results presented so far in this Section, although very insightful, are still
incomplete in the sense that they do not describe the radiative evolution of a 
single given orbit on the ($p,e$) plane. Instead, they provide local information
on the evolution of an orbit at a given point. An effort to ``paste'' together
a sequence of such points, in order to follow the full orbit is currently underway
\cite{paperII}. Meanwhile, we can use certain approximations to
foresee what the full inspiral trajectory will look like. As a starting point we
use the leading quadrupole-order expressions for $\dot{p}, \dot{e} $, 
i.e. setting $a=0$ in eqns (\ref{pdot5}),(\ref{edot5}) and derive,
\begin{equation}
p(e)= p_i  \left ( \frac{e}{e_i}  \right)^{12/19}
\left [ \frac{1 + \frac{121}{304}e^2} 
{1 + \frac{121}{304} e_{i}^{2} } \right ]^{870/2299} \;.
\label{inspN}
\end{equation}
Given some initial values $p_i, e_i$ this relation describes, in the weak-field
limit, the trajectory of the orbit on the ($p,e$) plane. Such curves for 
astrophysically relevant initial parameters, are shown in Fig.~\ref{inspfig}
(these curves remain essentially unchanged when the spin terms are retained
in eqns. (\ref{pdot5}), (\ref{edot5}) ).
One feature that is immediately seen in the 
Newtonian-order inspiral is the absence of the critical curve $\dot{e}= 0$ 
and the subsequent $\dot{e} >0 $ behaviour.
This should not come as a a surprise, as expression (\ref{inspN}) 
is not formally valid unless $ p \gg M $. It is not safe to use
weak-field approximations in strong field regimes. 

A simple way to make better predictions is by using the exact expressions
(\ref{pdot1}), (\ref{edot1}) for $\dot{p}, \dot{e} $, but still employing the 
weak-field formulae (\ref{PN1}), (\ref{PN2}) for the fluxes. The outcome of this 
trick is also shown in Fig.~\ref{inspfig}, for a set of orbits with initial 
parameters $r_p= 5, 10, 20M$ and $r_a= 10^6 M$ (this translates to 
$e= 0.99999, 0.99998, 0.99996$ and $p=10, 20, 40M$ respectively) for $a=0, 0.5M$. 
We also considered a set of retrograde orbits with initial $r_p= 7,10,20M $ and 
the same apastron as before, and $a=0.99M$ spin. Note that these curves, like the
ones given by eqn. (\ref{inspN}), are shape-invariant with respect to the 
mass ratio as long as $\mu/M \ll 1$. 
It is rewarding to see that these new trajectories do show the existence of the 
$\dot{e} > 0 $ region, and additionally are in good qualitive and quantitive agreement 
with the accurate numerical results \cite{inspiral_paper}. This is true as 
long as we do not attempt to evolve prograde orbits around rapidly spinning 
black holes, because the agreement quickly degrades when $p$ becomes small. 
Essentially this approach takes into account the correct 
form of the potential, which is the main cause behind the change in sign of 
$\dot{e}$, but the fact that the PN fluxes are increasingly inaccurate in 
strong-field regions precludes precise numerical agreement with the real 
trajectories.

The new curves clearly predict a higher residual eccentricity as compared to
the pure Newtonian curves. This approximate result strongly suggests that many
astrophysically relevant inspiralling orbits will have a significant amount of 
eccentricity left when they are close to plunging and will therefore be likely to 
exhibit zoom-whirl behaviour. Our full numerical results cannot at present be used 
to reproduce complete trajectories, but Fig.~\ref{planes}, which displays arrows 
which are tangential to these trajectories at individual points, certainly shows 
that if significant eccentricity remains at $p \sim 5 M$, that this eccentricity 
will not disappear in the last part of the inspiral before plunge.

The work of Freitag \cite{freitag} suggests that the initial periastra of scattered 
compact bodies which will eventually plunge into the black hole due to radiation 
reaction will be generally less than $40M$. However below that point the distribution 
of their periastra will be fairly flat, so that small initial periastra will be just 
as likely as large ones. As our figure shows,
one expects, in the case of a Schwarzschild black hole, that bodies with initial 
$r_p>20M$ will have $e<0.1$ by the time of plunge. But for initial 
$r_p<20M$ the final eccentricity will be $e>0.1$ and can easily be as great as 
$e \sim 0.7$ (see Fig.~\ref{inspfig}) or higher. For instance, for $r_p=10M$, the 
final eccentricity will be $e \sim 0.3$ (see Fig.~\ref{inspfig}).  
We know that retrograde orbits will have less time to 
circularize and a longer "de-circularizing" time, so eccentricities in this case 
should be greater. This is clearly seen in Fig.~\ref{inspfig}. 
In the case of prograde orbits we should generally expect smaller eccentricities 
before plunging  (as compared to orbits around Schwarzschild black holes) but still 
at a significant level.  Again Fig.~\ref{planes} 
suggests that the change in eccentricity will 
not be great, despite the longer circularizing and shorter de-circularizing times.
Moreover, near extreme holes allow a wider range of initial periastra (for prograde 
orbits), even down to $r_p \sim 2M$ and in these cases the residual eccentricity 
will be quite large. One concludes that eccentricity will play an important role 
in signal analysis for LISA.

The technique outlined here has been recently applied in constructing approximate,
radiation reaction-driven, inspirals of test-bodies in generic Kerr orbits (for which 
case only weak-field results are currently available \cite{ryan3}). 
For more details see \cite{inspiral_paper}.

Motivated by the results discussed above, we would like to obtain simple 
estimates of the total amount of eccentricity gain and the number of orbits the 
particle will spend in the $\dot{e} >0$ region. Such estimates may provide a 
useful guideline for assessing the observational importance of this phase.

The numerical radiation reaction evolution-arrows in Fig.~\ref{planes} show that
the gradient $de/dp$ grows (in the negative sense) monotonically as soon as
the critical curve is crossed. Hence, the maximum (negative) value is attained 
exactly at the separatrix. The approximate expression (\ref{epratio}) 
for $de/dp$ is expected to be very accurate there. We can use this fixed gradient
and extrapolate out to the critical curve, at some eccentricity $e_i$, 
for a given eccentricity $e_f$ at the separatrix. Then we have that  
$\delta e= e_f -e_i  \approx ( p_{s}(e_f) -p_{c}(e_f) ) (de/dp)_{p_{s}(e_f)} $.
This number should set an {\em upper} limit of the total increase in
eccentricity. Results for some representative cases are given in 
Table~\ref{tabgain}. From these numbers we deduce that, at best, there is
a fractional increase of 5-50 \% in eccentricity, the most favourable
case being low-eccentricity retrograde orbits around rapidly spinning
black holes. This gain decreases as we move upwards to larger final
eccentricities (basically due to the shrinkage of the $\dot{e} >0 $ region).
We therefore conclude that we should not expect any dramatic increase
in eccentricity when the orbit is about to become unstable.

A crude estimate on the number of orbits can be made by integrating
(\ref{pdot4}) to find the time required to cross the $\dot{e} > 0 $ region,
\begin{equation}
t_{c} \sim \frac{1}{\dot{L}} \int_{p_{crit}}^{p_s} 
\frac{H dp}{( L_{,e} \Omega_{\phi} -E_{,e} )} \;,
\label{tcross}
\end{equation}
where we have factored out the angular momentum flux as it can be
taken as constant within the integration interval (and recovered by our numerical
data). Similarly we have assumed a fixed eccentricity. The number of orbits is then 
calculated by dividing $t_c$ with a typical period $T_r$ (or $T_{\phi}$). 
We give some representative results in Table~\ref{tabNorb}. We need
to emphasise that these numbers should be viewed only as order-of-magnitude
estimates, as eqn. (\ref{tcross}) is a rough approximation. Nevertheless, we
can still draw some reliable conclusions. 
For a small eccentricity our numbers are in agreement with existing
results for nearly circular orbits around a Kerr black hole \cite{dk2}.
For $e=0.1$ we should typically have a few thousands revolutions in the 
$\dot{e} >0 $ region around a $a=0.99M$ hole and for a mass ratio
$\mu/M \sim 10^{-6}$. For the same parameters, but for retrograde orbits, there
is an order of magnitude increase in the number of orbits. On the other hand,
for all cases, there is an order of magnitude (or more) decrease as we move to 
eccentricity $e=0.5$. Note that it is possible to have a small number of full
orbits, but yet a significant number of azimuthal revolutions or "whirls"
(see for example
the $a=0.99, e=0.5$ case in Table~\ref{tabNorb}).


\subsection{\textbf{Waveforms and fluxes from zoom-whirl orbits}}

Let us now focus on the class of orbits we have named zoom-whirl. As we have 
shown, orbits located near the separatrix should radiate 
in accordance with (\ref{circflux}), as though they were nearly circular. 
In Table~\ref{tab3z} we list numerical fluxes for zoom-whirl orbits of various 
eccentricities. It is clear from these results that as the separatrix is 
approached, the analytic prediction (\ref{circflux}) is indeed confirmed. 
However, one has to be very cautious when applying (\ref{circflux}) to the 
study of a real astrophysical, extreme mass ratio, binary system. As our 
data reveal, in the region where this relation is fractionally accurate at 
the level of $\sim 10^{-2}$, the adiabaticity constraint
(\ref{adiab}) on the mass ratio is quite severe, typically 
$ \mu/M \ll 10^{-2}-10^{-3} $.

\begin{figure}[tbh]
\centerline{\epsfxsize=11cm \epsfbox{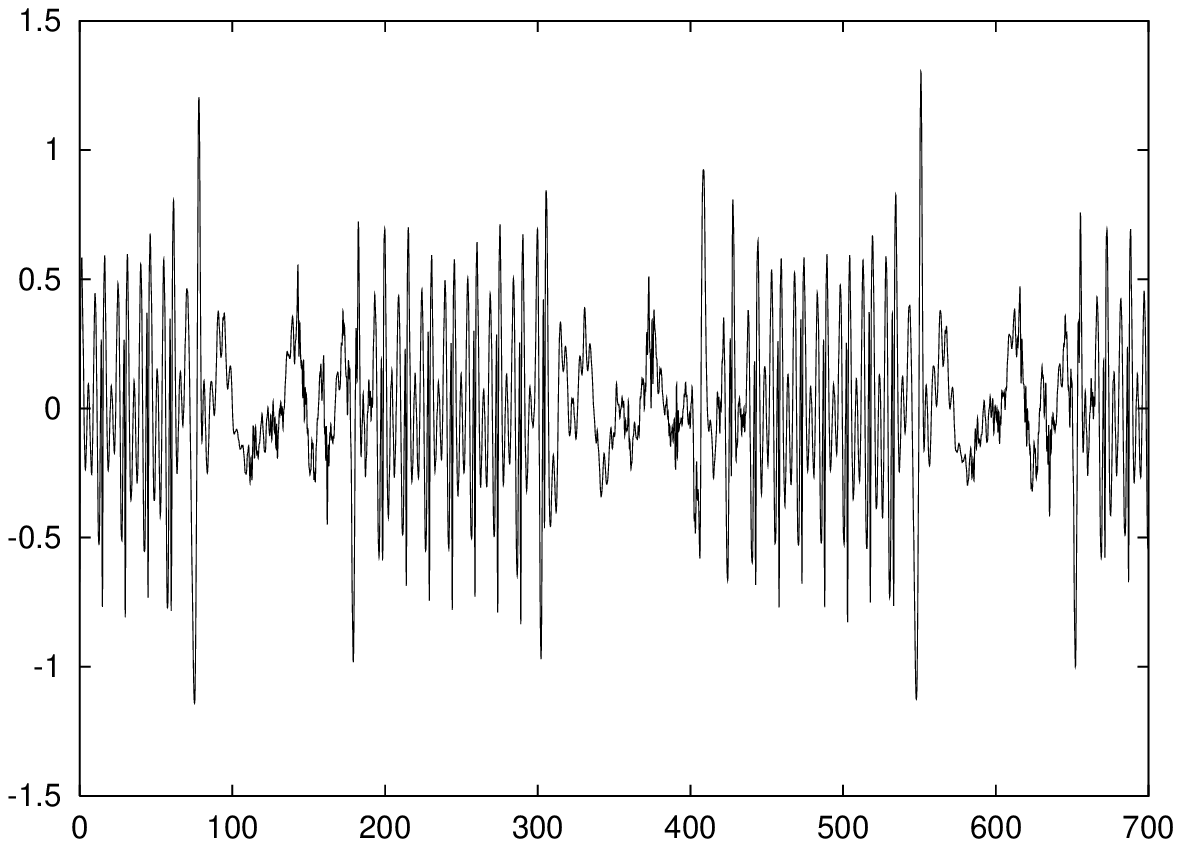}}
\centerline{\epsfxsize=11cm \epsfbox{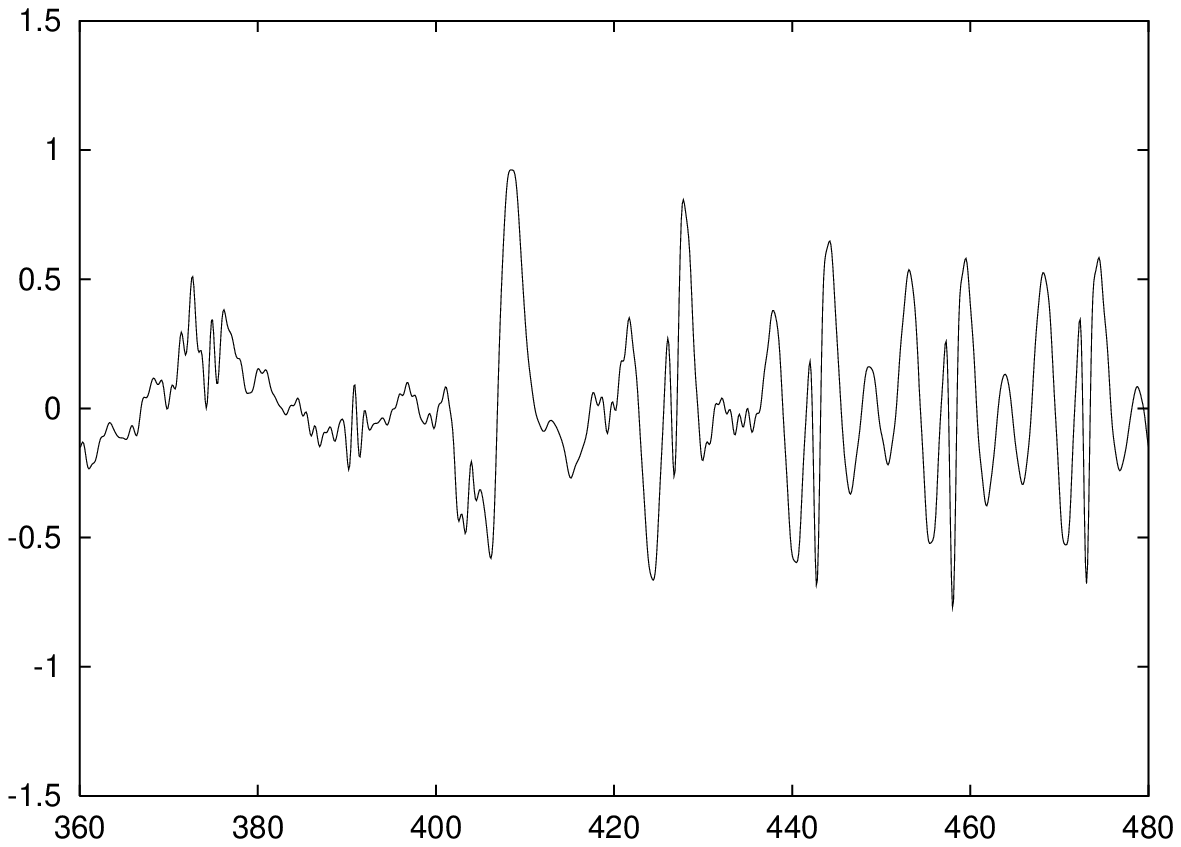}}
\caption{The waveform produced by a particle in a zoom-whirl orbit with 
parameters $p=2.11M, e=0.7$. Specifically, we graph the quantity 
$ (\mu/r) h_{+}  $ (where $r$ is the distance to the observation point, 
which is taken to be on the hole's equatorial plane) versus the 
retarded time $ t- r_{\ast}(r) $ (in units of $M$). We have set the black 
hole spin at $a=0.99M$ and included up to $\ell=18$ multipoles 
in order to generate this figure. The orbital period is $T_r= 236.8M$ and 
$N_r= 10.5$. Note the very characteristic shape of the waveform, 
which is a periodic succession of high-amplitude/high-frequency parts 
(coming from the whirling motion of the particle near the periastron) and 
intervening low-amplitude/low frequency parts (from the zooming in and out 
motion). On the bottom panel, the same waveform is graphed over a 
shorter time interval, offering a clearer view of its rich structure.}
\label{wfs1}
\end{figure}

\begin{figure}[tbh]
\centerline{\epsfxsize=11cm \epsfbox{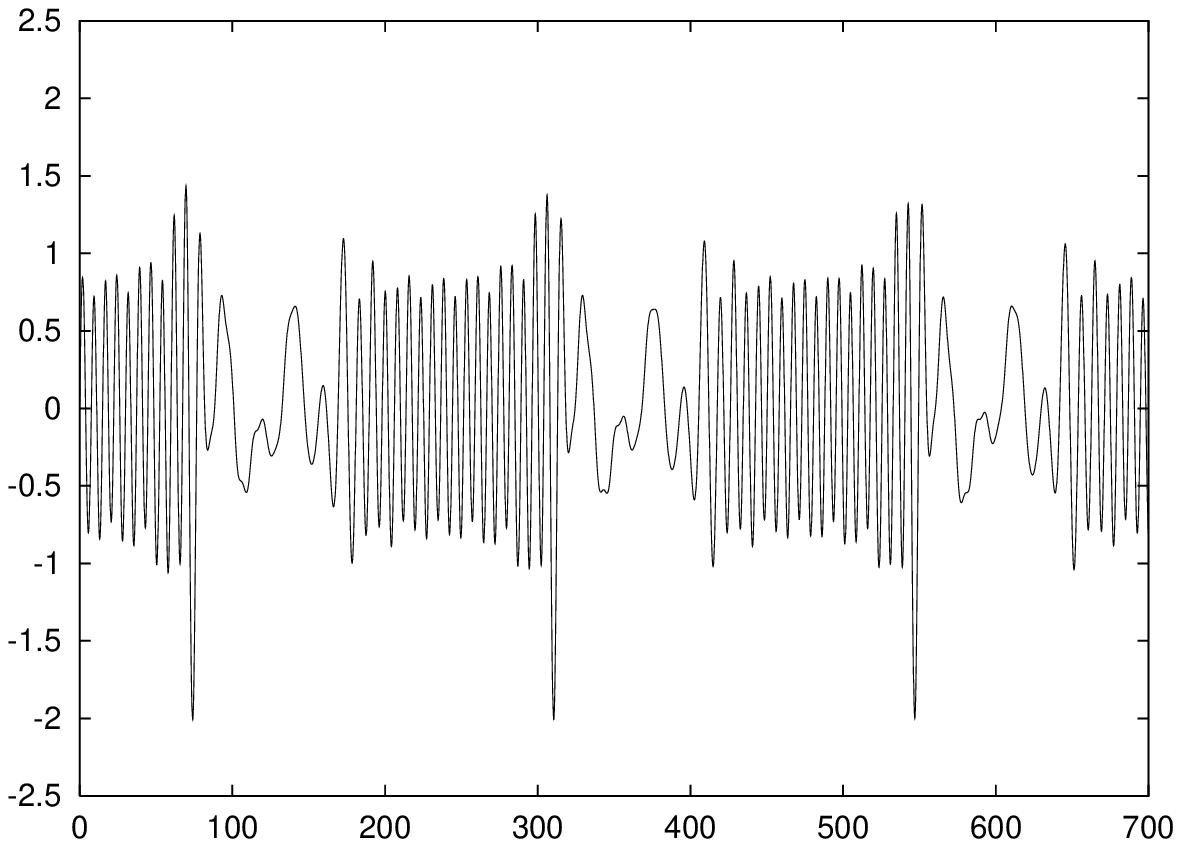}}
\centerline{\epsfxsize=11cm \epsfbox{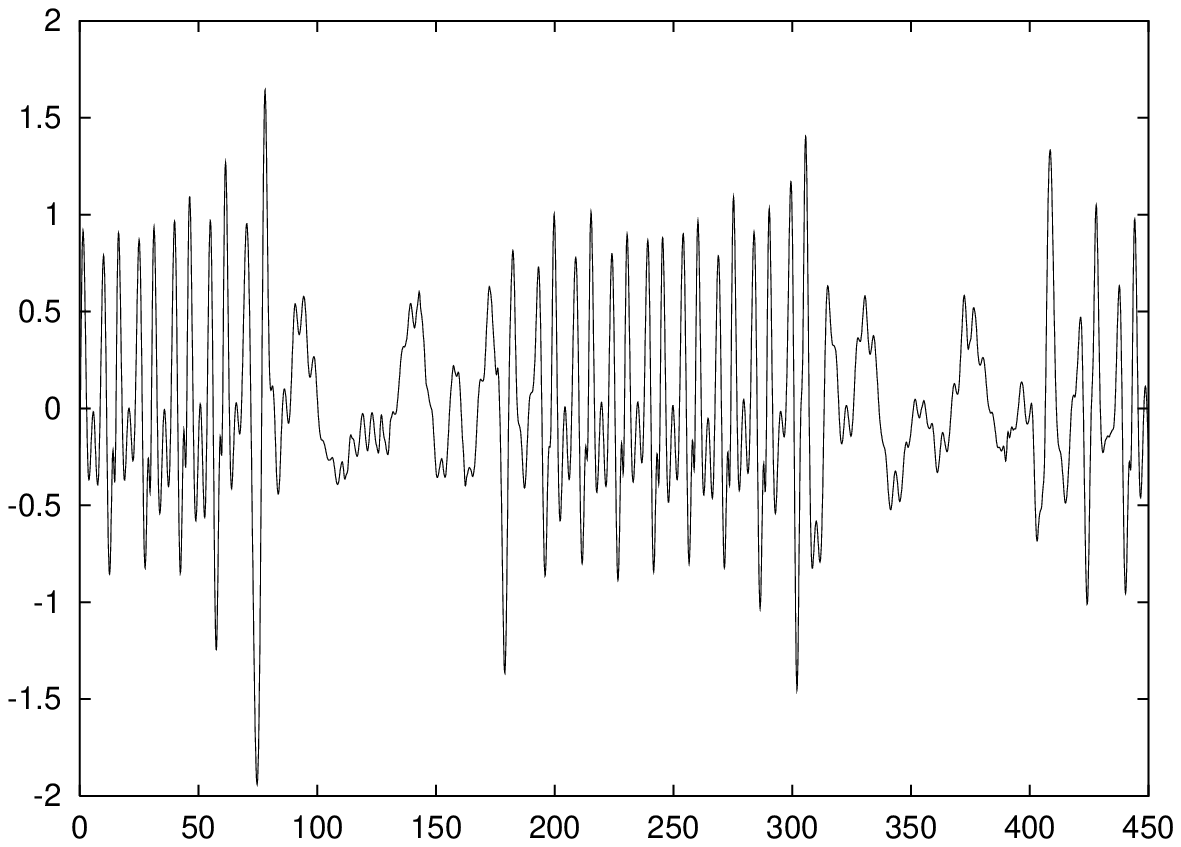}}
\vspace{0.1cm}
\caption{The same waveform as in Fig.~\ref{wfs1}, as seen by an 
observer along the hole's polar axis $\theta=0$ (top panel) and along 
$\theta= \pi/4$ (bottom panel). 
A comparison with the equatorial wave of Fig.~\ref{wfs1} 
reveals a substantial suppression of the high frequency features. 
This is a result of the fact that the wave's higher multipole components 
(which are responsible for the small-scale structure) are mainly ``beamed'' 
to directions close to the equatorial plane.}
\label{wfs2}
\end{figure}

\begin{figure}[tbh]
\centerline{\epsfxsize=10cm \epsfbox{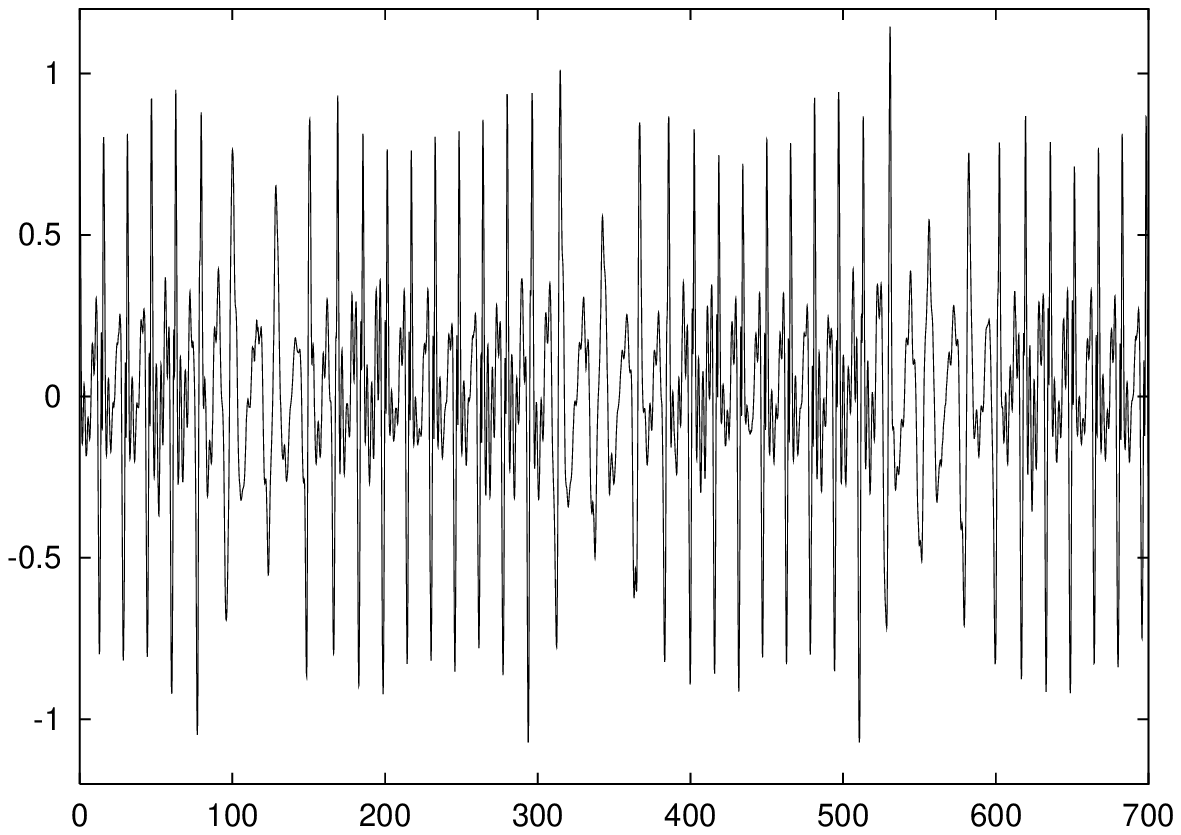}}
\centerline{\epsfxsize=10cm \epsfbox{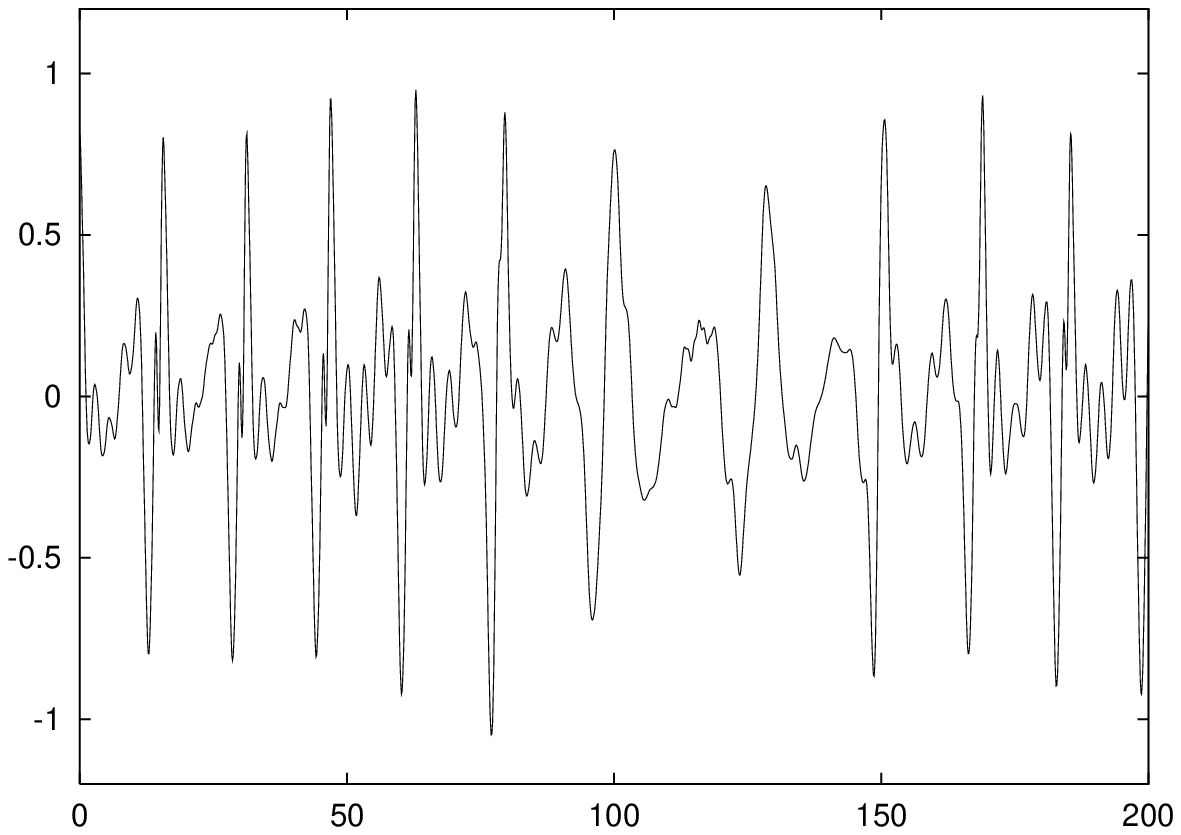}}
\centerline{\epsfxsize=10cm \epsfbox{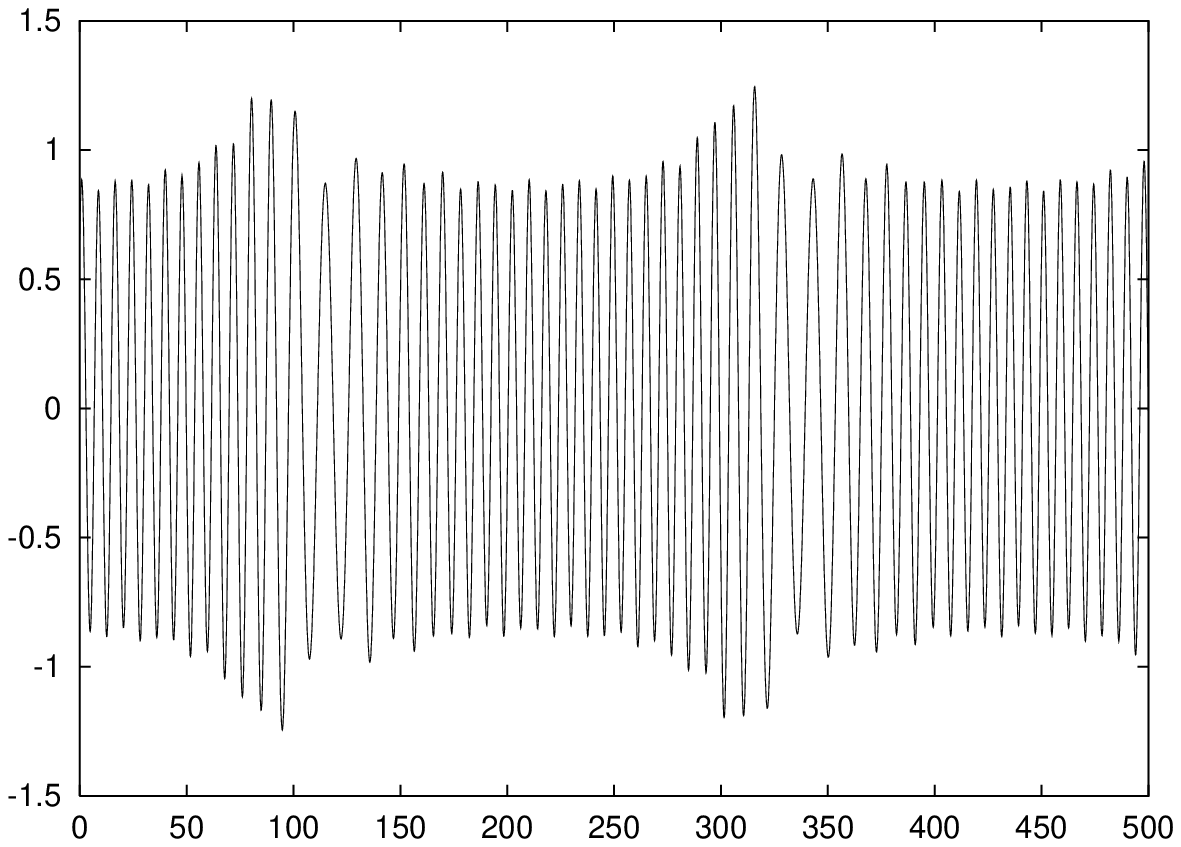}}
\vspace{0.1cm}
\caption{The waveform generated by a particle in a zoom-whirl orbit
with parameters $p=1.7M, e=0.3$ (we have again assumed a  black hole spin 
$a=0.99M$ and $\ell_{\rm max}=17$). The orbital period is $T_r= 221.36M$ 
and the number of
revolutions in one period is $N_r= 12.3$. The top and middle graphs show the 
signal seen by an equatorial observer, while the bottom graph corresponds to 
a polar observer. The same qualitive behaviour discussed in the caption of 
Fig.~\ref{wfs1} is also evident here.}
\label{wfs3}
\end{figure}

The zoom-whirl orbits are of interest for future detection efforts because of
the characteristic waveform they generate. In Figs.~\ref{wfs1} to  \ref{wfs3} 
we show such waveforms (in particular we plot the quantity $(\mu/r) h_{+} $ 
as a function of retarded time $t-r_{\ast}$) for a couple of strong-field  
zoom-whirl orbits ( $p=2.11M$ and $ e=0.7~$, $p=1.7M$ and $ e=0.3$) 
and for a rapidly spinning black hole with $a=0.99M$. 
Both orbits would evolve adiabatically if we consider a typical mass ratio 
$\mu/M \sim 10^{-6} $. The corresponding gravitational wave flux data can 
be found in Table~\ref{tab_num}. 

First we shall discuss the waveform as seen by an observer located on the black
hole's equatorial plane, see Fig.~\ref{wfs1} and top panel of 
Fig.~\ref{wfs3}. Clearly, these waveforms have a very distinct appearance. 
A rapidly oscillating, high amplitude signal is radiated during the 
whirling of the particle near periastron. In between these bursts one observes 
low-amplitude signals produced 
during the particle's zoom in and out from apastron. This 
contrast in amplitudes is greatest for larger eccentricities 
(compare Fig.~\ref{wfs1} and Fig.~\ref{wfs3}). 
An interesting feature of the equatorial waveform is the prominent 
high frequency ripples 
superimposed on the waveforms, associated with the 
higher multipoles components ($\ell=3$ and higher. The illustrated waveform 
includes all multipoles up to $\ell=18$) of the wave. It is noteworthy
that the high frequency features are prominent in both the zoom and the
whirl parts of the orbit, although if they are solely the result of 
beaming we might expect that they would be purely a whirl feature,
as the motion is fastest near periastron. However preliminary results
from a time domain code written by one of us (KG) suggest
that the high frequency features may be associated with quasi-normal
modes of the black hole, which are excited by the high frequency 
emissions from the orbit. The quasi-normal mode ringing results in a continuous
and time-delayed emission at these frequencies. 

We also show waveforms seen by an observer on the polar axis
of the black hole, in the top panel of Fig.~\ref{wfs2} and the bottom panel of 
Fig.~\ref{wfs3}. In this case both ``plus'' and ``cross'' polarisations are present 
(only $h_{\rm +}$ is nonzero for an equatorial observer) but we illustrate only 
$h_{\rm +}$ because the ``cross'' waveform is the same except for a phase lag.
The polar waveforms have the characteristic features of a high-amplitude,
multi-cycle whirl part and a low-amplitude two cycle zoom part, but the
high frequency features are absent. This suggests that the high frequency
features are associated with beaming due to the rapid motion of the particle
in the equatorial plane in the very strong field region, although the
high-frequency features are not associated only with the strongest-field
whirl part but are distributed throughout the whole cycle, including the
weaker field (larger radius) zoom part.
The polar waveforms are completely dominated by the quadrupole ($\ell=2$) 
emission. The $\ell=3$ and higher multipoles do not contribute significantly. 
In the equatorial waveforms the quadrupolar contribution is 
not much greater than that of the $\ell=3$ multipole and the fall off, in
terms of the amplitude of $h_{\rm +}$, for each subsequent multipolar waveform
is slow. The transition between the polar and equatorial waveforms can
by understood by looking at the waveform depicted in the bottom panel of
Fig.~\ref{wfs2} which corresponds to observation at angle $\theta =\pi/4$.  

One can get an idea of what is going on by looking at the $\theta$-dependence
of the energy flux from the system. In Fig.~\ref{antenfig} one sees that the 
$m=2$ flux (dominated by the $\ell=m=2$ contribution) is concentrated somewhat 
towards the pole ($\theta=0$), but there is a strong shift towards the equator 
($\theta=\pi/2$)
in the $m=3$ flux, and further concentrations in that direction for each
successively higher value of $m$. 
Therefore one sees the sort of beaming
of the higher multipoles observed in the waveforms, although because of
the dominance of the quadrupole the amplitude of the polar and equatorial
waveforms is similar in this case. 

\begin{figure}[tbh]
\centerline{\epsfysize=6.5cm \epsfbox{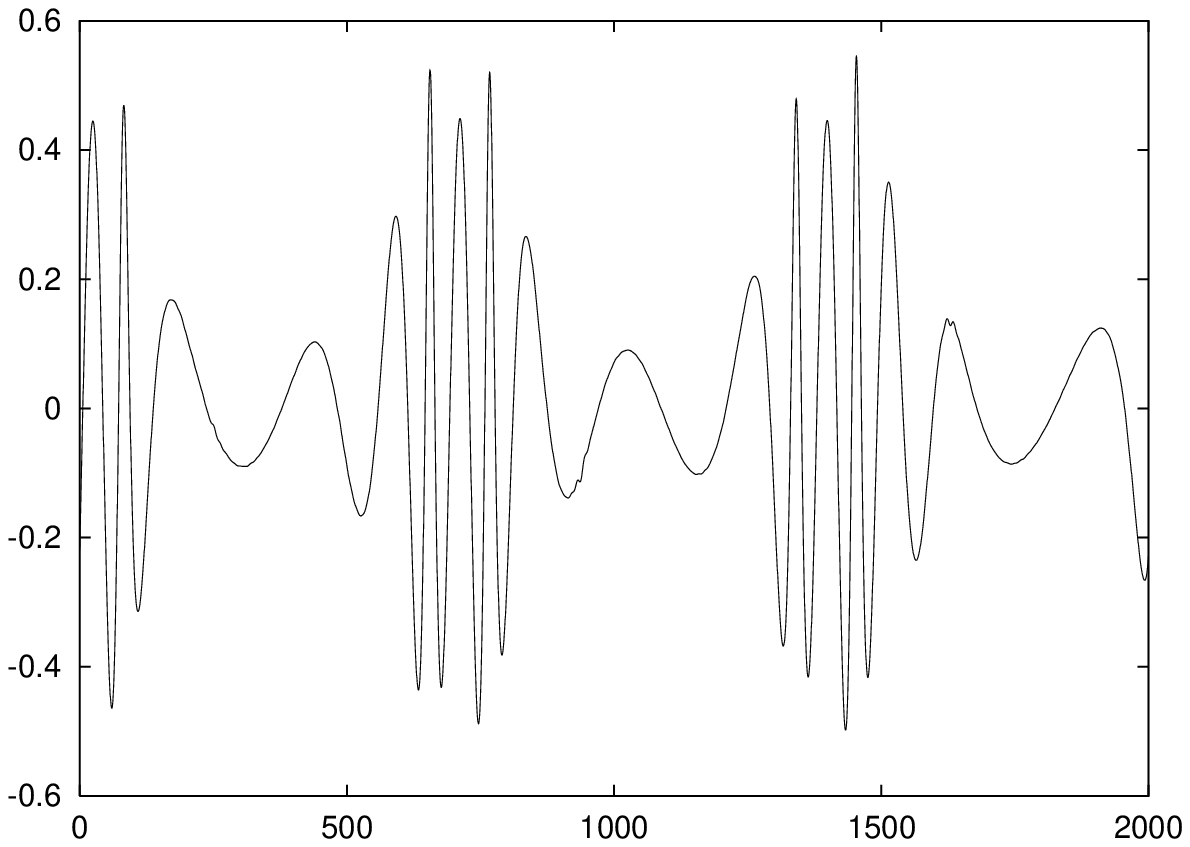}}
\centerline{\epsfysize=6.5cm \epsfbox{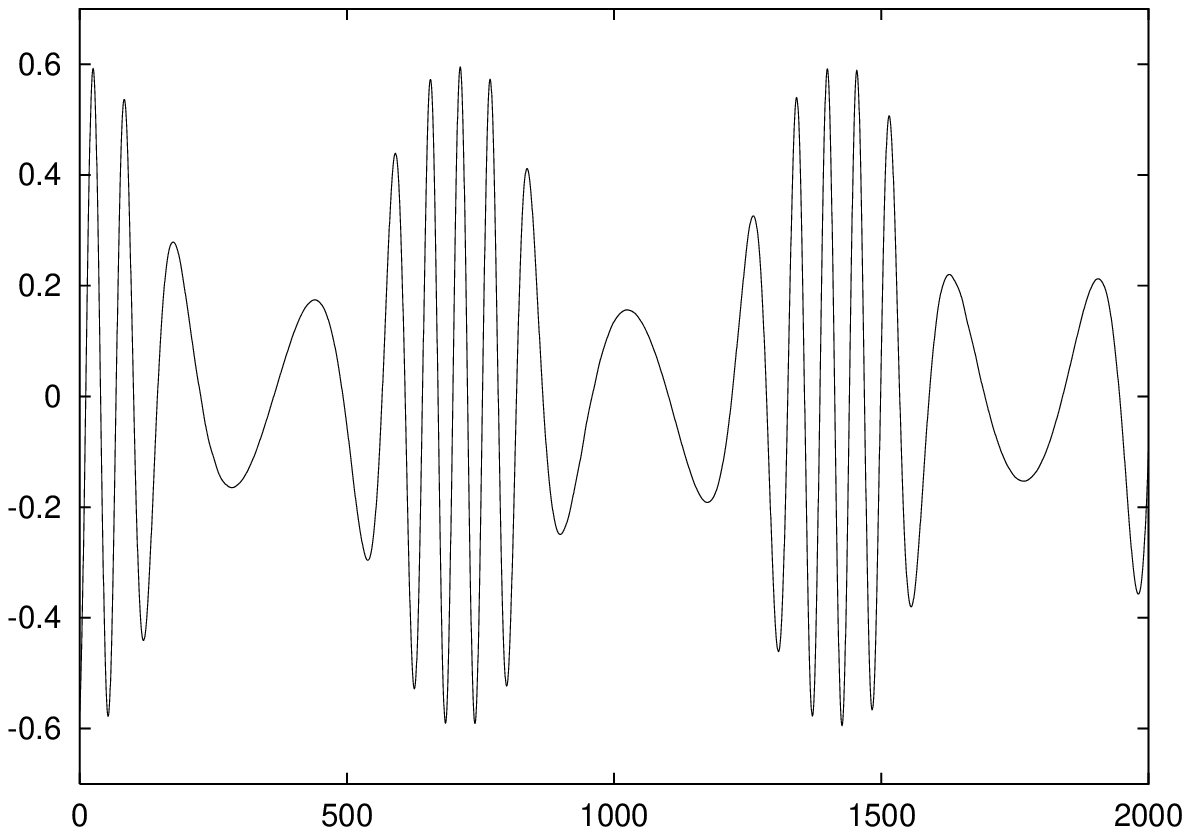}}
\vspace{0.5cm}
\caption{The waveforms generated by the retrograde zoom-whirl orbit
$p=10.5M$, $e=0.5$, $a=0.99M$ (and $T_r= 709M$, $N_r= 3.2$) as viewed 
by an equatorial observer (top panel) and by a polar observer (bottom panel). 
Note the absence of any small-scale structure and the similar appearance of 
the wave from different viewing angles, a clear evidence of small contribution 
from high $\ell$ multipoles.}
\label{wfs4}
\end{figure}

Next, in Fig.~\ref{wfs4}, we show the waveform from the retrograde zoom-whirl 
orbit $p=10.5M, e=0.5$, retaining $a=0.99M $ for the black hole spin. The 
familiar zoom-whirling pattern is clear also in this case. However, we do not
see any prominent high-frequency structure in these waveforms as the contribution
coming from higher multipoles is small (because the orbit does not reside in a 
very strong field regime). We are not surprised, in this case, that
the waveform seen when observing from a point along the hole's 
polar axis (see bottom panel of Fig.~\ref{wfs4}) is not very different.
    
\begin{figure}[tbh]
\centerline{\epsfxsize=8cm \epsfbox{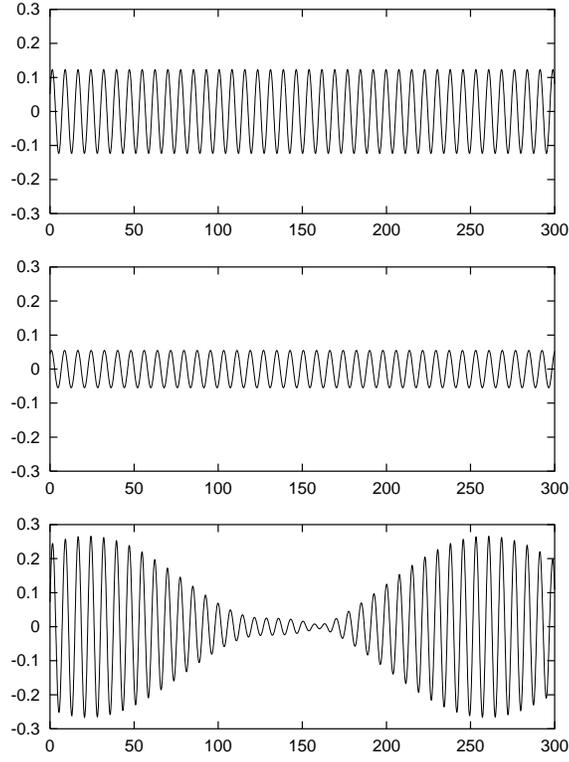}}
\caption{The waveform associated with the individual harmonics
(or ``voices'') $\ell=m=2$ and $k=k_{max}=10$ (top), $k=9$ (middle).
The combination of the $k=9-11$ voices gives the waveform at the bottom which 
already shows some zoom-whirl behaviour. The total $\ell=m=2$ signal finally 
resembles the waveform shown at the top panel of Fig.~\ref{wfs2}.}
\label{voicefig}
\end{figure}

\begin{figure}[tbh]
\centerline{\epsfysize=8cm \epsfbox{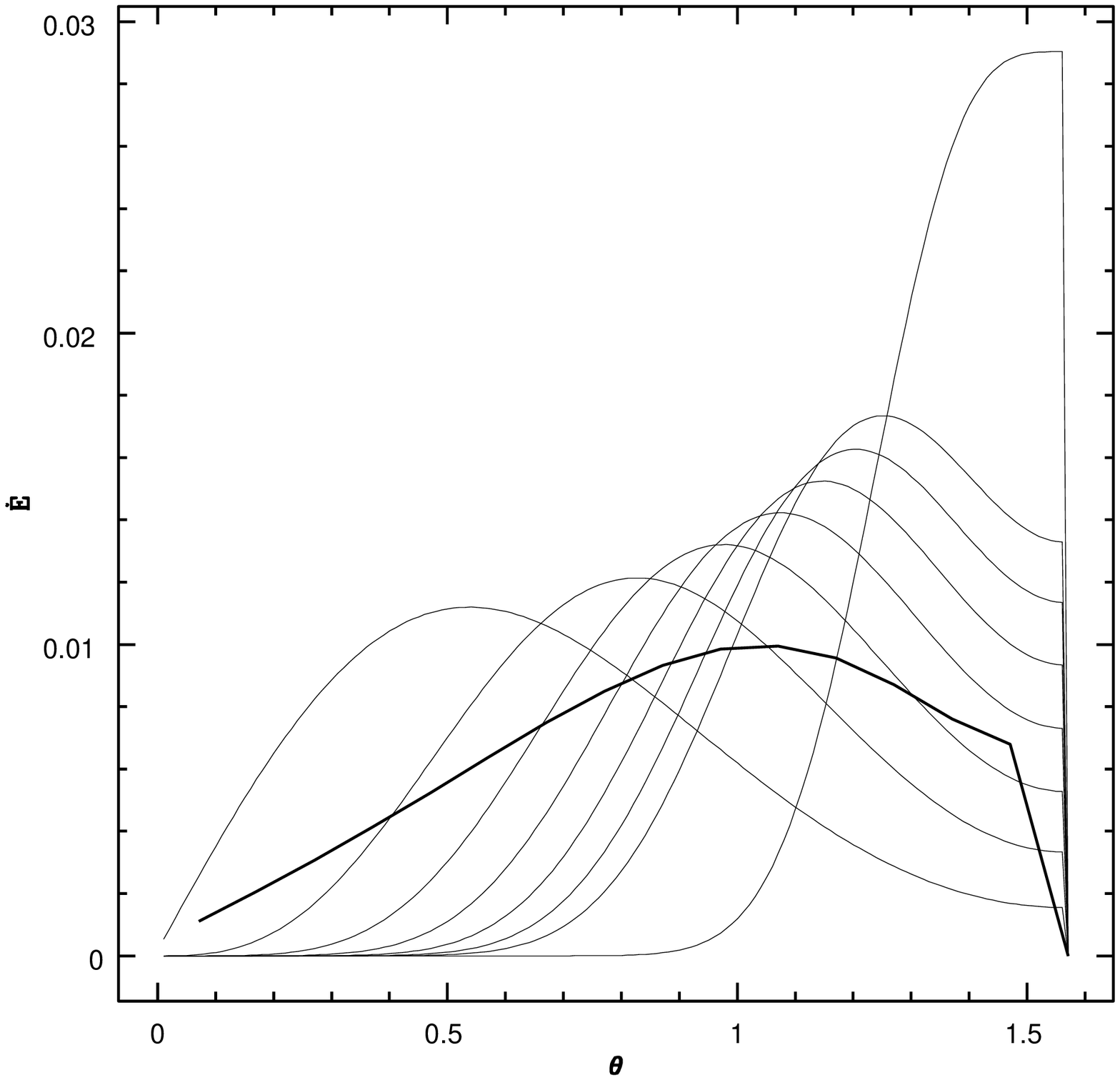}}
\caption{This figure shows the angular dependence of the energy flux for
the orbit with $p=2.11M$, $e=0.7$ and a black hole with $a=0.99 M$
(see Figs.~\ref{wfs1} and~\ref{wfs2}). 
The x-axis shows the coordinate $\theta$ in radians
and the y-axis shows the rate of energy emitted into a angle
0.01 radians wide, as a fraction of the total energy emitted (at
that multipole). Reading the different curves as they are peaked
from left to right (where the left hand side of the graph corresponds
to the pole of the black hole, and the right hand side to its
equator) we have the antennae pattern for m=2, m=3, m=4, m=5,
m=6, m=7, m=8 and at the extreme right m=18 (including the negative
m contributions in each case). One notes that above
m=2 the polar emission is greatly suppressed and the peak
direction tends ever more towards the equator. In bold, one sees
the curve for all multipoles at once (m= 2 to 18), but with the bins
0.1 radians wide, which peaks at around $\theta=\pi/3$.}
\label{antenfig}
\end{figure}

Although these zoom-whirl waveforms appear rather complex, and could present
a problem for matched filtering data analysis techniques, one can take comfort
in the fact that the number of harmonics of the spectrum which contribute
significantly to the waveform and overall flux is not that great. 
Scott Hughes \cite{scott_insp} has suggested that analysing individual "voices," which are 
monochromatic, of a complex wave may be an effective way to proceed in data analysis. 
Take the case of the zoom-whirl orbit with $p=2.11M$, $e=0.7$ and $a=0.99 M$.
If we look at Fig.~\ref{spec_full}, which shows the entire spectrum 
(up to $\ell=18$) of this signal, we see that the number of individual harmonics, or voices,
each corresponding to an instance of the three numbers $\ell$,$m$ and $k$,
which stand out are on the order of a dozen. Each of these voices
has a rather simple waveform, as can be seen from Fig.~\ref{voicefig}. 
It is only their superposition which is complicated. Therefore following the chirp 
of individual voices as the orbit evolves may not be such a formidable 
computational task.



\pagebreak

\subsection{\textbf{Spectra}}

The final part of our numerical results concerns the harmonic decomposition of 
the gravitational radiation fluxes. Specifically, we examine the $k$ 
distribution of the energy flux, at infinity and at the horizon, for a given 
multipole channel $\ell,m$ (the angular momentum flux spectrum exhibits a 
similar behaviour). In all the figures in this Section, 
expect Fig.~\ref{spec_full}, we plot $\dot{E}_{\ell m k}$ versus $k$.

\begin{figure}[tbh]
\centerline{\epsfysize=5cm \epsfbox{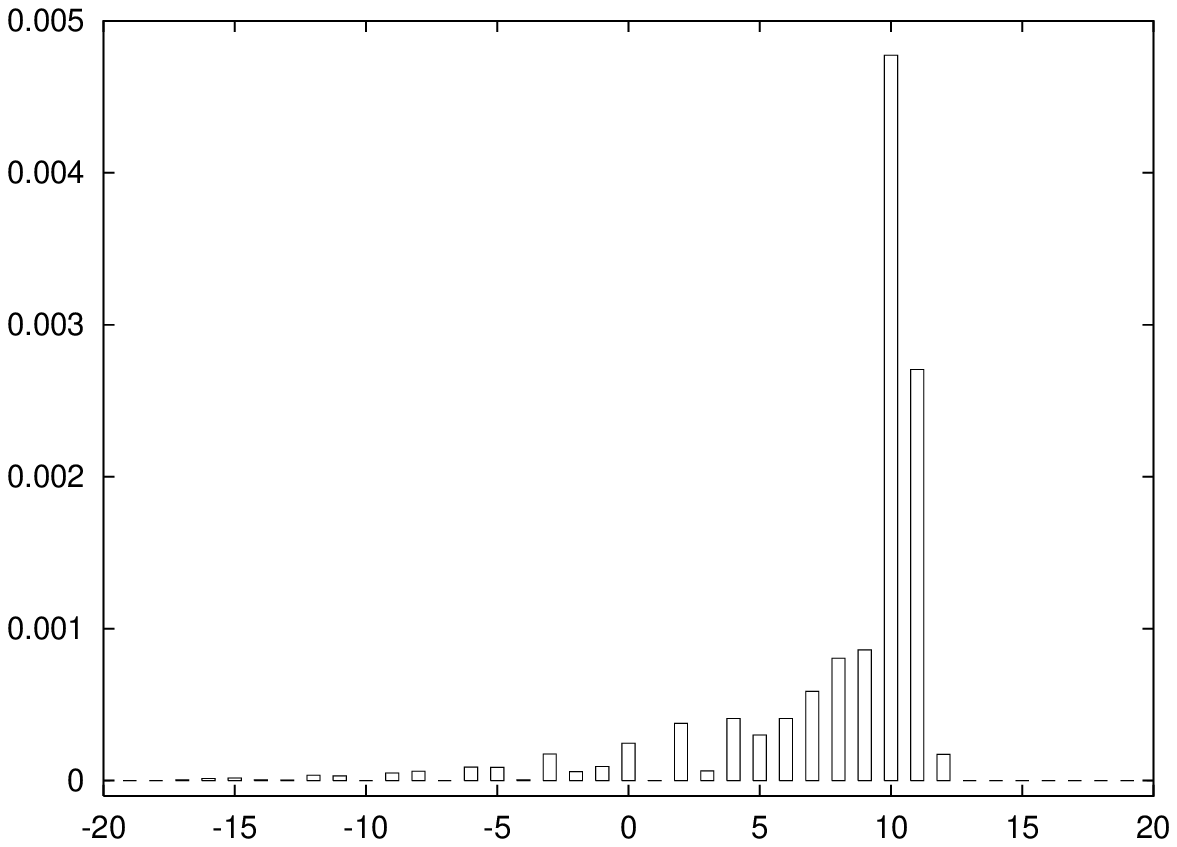}}
\centerline{\epsfysize=5cm \epsfbox{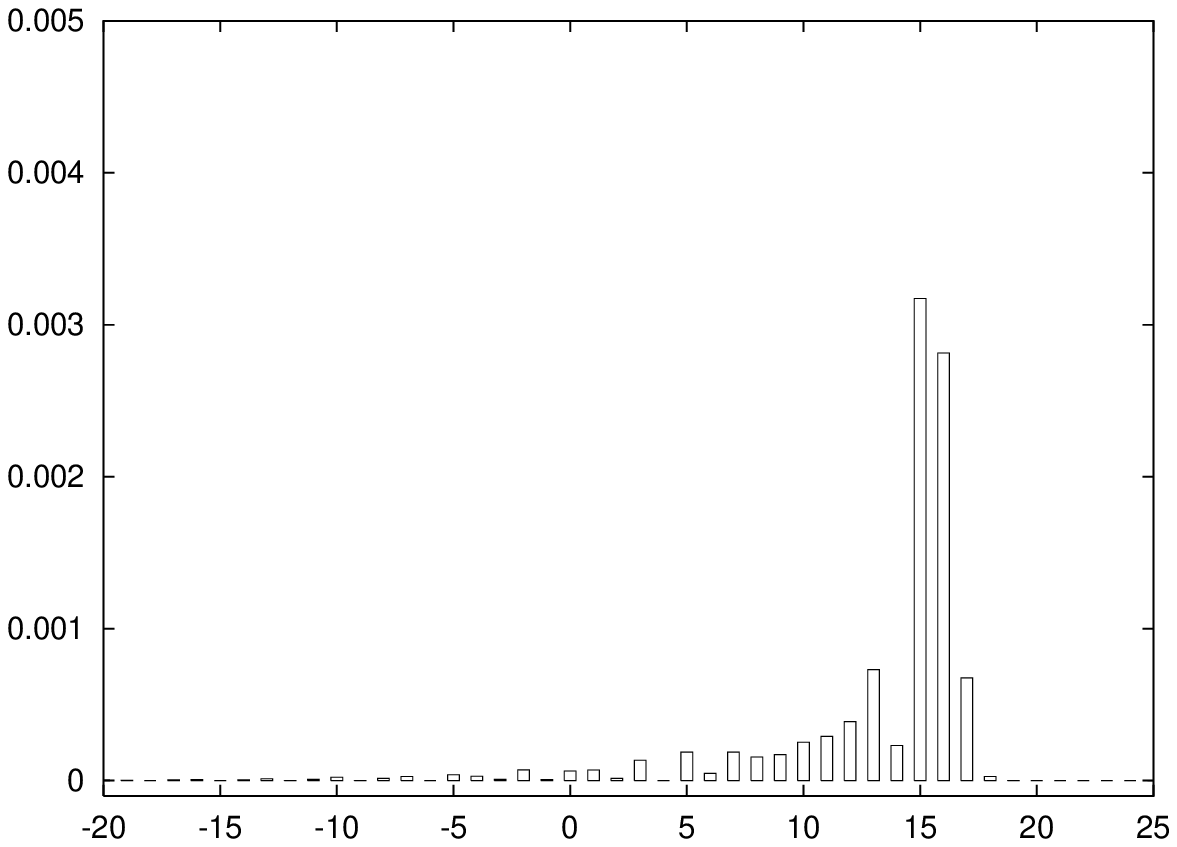}}
\centerline{\epsfysize=5cm \epsfbox{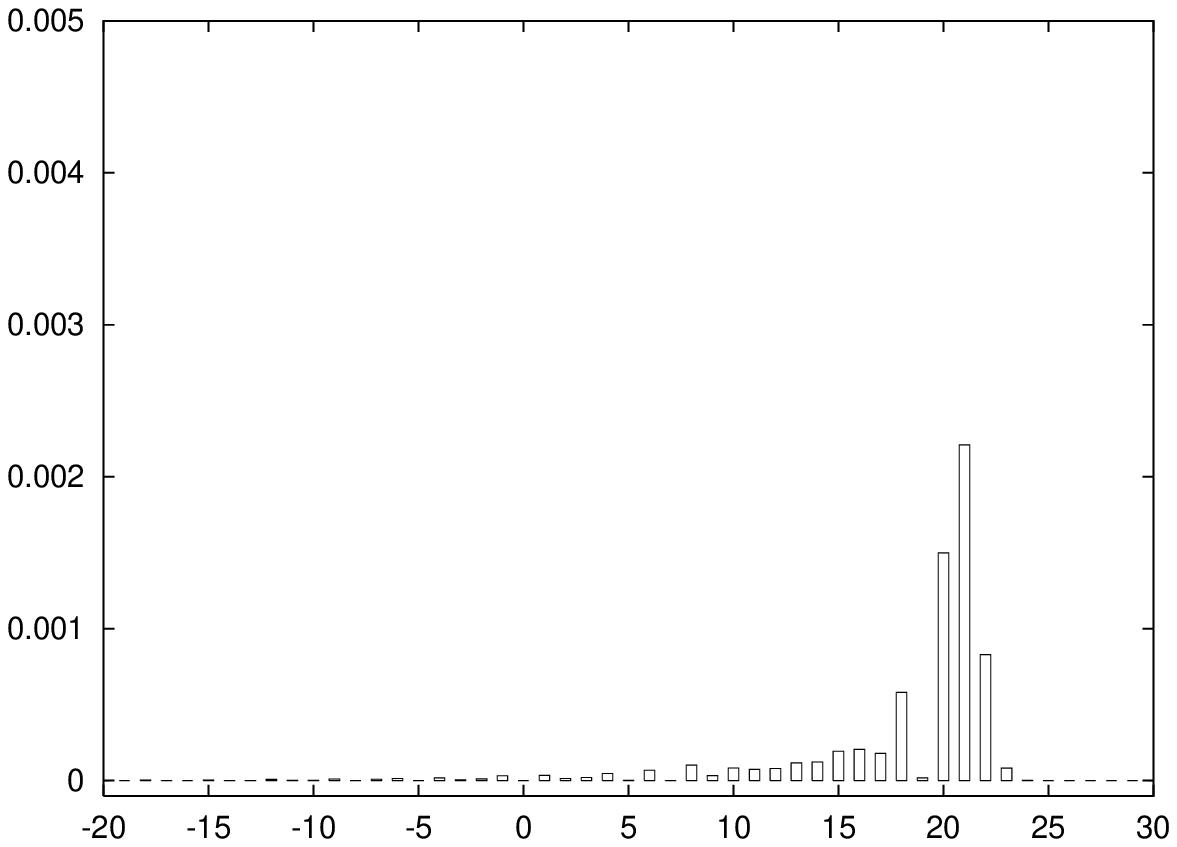}}
\centerline{\epsfysize=5cm \epsfbox{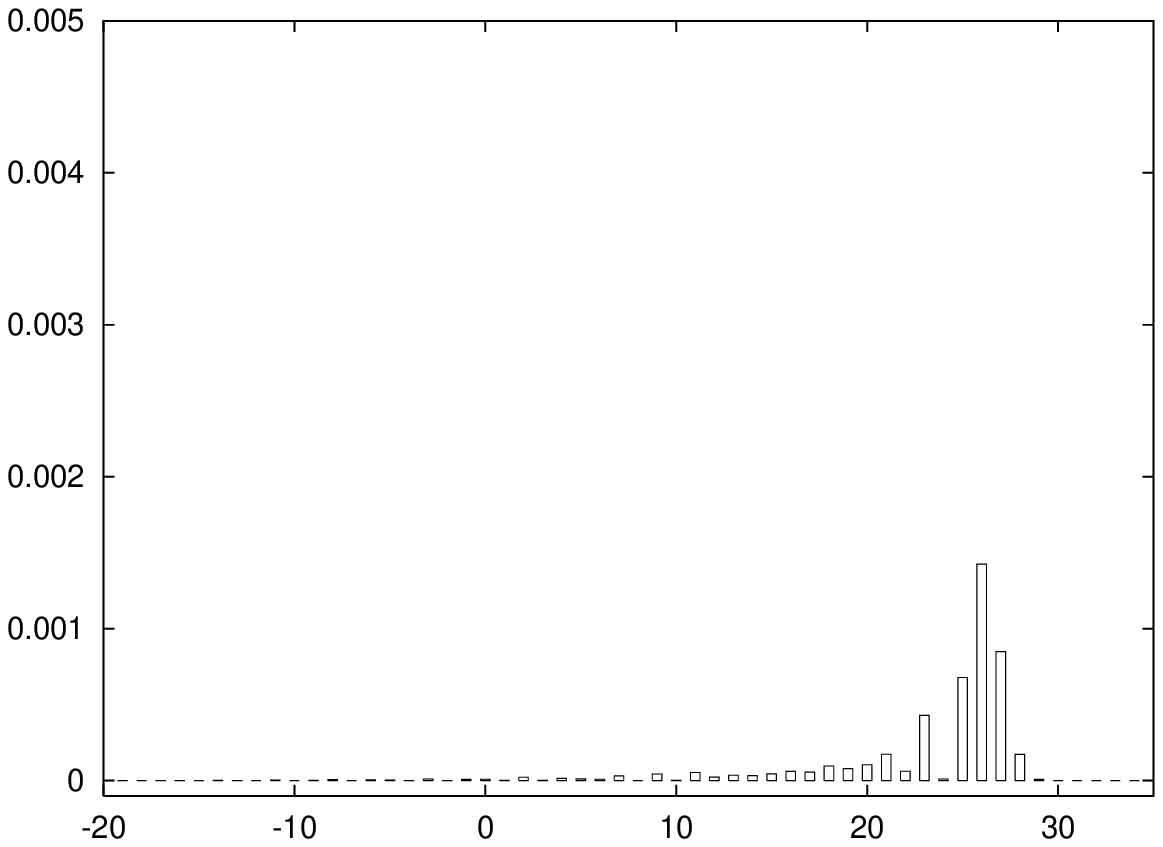}}
\vspace{0.1cm}
\caption{The distribution of the energy flux (in units of $\mu^2/M^2$) in 
terms of the $k$ harmonic number for an orbit with $p=2.11M , e=0.7 $ around an 
$a=0.99M$ Kerr black hole. From top to bottom we display the 
$\ell=m=2,~3,~4,~5$ multipoles.
Note the significant contribution of the higher multipole channels to the total
flux. The relevant orbital frequencies are $M\Omega_r= 0.02653$ and 
$M\Omega_{\phi}= 0.2791$.}
\label{spece07_if}
\end{figure}

\begin{figure}[tbh]
\centerline{\epsfysize=8cm \epsfbox{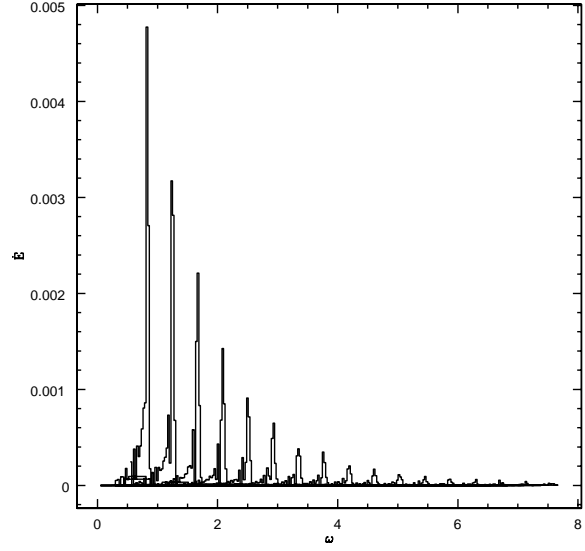}}
\caption{This figure shows all of the main peaks of the spectrum emitted by the
orbit $p=2.11M$, $e=0.7$ for a black hole with $a=0.99 M$ (see Figs.~\ref{wfs1}
and \ref{wfs2} for the waveform associated with this orbit). The flux of 
total energy emitted per unti time, on the y-axis,
and the frequency, on the x-axis, are both given in the geometrised
units of this paper (in which $5 \mu s$ is approximately unity, if the
black hole has mass of $10^6 M_\odot$). The main peaks in each multipole
are easily read from left to right as ($\ell=m=2, k=10$); ($\ell=m=3, k=15$);
($\ell =m=4; k=21$); ($\ell=m=5, k=26$); ($\ell=m=6, k=31$); ($\ell=m=7, k=37$); 
($\ell =m=8, k=42$);($\ell=m=9, k=47$); ($\ell=m=10, k=53$); ($\ell=m=11, k=58$); 
($\ell =m=12, k=63$); ($\ell=m=13, k=69$); ($\ell=m=14, k=74$); ($\ell=m=15, k=80$); 
($\ell =m=16, k=84$); ($\ell=m=17, k=90$); ($\ell=m=18, k=94$).} 
\label{spec_full}
\end{figure}

To begin with, in Fig.~\ref{spece07_if} we present the energy flux spectrum, 
at infinity, of the $\ell=m=2,~3,~4$ multipoles for an orbit with parameters 
$ p=2.11M,~ e= 0.7$, and for spin $a=0.99M$. As we have already mentioned, 
for such a strong-field orbit the higher multipoles give significant 
contributions to the total flux. The spectrum itself is composed of a series 
of ``humps'' which grow in height as $k$ increases, up to the point where 
the maximum harmonic is reached at $k=k_{max}$, after which the spectrum 
rapidly fades away. This behaviour closely resembles  the one found in the 
Schwarzschild case \cite{cutler}, although in the present case the  ``humps'' 
show a somewhat less regular structure. As we are 
not dealing with a very high eccentricity orbit, we find that $k_{max}$ is 
not very large. Moreover, the spectrum peak shifts to higher $k$ values as 
$\ell$ increases.

\begin{figure}[tbh]
\centerline{\epsfysize=5cm \epsfbox{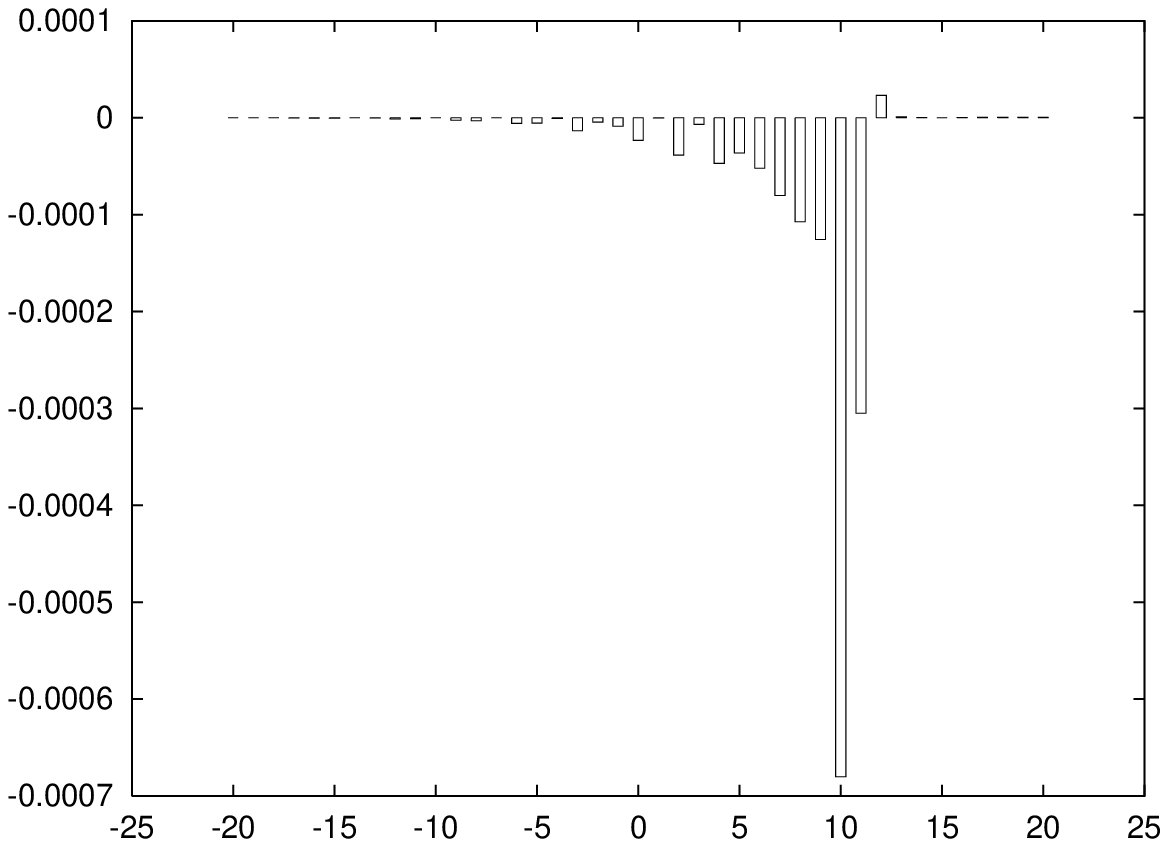}}
\centerline{\epsfysize=5cm \epsfbox{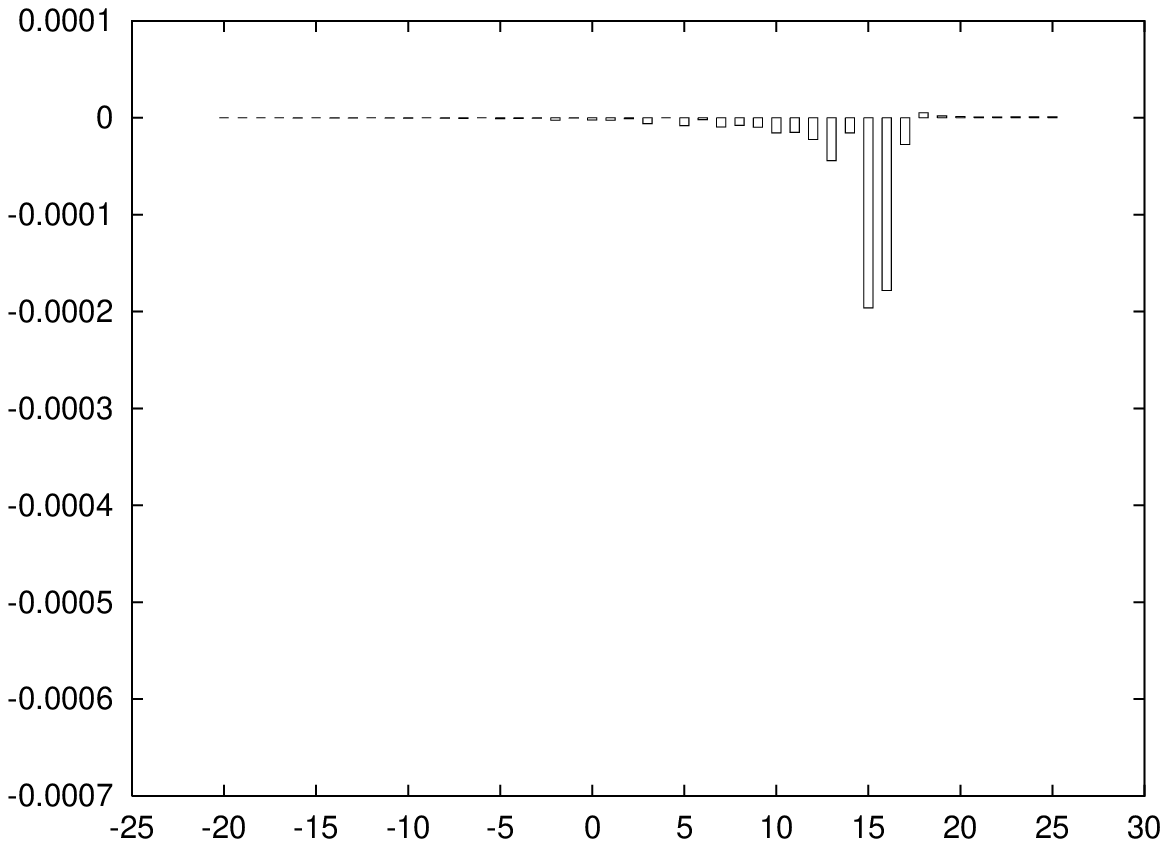}}
\centerline{\epsfysize=5cm \epsfbox{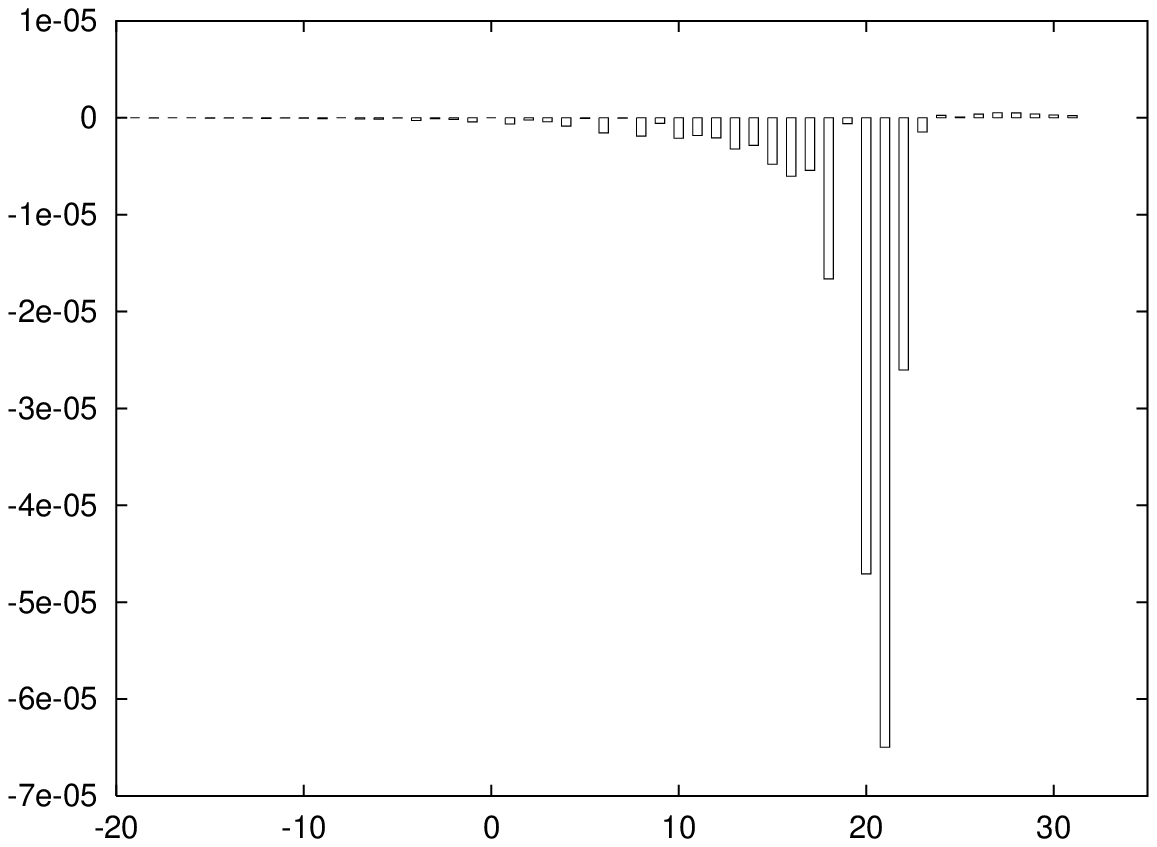}}
\caption{The horizon energy flux $k$-spectrum for the orbit 
$p=2.11M$, $e=0.7$, $a=0.99M$.
As in the previous figure, we graph, from top to bottom, the multipoles 
$\ell=m=2,~3,~4$. Note that both infinity and horizon fluxes peak at the same
$k$ harmonic, for given $\ell,m$. The latter spectrum, however, is strongly 
dominated 
by the quadrupole channel (which is roughly $10\%$ of the flux at infinity) 
while the higher multipoles quickly fade away. The fact that the horizon flux 
takes, almost entirely, negative values means that it represents superradiant 
radiation.} 
\label{spece07_h}
\end{figure}

In Fig.~\ref{spece07_h} we graph the corresponding horizon fluxes for 
the same orbit and same multipoles as in Fig.~\ref{spece07_if}. Both 
infinity and horizon fluxes peak at a common $k_{max}$ value. Since the 
selected orbit resides in a strong 
field regime, the horizon flux is a respectable fraction ($\sim 10\%$) 
of the flux at infinity, and moreover, is predominantly negative i.e. 
superradiant, as we should expect from the discussion in the previous 
Section. Note that the relative contribution of the quadrupole channel
is much larger in the case of horizon fluxes as compared to the
infinity fluxes.

The full energy flux spectrum, up to $\ell=18$, for the $p=2.11$,$~e=0.7$ orbit 
is shown in Fig.~\ref{spec_full} in terms of frequency rather that $k$.

\begin{figure}[tbh]
\centerline{\epsfysize=5.5cm \epsfbox{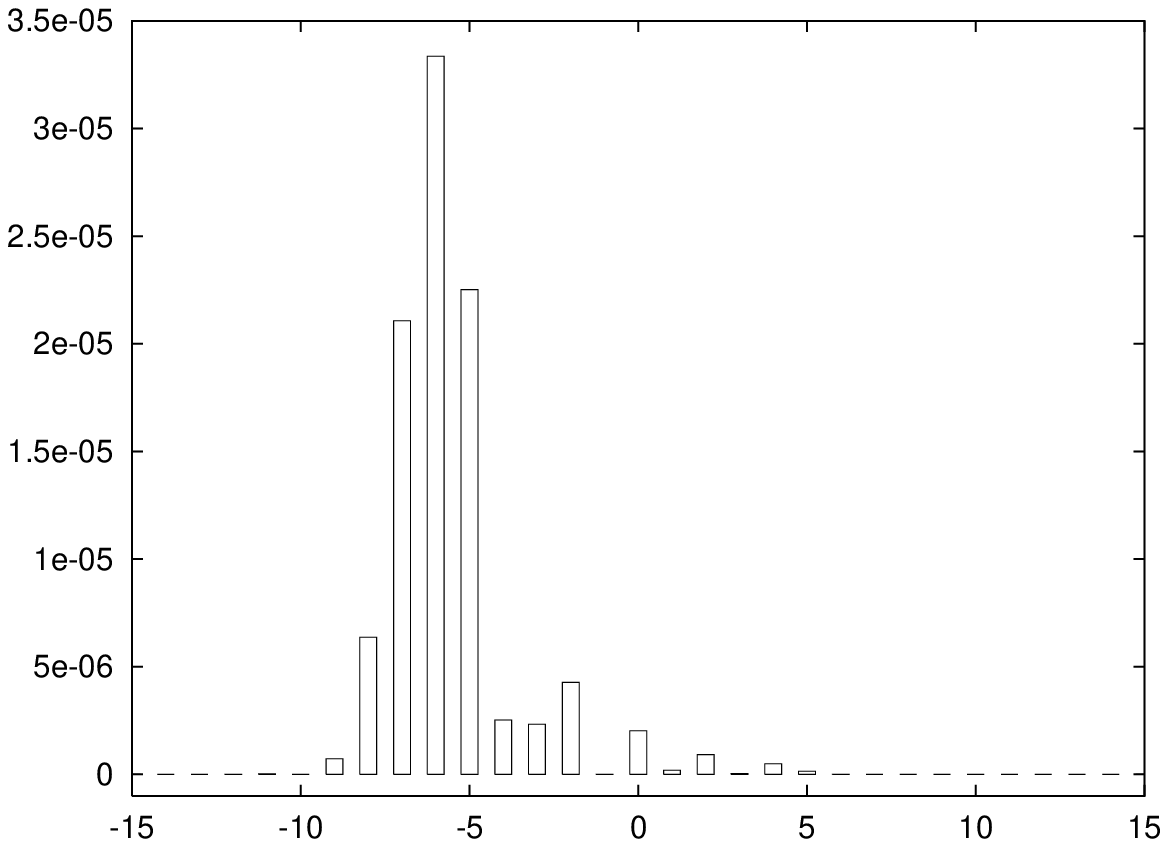}}
\centerline{\epsfysize=5.5cm \epsfbox{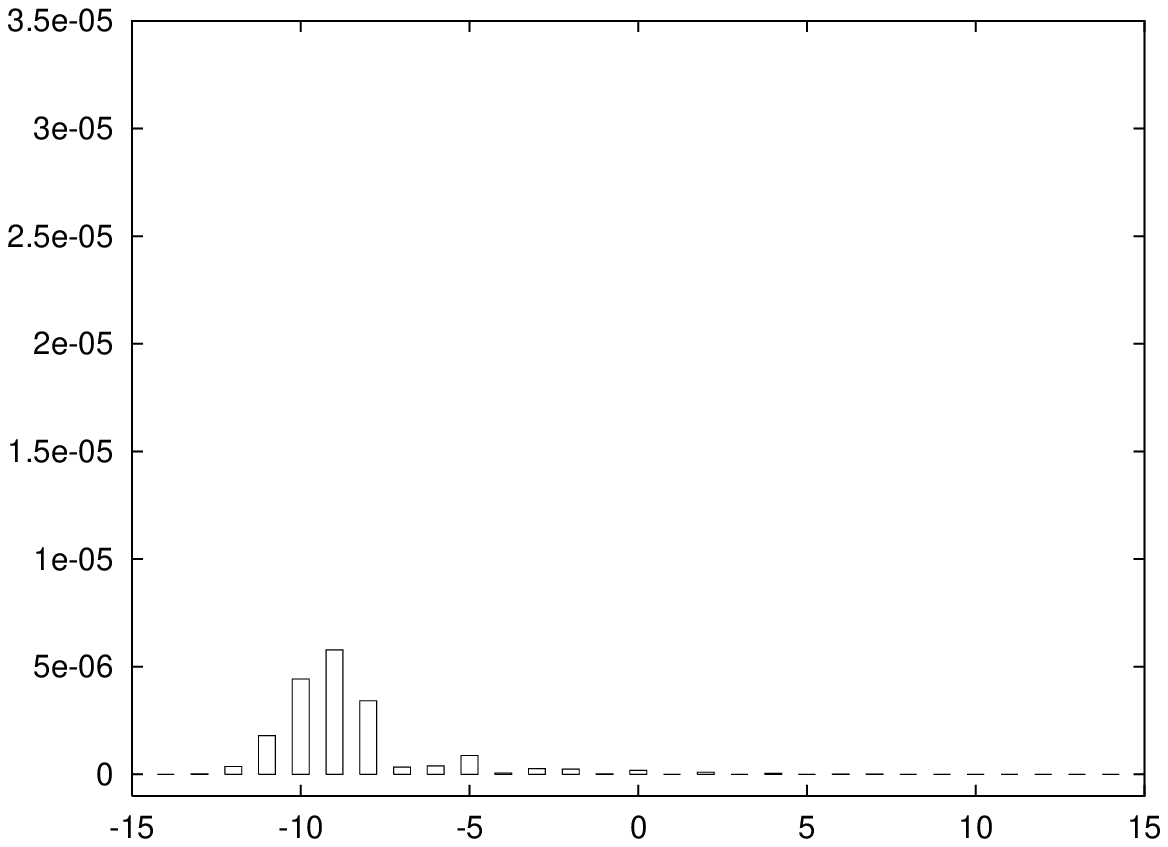}}
\vspace{0.1cm}
\caption{The $\ell=m=2$ (top) and $\ell=m=3$ (bottom) spectra for the 
$p=10.4$, $e=0.5$ retrograde orbit. The black hole spin is $a=0.99M$. 
Note the location of the maximum at some negative $k$ value and the 
dominance of the quadrupole component over the octapole (and higher) 
component.}
\label{spectretro}
\end{figure}

Spectra belonging to retrograde orbits appear similar to the prograde spectra 
with the difference that they peak at negative values of $k$.
For a typical situation see Fig.~\ref{spectretro} for the retrograde orbit of 
$p= 10.4M, e= 0.5$ and black hole spin $a=0.99M$.

Before closing this Section, we should point out that only the harmonics around 
$k_{max}$ (typically 10-30 of them) give a significant contribution to the total flux,
for given $\ell,m$. This statement applies for all eccentricities we examined, and 
could have an important impact on the computation of waveforms and fluxes from highly 
eccentric orbits in which case $k_{max}$ attains very large values. For example, 
if one could find by other means (for example a time-domain code, see discussion in the 
next Section) the location of the main peak, then the calculation of the surrounding harmonics 
should give an sufficiently accurate result for the flux. This strategy is far more economic, from
a computational point of view, than calculating all the harmonics between $k=0$ and
$k_{max}$.



\section{\textbf{Concluding discussion}}

We have examined gravitational waves from and radiation reaction of equatorial 
orbits of particles in the last stages of inspiral around a central black hole. 
We expect that such orbits will often have moderate eccentricities, not circular
but significantly less than $e=1$. As these orbits approach the point at which 
they will plunge into the black hole two things will happen. The first is that the
orbital eccentricity will begin to increase, having been decreasing throughout the
preceding inspiral. For retrograde orbits this eccentricity increasing
phase will last for many cycles, while for prograde orbits, especially ones which
retain fairly high eccentricities, this phase will be fairly brief (in
the case of rapidly rotating black holes). We have found a $\sim 10 \% $ 
fractional increase in eccentricity, for the most favourable situations. 
Secondly, as the orbit draws closer
to the unstable region it will tend to spend a longer and longer portion
of each orbital period near periastron.
For reasonably eccentric orbits this will promote the ``zoom-whirl'' behaviour we
have described, with its characteristic waveform. Although extreme cases involving
high eccentricities will presumably be rare,
it is still likely that LISA will be seeing signals with 
$5-10$ whirls, and therefore $10-20$ gravitational wave cycles between each 
apastron. Therefore templates of such signals will be very important. It is worth 
noting that because the orbital radius changes very little during these whirls, 
the whirl part of the waveform looks, in many respects, like a near circular 
waveform with harmonics of a single dominant frequency. This means, for instance, 
that for orbits with many whirls, the ratio of the fluxes of energy and angular 
momentum is close to $\Omega_ \phi$ (recall the celebrated relation for circular 
orbits $\dot{E}/\dot{L}=\Omega_{\phi}$). 

Considering inclined orbits (not confined to the equatorial plane) 
for a moment, we expect that zoom-whirl orbits will be found in cases with 
small inclination angles, because higher inclination angles feature greater 
plunge radii and the particle cannot ever have a very small periastron radius 
(as we see when the orbit is retrograde, which corresponds to the largest 
possible inclination angle). When the test-particle 
is in an inclined and eccentric orbit, close to crossing some separatrix of bound 
stable orbits, it will spend a considerable amount of time moving in a 
quasi-circular 
non-equatorial trajectory close to the periastron. It is plausible, by 
extrapolating the results of the present work, that during the ``whirling'' 
stage, the orbital energy, angular momentum and Carter constant will 
approximately evolve in such a way as if the orbit was circular and inclined. 
Assuming that 
most of the radiated energy, angular momentum and Carter constant are generated 
near the periastron (which in the present case is located in a strong field 
regime) then one might be able 
to estimate the, otherwise elusive, rate of change of the Carter constant for a 
near-to-plunge generic Kerr orbit. Such information could provide a very useful
test for the recently adopted assumption of obtaining the rate of
change of the Carter constant by keeping the orbital inclination angle fixed 
during the inspiral \cite{inspiral_paper}.

Waveforms from prograde orbits residing close to the horizon of the black hole 
will feature significant high frequency components when seen from a position on 
the equatorial plane of the system (i.e. when the orbit is observed ``edge on''). 
This seems to be due to beaming, resulting from the rapid motion of the orbiting 
particle along the line
of sight of the observer. When observed from on or near the polar axis the 
waveform is largely quadrupolar, dominated by a single nearly circular frequency.
A glance at the equatorial zoom-whirl waveforms presented here suggests that 
they can be very complex and not necessarily amenable to matched filtering 
methods following the full wavetrain.
But recall that a successful source identification may already have been made 
during the previous year of observation by LISA and from the source parameters 
deduced during
this period it may be possible to search for the whirl parts of the waveform 
individually. 
On the other hand, it is worth noting that this high 
frequency structure could make it possible for LISA
to detect late inspiral signals from very large black holes, with masses
above $10^7 M_{\odot}$, which would otherwise be too low frequency (in the low
multipole parts of the waveform) for detection. 
By contrast, if we are looking down on the
system from the pole then the signal is much ``cleaner'', without much 
contribution beyond the $\ell=m=2$ multipole. Obviously non-equatorial motion 
will introduce further harmonics and, as suggested in \cite{scott_insp} it may 
prove more useful to examine different harmonics or ``voices'' of the signal 
separately, rather than trying to model the entire complex signal as one template.

It has been recently realised \cite{phinney} that orbits during the long inspiral 
phase, before the radial orbital frequency is within the LISA waveband, will, 
in principle, emit 
detectable radiation. Recall that the periastron is rather close, 
$r_{p} < 20 M$, so that the frequency of the cycles in the whirl part of the 
waveform {\em does} fall within the LISA waveband. As the time between 
successive bursts (equal to the radial orbital period) will be very long 
(typically up to a century or even more) these bursts would not be expected 
to be detectable in 
practice. However, as the number of objects in this long inspiral stage will be 
rather great, one does 
have to take them into account as a background noise which will tend to hinder 
LISA's efforts to detect signals from those objects in the last stages of their 
inspiral. In fact, one could in principle expect to have proper zoom-whirl 
orbits even at this stage, provided the central hole is rapidly spinning and the 
orbit is retrograde. Then, as we have shown, the corresponding separatrix 
translates to a minimum periastron value, $r_{\rm isbo} \approx 6M $, which 
falls within the expected periastron distribution.

In a follow-up paper \cite{paperII} we will study orbits with larger
radii and larger eccentricities. In particular we are planning to produce
(using the method of \cite{scott_insp} which was applied for inspiralling 
circular inclined orbits) the full inspiral trajectory and the 
resulting waveform for bound equatorial orbits that could be of importance for 
LISA. It is of great interest to (a) produce full inspiral 
waveforms for the kind of orbits described above and (b) give a good estimate 
of the total inspiral time, and the residual eccentricity (taking under 
consideration effects like the sign reversal of $\dot{e}$) just before plunging. 
However, the calculation of fluxes and waveforms produced by bodies in 
$e \approx 1$ orbits, is currently ranked as a difficult task. 
When pursued in the frequency domain (as it was the case in the present work) one 
has to calculate an enormous number of individual harmonics, as it is obvious from 
the spectra we
presented. Moreover, the numerical computation of the 
integrals (\ref{Zlmk4}) is poorly convergent, a manifestation of the
fact that the source term in the Teukolsky equation 
(\ref{radTeuk}) diverges as $r \to \infty $ (that is, when the orbit
tends to become parabolic $e \to 1$). One way to cure this 
pathology would be to work with the inhomogeneous Sasaki-Nakamura equation 
\cite{shibata_ecc} which has a well behaved 
source term (in the sense that it decays at spatial infinity). 
However, the first difficulty outlined above will still be
present. A possible way to overcome it could be the calculation of the 
waveform/fluxes directly in the time domain by evolving the time-dependent 
Teukolsky equation without resorting to any separation of variables
apart from $\phi$. Conceivably, the required numerical code could be based on 
the Teukolsky codes used to study the dynamics of scalar and gravitational
perturbations in a Kerr background metric \cite{tcode}, see \cite{capra}
for a report on such an attempt. Looking further ahead, such time-domain
codes could be the only practical tool for computing the waveform/fluxes generated 
by bodies orbiting non-black hole massive compact objects (in which case
there is no known Teukolsky-like separable wave equation). 
We are currently working along both of these directions.      

\section{Acknowledgements}
The authors wish to thank Scott Hughes, Sterl Phinney, B.S. Sathyaprakash, 
Wolfram Schmidt and Lee Lindblom for very useful discussions and comments. 
In addition KG thanks the State Scholarships Foundation of Greece for 
financial support. He would also like to acknowledge support from 
PPARC grant PPA/G/0/1999/0214. DK thanks the Astronomy department of Oxford 
University for the generous use of their facilities and their unfailingly warm
hospitality which greatly aided this research. In particular he would like to 
thank Julia Kennefick and Gavin Dalton. He would also like to acknowledge 
support from NSF grant PHY-0099568, and the help of the Physics and Astronomy 
department at Cardiff University.


\section{Tables}
\begin{minipage}[t]{4in}
\begin{table}
\textwidth=7.0cm
\begin{tabular}{|c|c|c|c|}
$e$ & $ a=0.5M $ & $a=0.99M$ & $a= -0.99M$  \\  
\hline      
 0.10 & 4.377 (4.71)  &  1.516 (1.59) &  9.266 (10.03)  \\
 0.20 & 4.526 (4.77)  &  1.595 (1.64) &  9.552 (10.12) \\
 0.30 & 4.679 (4.85)  &  1.685 (1.71) &  9.830 (10.24) \\
 0.40 & 4.836 (4.96)  &  1.782 (1.79) & 10.102 (10.40) \\
 0.50 & 4.996 (5.08)  &  1.883 (1.89) & 10.367 (10.58) \\
 0.60 & 5.158         &  1.988        & 10.627   \\
 0.70 & 5.323         &  2.094        & 10.882   \\
 0.80 & 5.490         &  2.201        & 11.133   \\
 0.90 & 5.658         &  2.310        & 11.380   \\
 1.00 & 5.828         &  2.420        & 11.623   \\                                                          
\end{tabular}
\caption[Table]{The separatrix $p_{s}$ and the critical value 
$p_{crit}$ where $\dot{e}=$ (in parentheses, accurate to the decimals shown) 
for a variety of eccentricities and for three different black hole spins, 
$a=0.5M$,$~a=0.99M~$ and $a=-0.99M~$(retrograde orbits).}
\label{tab_sepax}
\end{table}
\end{minipage}


\begin{minipage}[t]{3in}
\begin{table}
\textwidth=7.0cm
\begin{tabular}{|c|c|c|} 
$q$ & $r_{crit}/M$ & corrected value \\ 
\hline
-0.9  & 9.64 & 9.74 \\ 
-0.5  & 8.37 & 8.43 \\
0.0   & 6.68 & 6.68 \\
0.5   & 4.70 & 4.69 \\ 
0.7   & 3.76 & 3.75 \\
0.9   & 2.56 & 2.54 \\ 
0.95  & 2.03 & 2.11 \\
.99   & 1.47 & 1.55 \\
1.0   & 1.0  & 1.0  \\
\end{tabular} 
\caption[Table]{The position of the critical radius, $r_{\text{crit}}$ 
in units of $M$, for different black hole spins $a$ and zero eccentricity. 
The parameter $q=a/M$ is defined here to be negative for retrograde 
orbits and positive for prograde orbits. This table is provided as an 
erratum to Table I of Ref. \cite{dk2}, which was incorrect due to a bug 
in the part of the code calculating the fluxes of energy and angular 
momentum radiated to the black hole horizon. This data was produced using
the corrected code from the previous paper, rather than with the code of
the present paper.}
\label{tabcorr}
\end{table}
\end{minipage}


\begin{minipage}[t]{4in}
\begin{table}
\textwidth= 7.0cm
\begin{tabular}{|c|c|c|c|}  
$a/M $ & $e$ & $p/M$ & $(M/\mu)^2\dot{E}^{\infty}_{GW} $ \\  
\hline
0.95   &      0  &  10.015 &  4.966452E-05     \\
       &         &         & (4.966247E-05)  \\ \hline     
0.95   &      0  &  40.795 &  5.277469E-08   \\    
       &         &         & (5.277415E-08)  \\ \hline
0.95   &      0  & 200.698 & 1.933592E-11    \\ 
       &         &         & (1.933573E-11)  \\ \hline
0.00   & 0.7641  &   8.754 &  1.57132E-04    \\
       &         &         &  (1.57131E-04)  \\ \hline
0.00   & 0.7446  &  13.198 &  1.43629E-05    \\ 
       &         &         &  (1.43632E-05)  \\  \hline   
0.90   & 0.3731  &  12.152 &  2.3570E-05    \\    
       &         &         &  (2.3893E-05)  \\   \hline
0.90   & 0.5634  &  50.513 &  2.1211E-08     \\
       &         &         &  (2.1192E-08)   \\   \hline
0.30   & 0.6519  &  19.969 &  2.1654E-06     \\
       &         &         &  (2.1375E-06)  \\                                                         \end{tabular}
\caption[Table]{Comparing results from our radiation reaction code with 
existing results found in the literature [42],\cite{shibata_circ},\cite{cutler} 
(data in parentheses). We find excellent agreement (at the predicted level) 
for equatorial circular Kerr and eccentric Schwarzschild orbits. On the other 
hand,  there is seems to be a $ \sim 1\%$ disagreement with Shibata's results 
\cite{shibata_ecc} for equatorial eccentric orbits.}
\label{tab_comp}
\end{table}
\end{minipage}

\begin{minipage}[t]{4in}
\begin{table}
\textwidth= 7cm
\begin{tabular}{|c|c|c|c|}
$ a/M $ & $e$ & $p/M$ & $ ( \dot{E}^{\infty}_{GW} /\dot{L}^{\infty}_{GW})/\Omega_{\phi} $  \\  
\hline     
0.50   & 0.3  & 4.70  & 1.071  \\    
0.50   & 0.4  & 4.90  & 1.128  \\
0.99   & 0.3  & 1.70  & 0.984  \\
0.99   & 0.3  & 1.80  & 0.896  \\
0.99   & 0.4  & 1.80  & 0.976  \\
0.99   & 0.7  & 2.11  & 1.138 \\  
\end{tabular}
\caption{Examining the validity of the approximate, near-separatrix, formula 
$\dot{E}= \Omega_{\phi} \dot{L} $ for various zoom-whirl orbits and 
for two black hole spins. Here, only the fluxes at infinity have been considered 
(the horizon fluxes yield similar results). Typically, this relation is found to be 
accurate to fractional accuracy $10^{-1}-10^{-2}$. In such cases the orbit is so 
close to the separatrix as to require $\mu/M \ll 10^{-2}- 10^{-3}$ 
for adiabaticity to hold.} 
\label{tab3z}
\end{table}
\end{minipage}

\begin{minipage}[t]{4in}
\begin{table}
\textwidth= 7cm
\begin{tabular}{|c|c|c|c|c|}  
$ a/M $ & $e_f$ & $ e_i $ & $ M(de/dp)_{p_{s}}$ & $ \delta e /e_i$  \\  
\hline     
0.50   & 0.1  & 0.086 & -0.0414 & 0.16  \\    
0.50   & 0.3  & 0.28  & -0.1342 & 0.073 \\
0.99   & 0.1  & 0.086 & -0.1555 & 0.16   \\
0.99   & 0.3  & 0.29  & -0.2186  & 0.019  \\
-0.99  & 0.1  & 0.066 & -0.0439 & 0.52   \\
-0.99  & 0.3  & 0.28  & -0.0546  & 0.083  \\    
\end{tabular}
\caption{Upper limits on the total eccentricity gain close to the
separatrix, for given final values $e_f$ for the eccentricity. Using the
gradient $de/dp$ at the separatrix we extrapolate to the critical curve. 
In this way we obtain the eccentricity $e_i$.} 
\label{tabgain}
\end{table}
\end{minipage}

\begin{minipage}[t]{4in}
\begin{table}
\textwidth= 7cm
\begin{tabular}{|c|c|c|c|c|c|c|}  
$ a/M $ & $e$ &  $(\mu/M^2) t_c$ & $T_{r}/M $ & $T_{\phi}/M$ &  $(\mu/M) N_r$ & 
$ (\mu/M) N_{\phi} $  \\  
\hline     
0.99   & 0.1  & 0.051 & 216.48 & 18.03      & 2.3E-4  & 2.8E-3    \\
0.99   & 0.5  & 0.0018  & 276.84  & 18.02   & 6.7E-6  & 1.1E-4    \\
-0.99  & 0.1  & 5.6     & 651.48  & 181.02  & 8.6E-3  & 3.1E-2    \\
-0.99  & 0.5  & 0.79    & 718.32  & 218.14  & 1.1E-3  & 3.6E-3   \\    
\end{tabular}
\caption{Approximate data for the number of orbits in the $\dot{e} >0$ regime.
The required crossing time is $t_c$ and we have defined $N_{r,\phi}= t_c /T_{r,\phi} $.
We have calculated the periods $ T_{r,\phi}$ at $p= (p_s + p_{crit})/2$.} 
\label{tabNorb}
\end{table}
\end{minipage}


\begin{table}
\begin{tabular}{|c|c|c|c|c|c|c|c|}  
\multicolumn{2}{c|}{$a=0.5M$ } & \multicolumn{2}{c}{ } \\
 $e$ & $p/M$ & $(M/\mu)^2\dot{E}^{\infty}_{GW} $ & $(M/\mu)^2\dot{E}^{H}_{GW}$ &  
 $ (M/\mu^2)\dot{L}^{\infty}_{GW}$ & $(M/\mu^2)\dot{L}^{H}_{GW}$ & 
 $ (M/\mu) \dot{p}$ & $ (M^2/\mu)\dot{e}$  \\  
\hline
0.10  & 4.60   & 2.88029E-3  & -6.41673E-6 & 2.88686E-2 & -6.47058E-5 & -5.79612E-1 &  9.61181E-3 \\
0.10  & 5.00   & 1.81736E-3  & -4.03021E-6 & 2.06724E-2 & -4.58643E-5 & -2.37079E-1 & -3.73062E-3 \\
0.10  & 6.00   & 7.10665E-4  & -1.27547E-6 & 1.05539E-2 & -1.88243E-5 & -8.45033E-2 & -2.00595E-3 \\
0.20  & 4.70   & 3.11812E-3  & -5.71886E-6 & 2.96692E-2 & -5.63759E-5 & -5.65181E-1 &  1.04172E-2  \\
0.20  & 5.00   & 2.09142E-3  & -4.27185E-6 & 2.21228E-2 & -4.58884E-5 & -2.56971E-1 & -7.19373E-3 \\
0.20  & 6.00   & 7.78541E-4  & -1.43776E-6 & 1.08487E-2 & -1.97306E-5 & -8.53506E-2 & -3.97940E-3 \\
0.30  & 4.70   & 4.96241E-3  & -2.34257E-6 & 4.07599E-2 & -2.62670E-5 & -2.55152E-0 &  2.64021E-1 \\
0.30  & 5.00   & 2.60439E-3  & -3.79849E-6 & 2.48384E-2 & -3.98856E-5 & -3.03745E-1 & -9.63732E-3  \\
0.30  & 6.00   & 8.88282E-4  & -1.63235E-6 & 1.12981E-2 & -2.06383E-5 & -8.65523E-2 & -5.87677E-3 \\
0.40  & 4.90   & 4.52598E-3  &  3.00302E-6 & 3.62936E-2 &  9.69843E-6 & -9.12063E-1 &  3.28050E-2 \\
0.40  & 5.00   & 3.53043E-3  & -5.18531E-9 & 2.98097E-2 & -1.10071E-5 & -4.41342E-1 & -6.78246E-3  \\
0.40  & 6.00   & 1.03261E-3  & -1.68472E-6 & 1.18257E-2 & -2.03399E-5 & -8.77285E-2 & -7.62622E-3 \\
0.50  & 5.10   & 4.21594E-3  &  9.19757E-6 & 3.26383E-2 &  5.01972E-5 & -5.47629E-1 & -4.96045E-3 \\
0.50  & 5.50   & 2.11797E-3  &  6.77254E-8 & 1.89546E-2 & -9.89420E-6 & -1.60976E-1 & -1.41868E-2 \\
0.50  & 6.00   & 1.19638E-3  & -1.22374E-6 & 1.22973E-2 & -1.65642E-5 & -8.82324E-2 & -9.10645E-3 \\
\end{tabular}

\begin{tabular}{|c|c|c|c|c|c|c|c|}  
\multicolumn{2}{c|}{$a=0.99M$ } & \multicolumn{2}{c}{ } \\
 
 $e$ & $p/M$ & $(M/\mu)^2\dot{E}^{\infty}_{GW} $ & $(M/\mu)^2\dot{E}^{H}_{GW}$ &  
 $ (M/\mu^2)\dot{L}^{\infty}_{GW}$ & $(M/\mu^2)\dot{L}^{H}_{GW}$ & 
 $ (M/\mu) \dot{p}$ & $ (M^2/\mu)\dot{e}$  \\  
\hline
0.10 &  1.55 & 9.26325E-2 & -7.85155E-3 & 2.63428E-1  & -2.23134E-2 & -1.51950E-0 &  1.74361E-1  \\
0.10 &  2.00 & 4.72325E-2 & -3.16550E-3 & 1.77532E-1  & -1.18650E-2 & -4.73445E-1 & -3.72486E-2 \\
0.10 &  3.00 & 1.12400E-2 & -4.14404E-4 & 6.83347E-2  & -2.50408E-3 & -2.26534E-1 & -1.26041E-2 \\ 
0.20 &  1.62 & 9.30011E-2 & -7.62204E-3 & 2.60464E-1  & -2.13080E-2 & -1.46506E-0 &  1.57196E-1  \\ 
0.20 &  2.00 & 5.06654E-2 & -3.43868E-3 & 1.82214E-1  & -1.22541E-2 & -4.74856E-1 & -7.36986E-2 \\
0.20 &  3.00 & 1.19893E-2 & -4.59376E-4 & 6.94427E-2  & -2.60733E-3 & -2.25311E-1 & -2.48251E-2 \\
0.30 &  1.70 & 9.56364E-2 & -7.52010E-3 & 2.59798E-1  & -2.03993E-2 & -1.64242E-0 &  1.74157E-1 \\
0.30 &  2.00 & 5.63412E-2 & -3.87323E-3 & 1.89941E-1  & -1.28649E-2 & -4.78881E-1 & -1.08512E-1 \\  
0.30 &  3.00 & 1.31541E-2 & -5.29365E-4 & 7.10070E-2  & -2.76123E-3 & -2.22638E-1 & -3.62186E-2 \\ 
0.40 &  1.80 & 9.53548E-2 & -7.06881E-3 & 2.55921E-1  & -1.89132E-2 & -1.28174E-0 & -5.85767E-3 \\ 
0.40 &  2.00 & 6.42861E-2 & -4.43924E-3 & 2.00778E-1  & -1.36388E-2 & -4.90607E-1 & -1.40712E-1 \\
0.40 &  3.00 & 1.45880E-2 & -6.16038E-4 & 7.25434E-2  & -2.93630E-3 & -2.17455E-1 & -4.61972E-2 \\
0.50 &  2.00 & 7.50848E-2 & -5.10737E-3 & 2.15885E-1  & -1.45086E-2 & -5.33508E-1 & -1.68200E-1 \\
0.50 &  2.50 & 3.29957E-2 & -1.83305E-3 & 1.22205E-1  & -6.59015E-3 & -3.02830E-1 & -9.41048E-2 \\
0.50 &  3.00 & 1.60427E-2 & -7.05424E-4 & 7.32601E-2  & -3.08563E-3 & -2.08134E-1 & -5.39259E-2 \\
0.70 &  2.11 & 9.29845E-2 & -5.01552E-3 & 2.39102E-1  & -1.30762E-2 & -9.57156E-1 & -1.64115E-1  \\                  
\end{tabular}

\begin{tabular}{|c|c|c|c|c|c|c|c|}  
\multicolumn{2}{c|}{$a=-0.99M$ } & \multicolumn{2}{c}{ } \\
 
 $e$ & $p/M$ & $(M/\mu)^2\dot{E}^{\infty}_{GW} $ & $(M/\mu)^2\dot{E}^{H}_{GW}$ &  
 $ (M/\mu^2)\dot{L}^{\infty}_{GW}$ & $(M/\mu^2)\dot{L}^{H}_{GW}$ & 
 $ (M/\mu) \dot{p}$ & $ (M^2/\mu)\dot{e}$  \\  
\hline
0.10 &  9.5 & 1.22528E-4 &  1.50991E-6 &  -3.31424E-3 & -3.98335E-5 & -1.93846E-1 &  4.94557E-3 \\
0.10 & 11.0 & 4.99506E-5 &  3.39590E-7 &  -1.72497E-3 & -1.13847E-5 & -3.05501E-2 & -2.63944E-4 \\
0.20 &  9.7 & 1.40484E-4 &  2.21408E-6 &  -3.53943E-3 & -5.25871E-5 & -2.20308E-1 &  7.70198E-3 \\
0.20 & 11.0 & 5.63927E-5 &  5.01279E-7 &  -1.80663E-3 & -1.47518E-5 & -3.17662E-2 & -5.24602E-4 \\
0.30 & 10.0 & 1.49711E-4 &  2.90957E-6 &  -3.53885E-3 & -6.33565E-5 & -1.70530E-1 &  4.06284E-3 \\ 
0.30 & 11.0 & 6.74024E-5 &  8.29938E-7 &  -1.94163E-3 & -2.11328E-5 & -3.40152E-2 & -7.76820E-4 \\
0.40 & 10.3 & 1.57135E-4 &  3.72816E-6 &  -3.47105E-3 & -7.49426E-5 & -1.33052E-1 &  1.49356E-3 \\ 
0.40 & 11.0 & 8.33663E-5 &  1.42912E-6 &  -2.12770E-3 & -3.18078E-5 & -3.77928E-2 & -1.00954E-3 \\
0.50 & 10.4 & 2.46763E-4 &  8.15567E-6 &  -4.73640E-3 & -1.46459E-4 & -5.62681E-1 &  1.91167E-2 \\
0.50 & 11.0 & 1.04907E-4 &  2.47885E-6 &  -2.36298E-3 & -4.88973E-5 & -4.44577E-2 & -1.19582E-3 \\

\end{tabular}
\vspace{0.1cm}
\caption{Numerical data for the rate of change, under radiation reaction,
of $E,L$ (separately for infinity and the horizon) and $p,e$ (total amount) 
for a selection of strong-field orbits and for two black hole spins,
$a=0.5M$ (top), $a=0.99M$ (middle) and $a=-0.99M$ (bottom). 
Most of these data were used to generate the vectors in Fig.~\ref{planes}. 
In the computations we have used $ \ell_{\rm max}= 10-17 $.}
\label{tab_num}
\end{table}


\newpage

\appendix

\section{Functions that appear in the solution for $ x^2$.}

The quantity $x= L_z -aE $ satisfies the quartic equation,
\begin{equation}
 F(p,e)x^4 + N(p,e)x^2 + C(p,e)=0
\label{xeq}
\end{equation}
where
\begin{eqnarray}
F(p,e)&=& \frac{1}{p^3}[p^3 -2M(3 +e^2)p^2 + M^2(3 +e^2)^2p -4Ma^2(1-e^2)^2 ] 
\\
\nonumber \\
N(p,e)&=& \frac{2}{p}[-Mp^2 + (M^2(3+e^2) -a^2)p -Ma^2(1 +3e^2) ] 
\\
\nonumber \\
C(p,e)&=& (a^2 -Mp)^2
\label{xcoeff}
\end{eqnarray}
Defining the discriminant
\begin{equation}
\Delta_x(p,e)= N^2 -4FC= \frac{16a^2M}{p^3} [ p^4 -4M p^3 + 2 \{ 2M^{2}(1-e^2)
+ a^2 (1+e^2) \} p^2 -4Ma^2(1-e^2) p + a^4 (1-e^2)^2 ]
\label{discr}
\end{equation}
the solution for $x^2$ is,
\begin{equation}
x^2= \frac{ -N \mp \Delta^{1/2}_x}{2F}
\end{equation}
where the upper (lower) sign corresponds to prograde (retrograde) motion.


\section{Fundamental frequencies for bound equatorial orbits in Kerr geometry.}

The fact that the function $r(t)$ is periodic in time 
(with period $T_{\rm r}$) implies that the function
\begin{equation}
\frac{d\phi}{dt}= \frac{aT + \Delta x}{(r^2+a^2)T + a x\Delta}
\end{equation}
is also periodic and with the same period.
Hence, it can be expanded in a Fourier series,
\begin{equation}
\frac{d\phi}{dt}= \sum_{k=-\infty}^{+\infty} \beta_{\rm k} 
e^{-ik\Omega_{\rm r}t}
\end{equation}
By integrating this relation we get,
\begin{equation}
\phi(t)= \beta_{\rm 0} t + \sum_{k \neq 0} c_{\rm k} 
e^{-ik\Omega_{\rm r} t} + \mbox{(const)}
\end{equation}
where $c_{\rm k}= i\beta_{\rm k}/k\Omega_{\rm r} $.
This expression clearly shows that the $\phi$-motion is of ``rotation''
type \cite{goldstein}.
Making use of the integration constant to define the term $c_{\rm 0} $, 
we have
\begin{equation}
\phi(t) -\beta_{\rm 0}t= \sum_{\rm k} c_{\rm k} e^{-ik\Omega_{\rm k} t}
\end{equation}
That is, the function $ F(t)=\phi(t) -\beta_{\rm 0}t $ is periodic with
period equal to $T_{\rm r}$. As we have defined 
$\Delta\phi= \phi(t+T_{\rm r}) -\phi(t)$ we find that
\begin{equation}
 \beta_{\rm 0}= \Omega_{\phi} \equiv \frac{\Delta\phi}{T_{\rm r}}
\end{equation}
 

\section{Potentials of the Sasaki-Nakamura equation}

We give an explicit listing of the potentials $F(r)$,$U(r)$ that appear
in the Sasaki-Nakamura equation (\ref{sneq}):
\begin{eqnarray}
F(r) &=& \frac{\eta_{,r}}{\eta} \frac{\Delta}{r^2+a^2}
\\
U(r)&=& \frac{\Delta U_{\rm 1}(r)}{(r^2+a^2)^2} + G(r)^{2} + 
\frac{\Delta G_{,r}}{r^2+a^2} - F(r)G(r)
\end{eqnarray}
where the function $\eta(r)$ is given by
\begin{equation}
\eta(r)= c_{\rm 0} + c_{\rm 1}/r + c_{\rm 2}/r^2 + c_{\rm 3}/r^3+
c_{\rm 4}/r^4
\end{equation}
with the following coefficients
\begin{eqnarray}
c_{\rm 0} &=& -12i\omega M + \lambda(\lambda +2) -12a\omega(a\omega -m)
\\
c_{\rm 1} &=& 8ia[ 3a\omega -\lambda(a\omega -m)]
\\
c_{\rm 2} &=& -24iaM(a\omega -m) + 12a^2[1- 2(a\omega -m)^2]
\\
c_{\rm 3} &=& 24ia^3(a\omega -m) -24Ma^2
\\
c_{\rm 4} &=& 12a^4.
\end{eqnarray}
In addition, the functions $G(r)$ and $U_1(r)$ are 
\begin{eqnarray}
G(r) &=& -\frac{2(r-M)}{r^2+a^2} + \frac{r\Delta}{(r^2+a^2)^2}
\\
U_{\rm 1}(r) &=& V(r) + \frac{\Delta^2}{\beta} \left 
[ \left ( 2\alpha + \frac{\beta_{,r}}{\Delta} \right )_{,r} -
\frac{\eta_{,r}}{\eta} \left ( \alpha + \frac{\beta_{,r}}{\Delta}
\right ) \right ]
\\
\end{eqnarray}
with 
\begin{eqnarray}
\alpha &=& -\frac{i\beta K}{\Delta^2} + 3iK_{,r} + \lambda  + 
\frac{6\Delta}{r^2}
\\
\nonumber \\
\nonumber \\
\beta &=& 2\Delta ( -iK + r -M -2\Delta/r )
\end{eqnarray}
The functions $K(r),V(r)$ appear in the Teukolsky equation (\ref{radTeuk}).



\end{document}